# Direct Joule-Heated Non-Equilibrium Synthesis Enables High Performing Thermoelectrics

Chenguang Zhang[1]*, Jose Recatala-Gomez[1]*#, Zainul Aabdin[2]*, Yi Jiang[3]*, Luyang Jiang[1], Sze Yu Tan[2], Hong Liu[4,5], Yuting Qian[4,5,6], Coryl Jing Jun Lee[2], Sabrine Hachmioune[7], Vaishali Taneja[8], Anqi Sng[2], Pawan Kumar[2], Haiwen Dai[1], Zhiqian Lin[1], Weng Weei Tjiu[2], Fengxia Wei[2], Qianhong She[4,5], D. V. Maheswar Repaka[2]#, David Scanlon[9], Kanishka Biswas[8], Yee Kan Koh[3]#, Kedar Hippalgaonkar[1,2]#

*These authors contributed equally to this work.

#Corresponding author. Email: kedar@ntu.edu.sg, joserecatala.gomez@ntu.edu.sg, repakadvm@imre.a-star.edu.sg, kohyeekan@nus.edu.sg

[1]School of Materials Science and Engineering, Nanyang Technological University, 50 Nanyang Avenue, Singapore 639798, Singapore

[2]Institute of Materials Research and Engineering, Agency for Science Technology and Research (A*STAR), Singapore 138634, Singapore

[3]Department of Mechanical Engineering, National University of Singapore, 117576, Singapore

[4]School of Civil and Environmental Engineering, Nanyang Technological University, 50 Nanyang Avenue, Singapore 639798, Singapore

[5]Singapore Membrane Technology Centre, Nanyang Environment and Water Research Institute, Nanyang Technological University, 1 Cleantech Loop, Clean Tech One, Singapore 637141, Singapore

[6]Interdisciplinary Graduate Programme, Nanyang Technological University, 50 Nanyang Avenue, Singapore 639798, Singapore

[7]Department of Chemistry, University College London, London WC1H 0AJ, U.K.

[8]New Chemistry Unit, International Centre for Materials Science and School of Advanced Materials, Jawaharlal Nehru Centre for Advanced Scientific Research (JNCASR), Jakkur P.O., Bangalore 560064, India

[9]The School of Chemistry, University of Birmingham, Birmingham B15 2TT, U.K.

## Abstract

High-throughput synthesis of bulk inorganic materials is crucial for accelerating functional materials discovery but is hindered by slow, energy-intensive solid-state methods. We introduce Direct Joule-Heated Synthesis (DJS), a rapid, single-step and scalable solid-state synthesis technique achieving a $10^5$-fold speedup and 20,000× energy efficiency improvement over conventional synthesis. DJS enables the synthesis of dense, bulk chalcogenides ($Bi_{0.5}Sb_{1.5}Te_3$, $AgSbTe_2$), achieving a zT of 2.3 at 573 K in optimally Cd/Se co-doped $AgSbTe_2$, one of the highest for polycrystalline materials at this temperature. DJS enables optimal co-doping and rapid, non-equilibrium solidification, producing lamellar microstructures, interfacial regions, and cation-ordered nanodomains that scatter all-scale phonons, achieving ultralow lattice thermal conductivity (~0.2 W m$^{-1}$K$^{-1}$ at 573 K). DJS establishes a new benchmark for scalable and fast synthesis, accelerating functional material discovery.

## Main

Inorganic bulk materials are central to various energy and environmental friendly sustainable technologies, such as high performance electronic devices[1], efficient batteries[2,3], and solar cells[4,5] among many others. Conventional solid-state synthesis is still primarily used as a synthesis approach for bulk materials.[6–9] Obtaining single phase with high densities involves several long annealing and sintering steps as they drive through the diffusion of constituent elements. Typically, the preparation of fully dense materials involves two unconnected, and serially applied steps: 1) annealing above a threshold temperature (for instance the Tammann temperature, two-thirds of the melting point of the lowest-melting precursor) to enable formation of desired materials and 2) densification at high temperature/pressure to obtain a dense bulk material.[8] To circumvent this slow process, non-equilibrium synthesis strategies have been investigated to achieve controllable accelerated synthesis of inorganic materials.[10,11] Ultrafast high-temperature sintering (UHS) developed by Wang *et al.* consists of short sintering (tens of seconds) cold-pressed powder precursors sandwiched in between heating elements, usually carbon paper.[12] While applicable to any material (metal, insulator, semiconductor), UHS requires a pressed ingot/pellet as starting material. Flash Joule heating (FJH) has been used to rapidly transform amorphous carbon into high-quality graphene by heating it to over 3,000 K using a high-voltage electric discharge, achieving rapid transformation in less than 100 milliseconds.[13–15] Photonic sintering densifies soft material mixtures under ambient conditions by subjecting them to light flashes. Despite a variety of applications these techniques enable, none achieve dense, bulk inorganic materials, which thereby prevents them from being used in the synthesis of pristine and doped materials.

One application that could benefit from the fast synthesis of dense materials is thermoelectrics (TE). TE materials convert heat-into-electricity and vice versa, which makes them a promising technology for sustainably scavenging waste heat or for solid-state cooling.[16] The efficiency of a TE material is related to the dimensionless figure-of-merit ($zT$) defined as $zT = S^2\sigma T/\kappa$, where $S$ is the Seebeck coefficient, $\sigma$ is the electrical conductivity and $\kappa$ is the thermal conductivity, all defined at a thermodynamic temperature $T$.[16] Several materials have been explored for TE applications at intermediate temperatures (500 - 900 K), such as high entropy GeTe[17], SrTe-doped PbTe[18], In-Te-doped Bi-alloyed $Mg_3Sb_2$[19], Cl-doped and Pb-alloyed SnSe[20]. Performance enhancement strategies rely on doping/alloying to (a) optimize carrier concentration and electronic structure for improved power factor ($S^2\sigma$) and (b) introduce mass/strain fluctuations and nanostructuring to lower lattice thermal conductivity. These two interconnected knobs make TE a challenging problem. Electronic transport properties ($S$, $\sigma$, $\kappa_e$) are highly sensitive to carrier concentration, which translates into extensive, time consuming and costly sample preparation until the optimal value is achieved through doping. On the other hand, synthesis must produce nanostructuring that effectively scatters both short and midrange mean free path phonons. Mesoscale morphology such as grain boundaries, dislocations and precipitates can scatter heat-carrying phonons, but also scatter charge carriers, reducing carrier mobility and therefore resulting in a decrease in power factor. Ideally, maintaining electronic mobility in a material with a large asymmetry of density of states at the Fermi Level (at optimal doping) ensures high power factor; the microstructure produced from synthesis must not affect the electron mean free path, but scatter phonons effectively to make a phonon-glass-electron-crystal-system.[21]

At intermediate temperatures, state-of-the-art performance has been attained by enhanced atomic ordering in thermodynamically grown Cd-doped $AgSbTe_2$, with each sample undergoing a total of ~46 hours of heating and cooling steps, achieving a $zT$ of ~2.6 at ~600 K.[22] Pristine $AgSbTe_2$ is a semi-metallic disordered system with Ag and Sb randomly occupying cationic sites. When 6 mol % Cd is doped, the system undergoes partial cation ordering in the local nanoscale in the cation disordered bulk matrix. This results in the formation of nanoscale superstructures (~2-4 nm), which act as scattering centers for low lying acoustic phonons with mean free paths of the same order, therefore purportedly reducing lattice thermal conductivity down to the amorphous level without reducing carrier mobility.[22] Partial cation ordering has also been observed in Yb-doped $AgSbTe_2$[23] and Hg-doped $AgSbTe_2$.[24] A challenge also lies in computationally modelling this system: density functional theory (DFT) can only predict the band structure of fully ordered $AgSbTe_2$ ($Fd\bar{3}m$), while disordered $AgSbTe_2$ ($Fm\bar{3}m$) requires large supercells, with its ground state not being established. Thus, a clear picture of the band structure evolution with doping in disordered $AgSbTe_2$ cannot be easily obtained. Finally, partial cationic ordering seems to be synthesis-dependent, observed exclusively in thermodynamically grown single crystalline $AgSbTe_2$ and polycrystalline ingots of Cd/Hg/Yb doped $AgSbTe_2$. Currently, there is no understanding of the possible correlation between synthesis conditions and partial cationic ordering.

Here, we introduce Direct Joule-Heated Synthesis (DJS) a fast, single-step solid-state synthesis technique that results in dense, large grain polycrystalline materials (Fig. 1A), and use it to systematically address these challenges. The high-throughput capability of DJS enables us to rapidly explore the microstructure and transport properties through diverse co-doping, investigate the origin of the low lattice thermal conductivity in $AgSbTe_2$, and obtain a state-of-the-art thermoelectric performance, achieving a maximum $zT$ of 2.3 at 573K for co-doped $AgSb_{1.01}Cd_{0.04}Te_{1.86}Se_{0.20}$, which matches the highest values among all reported polycrystalline doped-$AgSbTe_2$ within error bar (Fig. 1B and Table S2). Advanced materials characterization coupled with electronic and thermal transport demonstrates that DJS results in high-quality solid-state materials with unique microstructures composed of large single crystalline grains separated by interface layers. Apart from optimized $AgSbTe_2$, which is the focus of the current work, we also performed synthesis of optimized $Bi_{0.5}Sb_{1.5}Te_3$ (BST, Fig. 1B), with complementary microstructure analysis, showing the adaptability of DJS to synthesize different materials. DJS sets the benchmark for high-throughput solid-state inorganic synthesis, with broad potential across materials science, extending beyond the field of thermoelectrics and ultimately with potential to advance industries engaged in bulk material production.

**Table 1. Benchmark of rapid solid-state synthesis of inorganic materials.**

| Technique | Physical Principle | Throughput | Densified material |
|---|---|---|---|
| Traditional [8] | Radiative heating in a furnace, in air or inert atmosphere or vacuum. | 0.07 g/min | No |
| UHS [12] | A pellet of pressed precursor powders is sandwiched in between carbon strips under inert atmosphere. Reaction occurs through radiative and conductive resistive Joule heating (~10-20 A, 10s). | 1-10 g/min | ~97% (pre-pressed) |
| FAST [25] | Combination of pulsed electrical discharges, resistive heating (10 V, 600-1000 A) and pressure application (< 100 MPa) in vacuum. | 0.04 g/min | 50-80% |
| FJH [13] | Rapid (~100 ms) discharging high-voltage (400V) electric pulses of 1,000 A through a carbon element at ambient conditions. | < 1g/s | No |
| FWF [14] | Pulse current is transferred to an outer conductive vessel, which transfers heat to an inner vessel containing the target reagents at ambient conditions. | 12 g/min | No |
| DJS (this work) | Current passes through the powder mixture packed in the quartz tube at spring load force. | 12 g/min | ≥ 95% |

**Results and Discussion**

In contrast to previously reported methods (Table 1), in a DJS reaction electrical current passes directly through a reagent powder mixture, sandwiched between spring compressed carbon disks to create an electrically conducting path (Supplementary Section S1). Graphite electrodes were used to establish electrical connections with copper tape, which is connected to a power supply unit (PSU). The overall series resistance of the powder, carbon felt disks and graphite electrodes is <2 Ω. Supplementary section S1 (Fig. S2) contains a full list of components and detailed explanation of the process, shown in Supplementary Video S1. The high throughput of DJS allows us to explore several samples with different doping conditions in a week, including precursor preparation, synthesis, materials characterization and TE transport. This enables us to achieve optimal carrier concentration at a higher rate relative to the slow traditional synthesis approaches.

**Insights into a DJS reaction**

Fig. 1C presents a temperature-time ($T$-$t$) curve alongside schematics illustrating evolution of atomic rearrangement as the reaction proceeds. Joule heating initiates at the contact points between the carbon disk and powder, rapidly increasing the temperature within 5–10 seconds as current flows. Heat propagates from both sides, as captured by IR imaging (red spots, Fig. 1C). At ~450°C (~15 s), the powder transitions from dark grey to light grey, signalling phase formation (~35 s, Fig. S3). Complete melting occurs at ~60 s, at which point the current is stopped and the sample is left to cool down at ambient conditions, with a calculated average cooling rate of ~5°C/s. Compared to conventional $AgSbTe_2$ synthesis (~850°C for ~46 hours), DJS achieves bulk ingots in just one minute at ~670°C. The as-synthesized ingot is removed from the quartz tube and cut into different shapes using a diamond coated wire cutter to perform subsequent characterizations: thermoelectric measurements, X-Ray Diffraction (XRD), Electron Backscatter Diffraction (EBSD) and Transmission Electron Microscopy (TEM). Further details can be found in Supplementary Sections S1 and S3.

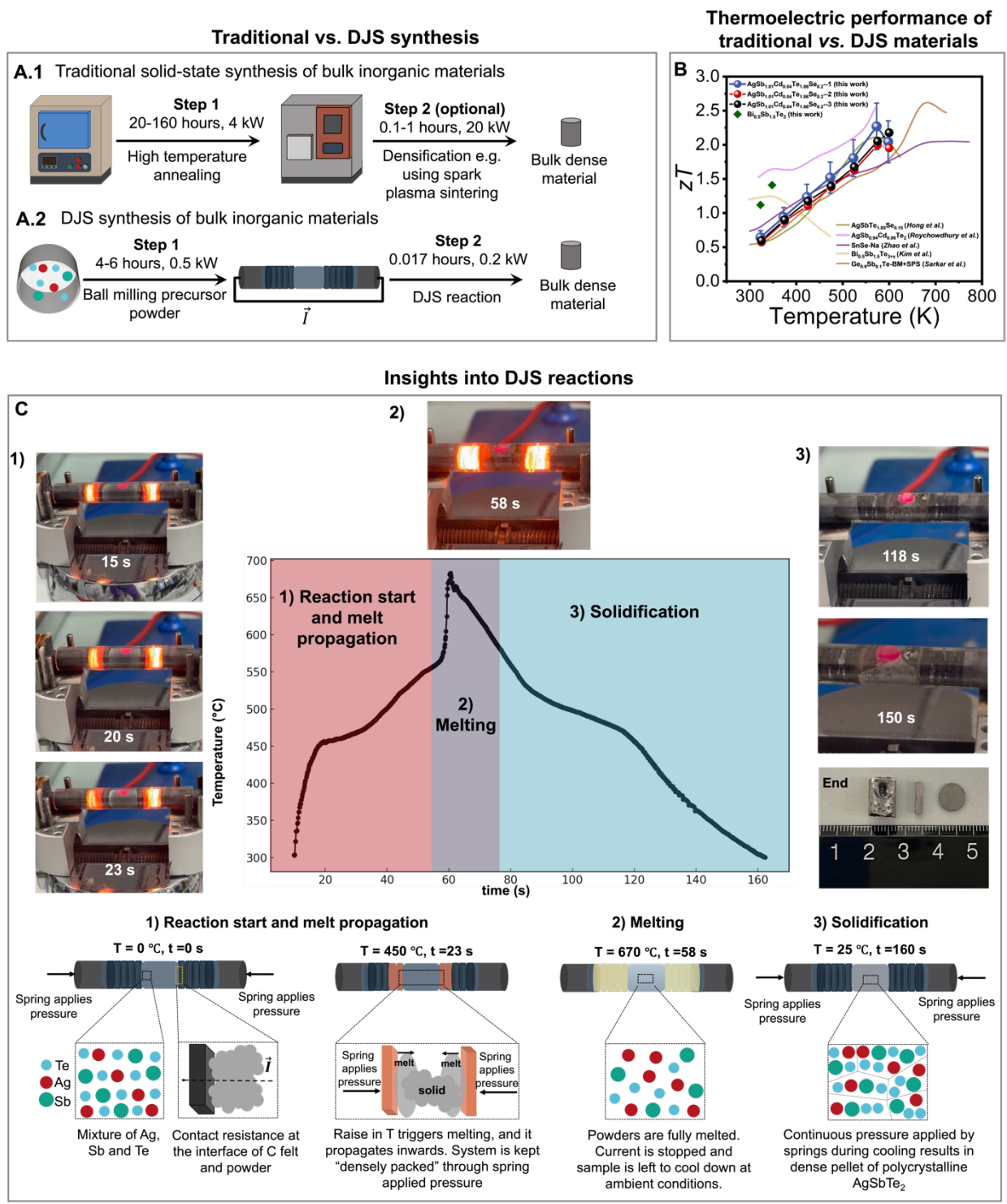

**Fig 1. Overview of Direct Joule Synthesis (DJS). (A)** Traditional vs. DJS synthesis. DJS is 20,000 times more energy efficient and $10^5$ times faster. **(B)** Dimensionless figure-of-merit ($zT$) for a series of Se co-doped AgSb$_{1.01}$Cd$_{0.04}$Te$_{1.86}$Se$_{0.20}$ samples, benchmarked against state-of-the-art previously published materials: AgSb$_{0.94}$Cd$_{0.06}$Te$_2$[22], GeTe[26], BST[27], Na-SnSe[28]. **(C)** Representative temperature vs. time curve (*T-t*) collected for AgSbTe$_2$ along with schematics of the melt-crystallization mechanism in a DJS reaction. The optical images capture important moments during a typical DJS reaction.

**Materials characterization: Observation of nanostructures and Interface layer formations**

DJS operates with ball milled powder inside a non-sealed quartz tube, causing minute amount of element loss, such as Te and Sb. To compensate, Inductively Coupled Plasma Optical Emission Spectroscopy (ICP-OES) is used to establish a baseline $AgSbTe_2$ composition with excess Sb and Te (Supplementary Section S2). Subsequently, we tune the TE properties by substituting Cd in the Sb site. The best performing Cd-doped samples were chosen, and we further enhanced the transport properties by Se co-doping at the Te site to enhance the thermoelectric performance (Methods).

Fig. 2 (Column 1) shows the structural characterization and morphology observed across different length scales for the best performing samples, namely: undoped $AgSb_{1.05}Te_{2.06}$, Cd-doped $AgSb_{1.01}Cd_{0.04}Te_{2.06}$, Se co-doped $AgSb_{1.01}Cd_{0.04}Te_{1.86}Se_{0.20}$ and undoped $AgSb_{1.05}Te_{2.06}$ densified using spark plasma sintering (SPS). Fig. 2A shows the powder XRD patterns and Rietveld refined data for the undoped $AgSb_{1.05}Te_{2.06}$. Most peaks are indexed to rock salt $AgSbTe_2$ ($Fm\bar{3}m$, PDF 00-015-0540), with a low content (8.60 %) of impurity Ag-Te binary phase precipitates indexed to monoclinic $Ag_2Te$ (*P21/c*, PDF 00-034-0142), in stark contrast with previously reported pristine $AgSbTe_2$.[22–24] As CdTe is introduced, the lattice parameter changes linearly (Fig. S4) according to Vegard's Law until the solubility limit is achieved at 4 mol % CdTe. Further doping results in CdTe precipitates, as determined using SEM-EDX (Fig. S9D). The introduction of Se in $AgSb_{1.01}Cd_{0.04}Te_{2.06-2y}Se_{2y}$ samples follows Vegard's Law for the whole doping range (Fig. S5) (y = 0.05 to 0.1), indicating good solubility of Se at Te lattice sites. Small peaks centered at 2θ 34.09º and 34.96º that observed in undoped $AgSb_{1.05}Te_{2.06}$ (SPS), 4 mol % Cd-doped, 6 mol % Cd-doped (Fig. S4A), as well as for all Se co-doped $AgSb_{1.01}Cd_{0.04}Te_{1.86}Se_{0.20}$ samples (Fig. S5A) correspond to Ag-Te binary phase precipitates indexed to $Ag_{4.96}Te_3$ (ICSD 244381). The different Ag-Te binary phases across samples may be attributed to varying composition-dependent and redissolution dynamics during non-equilibrium synthesis.[29,30] Extended characterization of these samples can be found in Supplementary Section S3.

Fig. 2 (Column 2-6) shows the detailed grain morphology and interfaces between large grains formed by DJS synthesis, *via* electron backscattered diffraction (EBSD) scans for undoped

AgSb$_{1.05}$Te$_{2.06}$ (Fig. 2B to Fig. 2F), Cd-doped AgSb$_{1.01}$Cd$_{0.04}$Te$_{2.06}$ (Fig. 2H to Fig. 2L), Se co-doped AgSb$_{1.01}$Cd$_{0.04}$Te$_{1.86}$Se$_{0.2}$ (Fig. 2N to Fig. 2R) and undoped AgSb$_{1.05}$Te$_{2.06}$ (SPS, Fig. 2T to Fig. 2X). At low magnification, the band contrast image of undoped AgSb$_{1.05}$Te$_{2.06}$ (Fig. 2B) reveals large grains (typically exceeding 20 μm), a trend consistently observed across all measured DJS samples (Fig. 2B, Fig. 2H, and Fig. 2N). These grains are separated by distinct Interface Layers (ILs), defined here as boundary regions corresponding to different phases/grains, resembling the microstructure seen in melt-crystallized samples.[31]

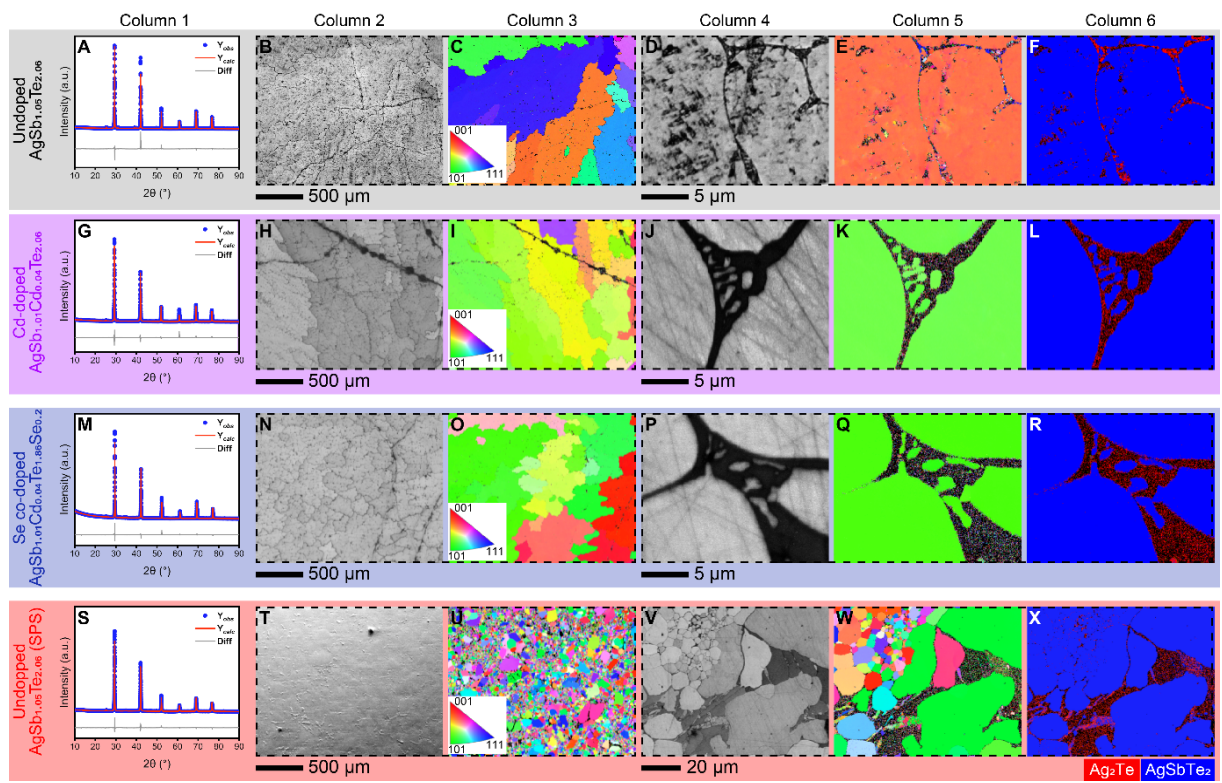

**Fig. 2. Structural and morphological characterization of DJS samples.** Rows correspond to undoped AgSb$_{1.05}$Te$_{2.06}$ (**A–F**), Cd-doped AgSb$_{1.01}$Cd$_{0.04}$Te$_{2.06}$ and (**G–L**), Se co-doped AgSb$_{1.01}$Cd$_{0.04}$Te$_{1.86}$Se$_{0.20}$ (**M–R**) and Undoped AgSb$_{1.05}$Te$_{2.06}$ (SPS) for comparison (**S-X**). **Column 1:** Rietveld-refined diffraction patterns. The respective values of $R_{wp}$ were 7.45 for A, 6.87 for G, 6.28 for M and 5.81 for S, respectively; **Column 2:** low magnification band contrast image; **Column 3:** inverse pole figure (IPF) map corresponding to Column 2; while the DJS samples show large grains (**C, I, O**), SPS samples (**U**) show an order of magnitude smaller grain sizes; **Column 4:** high-magnification band contrast image focusing on interface regions between large grains; **Column 5:** IPF map corresponding to Column 4; **Column 6:** phase map corresponding to Column 4. The diffraction peaks in (A, G, M, S) confirm the AgSbTe$_2$ rocksalt structure (PDF 00-015-0540). Band contrast images (B, H, N, T) show large grains, separated by interfacial layers. IPF maps (C, I, O, U) highlight the different crystallographic orientation of the grains. High-magnification images (D, J, P, V) reveal Ag-Te binary

nanoprecipitates in undoped AgSb$_{1.05}$Te$_{2.06}$ matrix and decreasing number of Ag$_2$Te nanoprecipitates for doped samples, while IPF maps (E, K, Q, W) confirm a nearly single-crystal AgSbTe$_2$ matrix for all samples. The phase maps in (F, L, R, X) reveal that large grains are indexed to rocksalt AgSbTe$_2$ (blue phase), while the precipitates at the interfacial layers are preferentially indexed to monoclinic Ag$_2$Te (red phase). The black phase corresponds to non-indexed solutions.

The ILs are typically ~100 nm to a few μm wide. The inverse pole figure (IPF) maps show grains with different crystallographic orientation, as shown in Fig. 2C, Fig. 2I and Fig. 2O, indicating that at the millimeter scale all samples are polycrystalline. Higher magnification images of undoped AgSb$_{1.05}$Te$_{2.06}$ (Fig. 2D) reveal large grains corresponding to rock salt AgSbTe$_2$ with Ag-Te binary nano-precipitates forming lamellar patterns (Fig 2F). The lamellar Ag-Te binary precipitates are also observed for Cd-doped AgSb$_{1.01}$Cd$_{0.04}$Te$_{2.06}$ (from Fig. 2J to Fig. 2L) and Se co-doped AgSb$_{1.01}$Cd$_{0.04}$Te$_{1.86}$Se$_{0.20}$ (from Fig. 2P to Fig. 2R), albeit in reduced quantity. The EBSD calculated phase fractions are 5.24%, 2.36%, 2.32% and 5.53 % for undoped AgSb$_{1.05}$Te$_{2.06}$, Cd-doped AgSb$_{1.01}$Cd$_{0.04}$Te$_{2.06}$, Se co-doped AgSb$_{1.01}$Cd$_{0.04}$Te$_{1.86}$Se$_{0.20}$ and undoped AgSb$_{1.05}$Te$_{2.06}$ (SPS), respectively (Fig. S11, Table S4). The lamellar structures observed in DJS samples have also been observed in the pseudo-binary systems Ag$_2$Te–Sb$_2$Te$_3$ and PbTe–Sb$_2$Te$_3$.[32,33] The Ag-Te binary precipitates comprising the lamellar features have low crystallinity and small size, around tens of nanometers, which could not be indexed due to EBSD resolution limit (~40 nm).[34] The microstructure of DJS samples markedly differs from that of undoped AgSb$_{1.05}$Te$_{2.06}$ densified using SPS (Fig. 2T to 2X). In the later, the microstructure is comprised of smaller grains (less than 20 μm) with different crystalline orientations. These observations highlight the distinct nature of DJS synthesis relative to previous approaches.

In Fig. 3 TEM/STEM imaging, selected area electron diffraction (SAED) and STEM-EDX maps for undoped AgSb$_{1.05}$Te$_{2.06}$, Cd-doped AgSb$_{1.01}$Cd$_{0.04}$Te$_{2.06}$ and Se co-doped AgSb$_{1.01}$Cd$_{0.04}$Te$_{1.86}$Se$_{0.20}$ are conducted for direct visualization of microstructures and defects, phase analysis and elemental distribution homogeneity. The HAADF-STEM images in Figs. 3A, 3H, 3O confirm the presence of large grains with ILs of size between 0.5 up to 5 μms as seen earlier in the EBSD images (Fig. 2). First, looking at the undoped AgSb$_{1.05}$Te$_{2.06}$ sample, high-magnification TEM/STEM images (Fig. 3B-C, Fig. S15, S16, S21 and S22) show large grains indexed to rock salt AgSbTe$_2$ phase ($Fm\bar{3}m$, PDF 00-015-0540) (Fig. 3E and 3F, Fig.

S12 and S15), and some secondary Ag-Te binary precipitates indexed to monoclinic $Ag_2Te$ (PDF 00-034-0142, Fig. 3F and Fig. S15, Fig. S21), in good agreement with XRD results (Fig. 2A, Table S6). Interestingly, only for this undoped sample, high-magnification TEM images (Fig. S15) show that the large grains found in EBSD are comprised of smaller grains (~120 nm) where twin boundaries, dislocations, Moiré fringes and stacking faults are observed. (Fig. 3D and Fig. S22) HR-TEM shows the matrix phase of $AgSbTe_2$ embedded with secondary phase of nanoscale $Ag_2Te$ particles. This is also confirmed by following STEM-EDX maps of the IL show monoclinic $Ag_2Te$, with Sb nanoparticles (~100 nm) decorating the edges of IL (Fig. 3G and Fig. S16).

Next, we look at the Cd-doped sample, Cd-doped $AgSb_{1.01}Cd_{0.04}Te_{2.06}$ (Fig. 3I-J, Fig. S17, S18, S21 and S22), which exhibits larger grains and fewer dislocations compared to the undoped samples (Fig. S21, S22). Moiré fringes are significantly reduced, likely due to suppressed growth of Ag-Te binary particles. The IL consists of lamellar Ag-Te binary precipitates whose composition preferentially matches $Ag_2Te$ (Fig. S17 and S18), while enlarged EDX maps reveal embedded CdTe nanoprecipitates (~34±10 nm). Notably, Cd-rich precipitates appear at the IL edge (Fig. 3N, Fig. S18).

Finally, the Se co-doped $AgSb_{1.01}Cd_{0.04}Te_{1.86}Se_{0.20}$ sample, similar to Cd-doped $AgSb_{1.01}Cd_{0.04}Te_{2.06}$, (Fig. 3P-Q, Fig. S19, S20, S21 and S22) also exhibits large grains. At the nanoscale, low-angle grain boundaries and dislocations are present, but Moiré fringes are absent (Fig. S21 and S22), indicating no secondary Ag-Te precipitate phases, consistent with EDX maps (Fig. S21 and S22). The IL structure contrasts sharply with the Cd-doped $AgSb_{1.01}Cd_{0.04}Te_{2.06}$, as Se preferentially incorporates into the lattice (Fig. S5), therefore being present in the large grains comprising the matrix. Relative to Cd-doped $AgSb_{1.01}Cd_{0.04}Te_{2.06}$, in this sample nanoprecipitates of CdTe disappear. Importantly, STEM-EDX quantification determines that Cd concentration increases in both the large-grained matrix (from ~1at. % to ~2at. %) and ILs (from 0~0.8at. % to 1.2~4.8at. % of $Ag_2Te$), suggesting Se enhances Cd solubility. This is well-aligned with Rietveld-refined lattice parameters vs. dopant concentration. The lower Vegard regime in Fig. S4B suggests limited Cd lattice incorporation when the dopant concentration is larger than 2 mol %, while Se fully occupies lattice positions for the explored doping range (Fig. S5B). Cd segregation at ILs may result from phase stability [33,35] and factors specific to the DJS reaction. Phase stability and microstructure are tied to dopant segregation to grain boundaries because it compensates for lattice mismatches and lowers the overall strain energy.[36] DJS is a non-equilibrium synthesis, and hence solubility limits, formation energies, along with differences in chemical potential and atomic mobility across phases and interfaces are expected to markedly differ from those observed in thermodynamic synthesis.[31,37,38]

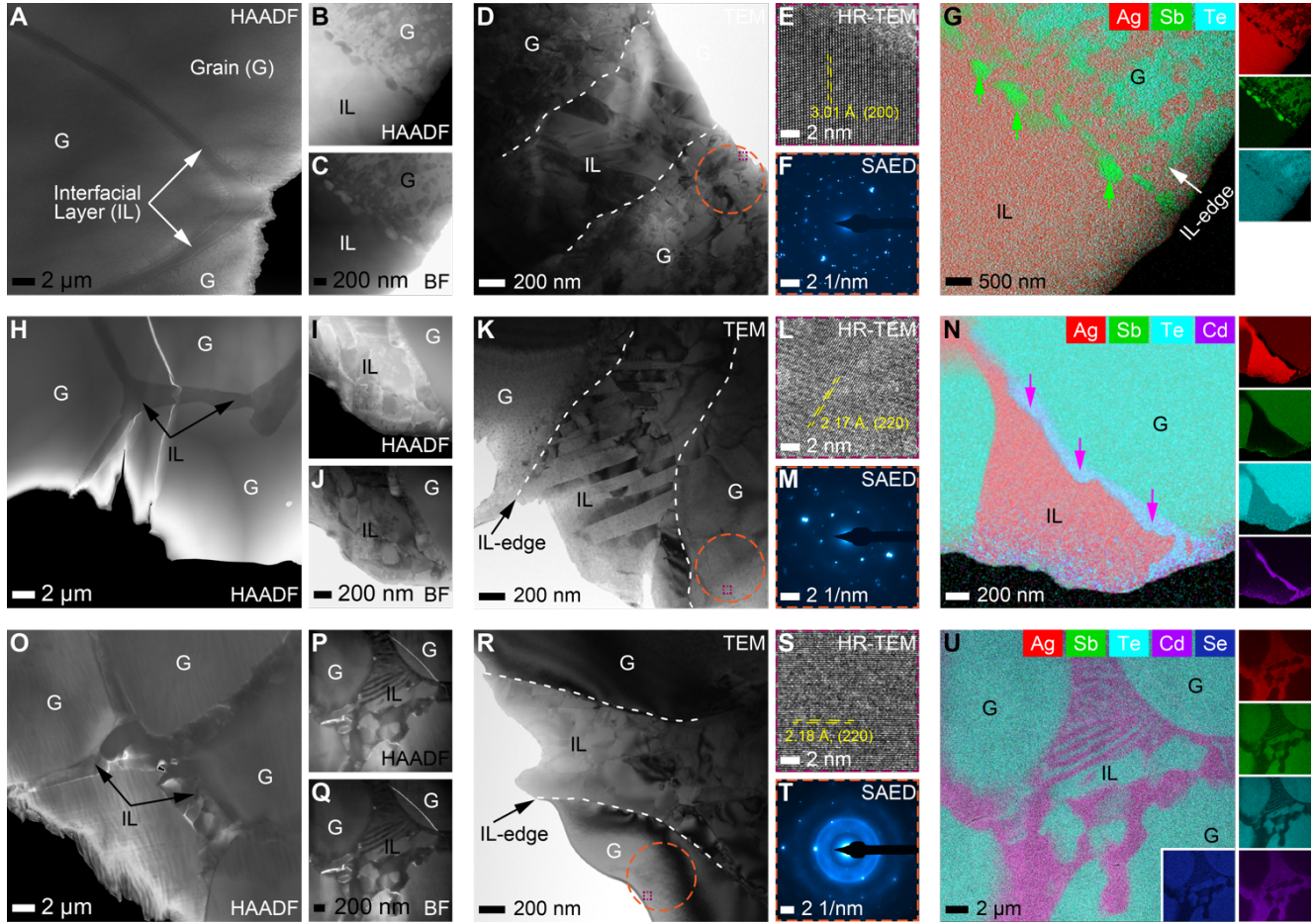

**Fig. 3. Detailed microstructural characterization of undoped, Cd-doped and Se co-doped samples.** (**A**) Low-magnification HAADF-STEM image of undoped $AgSb_{1.05}Te_{2.06}$ sample showing the interfacial-layer (IL) and grain structure. (**B**) and (**C**) High-magnification HAADF-STEM and BF-STEM images, respectively, showing the microstructure of interface. (**D**) Low-magnification TEM image of the interface. (**E**) High-resolution TEM image (red box in D) of the grain showing lattice-fringes ($d_{200} = 3.01$ Å), and (**F**) SAED of grain (orange box in D) confirming the lattice match with $AgSbTe_2$. (**G**) STEM-EDX elemental mapping overlay image showing the chemical composition of interface. Second row (**H-N**) and third row (**O-U**) shows exactly the same analysis under similar conditions for Cd-doped $AgSb_{1.01}Cd_{0.04}Te_{2.06}$ and Se co-doped $AgSb_{1.01}Cd_{0.04}Te_{1.86}Se_{0.2}$ samples, respectively. Inserts in (G, N, and U) show elemental maps of Ag (*red*), Sb (*green*), Te (*cyan*), Cd (*purple*), and Se (*blue*).

As previously discussed, Cd-rich precipitates are observed at the edges of the IL of the Cd-doped $AgSb_{1.01}Cd_{0.04}Te_{2.06}$. Interestingly, HR-TEM acquired in the vicinity of the edges of the IL showed an ordered pattern (Fig.S22B and S23) and FFT measurements show satellite spots indicating partial cationic ordering. These were not observed in the matrix of both Cd-doped AgSb$_{1.01}$Cd$_{0.04}$Te$_{2.06}$ and Se co-doped AgSb$_{1.01}$Cd$_{0.04}$Te$_{1.86}$Se$_{0.20}$ samples. The absence of cationic ordering (Fig. 3M and 3T) could be attributed to the relatively low concentration of Cd in the matrix (~1 at.% for AgSb$_{1.01}$Cd$_{0.04}$Te$_{2.06}$ and ~2 at.% for AgSb$_{1.01}$Cd$_{0.04}$Te$_{1.86}$Se$_{0.20}$ instead of the nominal 4 at.%, Fig. S18 and S20), compared to 6 mol % Cd doped AgSb$_{0.94}$Cd$_{0.06}$Te$_2$ by Roychowdhury et al.[22]

**High-throughput maximization of thermoelectric figure-of-merit (*zT*)**

First, the TE properties of compensated undoped AgSb$_{1.05}$Te$_{2.06}$ are evaluated (Supplementary Section S5, Fig. S24). Notably, the temperature dependence of the electrical conductivity follows a non-monotonic trend, with a minimum at 425 K (Fig. S24B), which indicates ambipolar behavior (detailed theoretical calculations of ordered and disordered phases in Supplementary Section S8), consistent with literature. Cd was introduced with increasing concentration into undoped AgSb$_{1.05}$Te$_{2.06}$. As Cd doping increases, Seebeck coefficient decreases, and electrical conductivity increases with increasing temperature (Fig. S26). This indicates that the carrier concentration increases, in line with previous reports.[22] Overall, Cd doping increases power factor while reducing lattice thermal conductivity, resulting in an increase in *zT*, achieving a maximum value of 1.5 at 573 K for Cd-doped AgSb$_{1.01}$Cd$_{0.04}$Te$_{2.06}$.

Selenium is substituted in the Te site to further increase the TE performance, resulting the first instance of cation-and-anion co-doped AgSbTe$_2$. Fig. 4 shows the temperature dependent thermoelectric transport properties of AgSb$_{1.01}$Cd$_{0.04}$Te$_{2.06-y}$Se$_y$ (y = 0 to 0.20). At room temperature, an increase in the Seebeck coefficient for y = 0.05, 0.1 and decreases for y > 0.1 (Fig. 4A) is observed.[39] The Seebeck coefficient also increases with increasing temperature, observing an upturn at ~420 K. The peak in the Seebeck shifts with Se co-doping, indicating a change in the quasiparticle bandgap.[40,41] The electrical conductivity shows a monotonic increase with temperature, indicative of semiconducting behavior. The co-doped samples show further increase in the power factor and decreased thermal conductivity. The power factor (Fig. 4C) reaches a maximum value of ~17 μW cm$^{-1}$ K$^{-2}$ at 573 K for AgSb$_{1.01}$Cd$_{0.04}$Te$_{1.86}$Se$_{0.20}$. The introduction of Se lowers the total thermal conductivity (Fig. 4D), achieving a minimum of ~0.4 Wm$^{-1}$K$^{-1}$ at the same temperature. Electronic thermal conductivity is extracted using an effective value of the Lorentz number obtained solving the corresponding Boltzmann Transport Equation (Eq. S4, Methods), and the results are shown in Fig. 4C. We observe a reduction of lattice thermal conductivity, $\kappa_{lat}$ to values close to the amorphous limit, 0.2 Wm$^{-1}$K$^{-1}$ when the temperature is higher than 500K, well aligned with reported values by Roychowdhury et al.[22], Taneja et al.[23] and Bhui et al.[24] The increase in power factor, accompanied by a reduction in thermal conductivity resulting in a state-of-the-art zT~2.3 at 573 K for polycrystalline Se co-doped AgSb$_{1.01}$Cd$_{0.04}$Te$_{1.86}$Se$_{0.20}$. Thermoelectric transport was shown to be highly reproducible by measuring three samples (Fig. 1B and Fig. S29). The stability of the samples with respect to thermal cycles was also tested, and we found that Seebeck values are highly repeatable, whereas electrical conductivity values tend to slightly decrease (~10%) with number of thermal cycles below 450 K but are consistent from 450-600 K (Fig. S30).

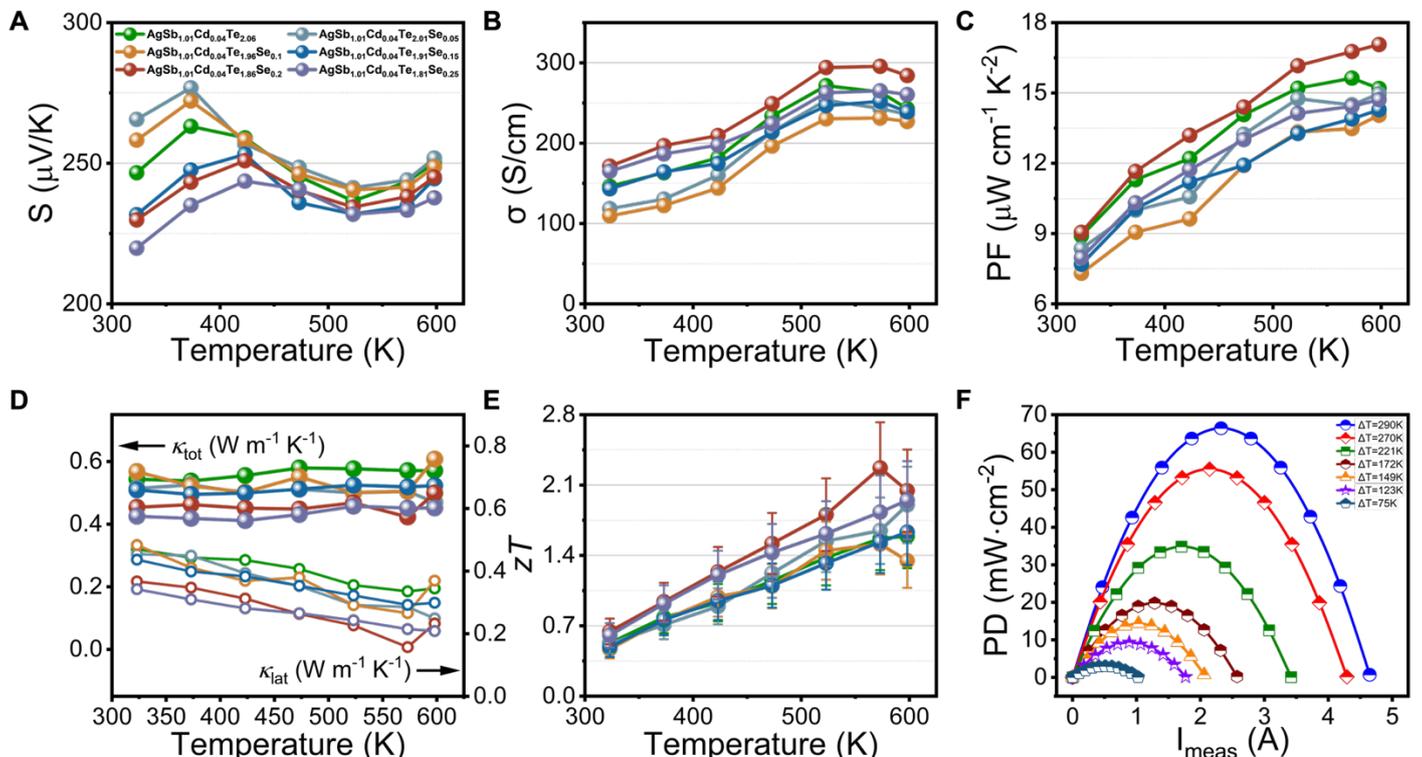

**Fig. 4. Thermoelectric transport of Se co-doped AgSb$_{1.01}$Cd$_{0.04}$Te$_{2.06-y}$Se$_y$ (y = 0 to 0.20).** (A to F) Temperature-dependent **(A)** Seebeck coefficient (S), **(B)** electrical conductivity (σ), **(C)** Power factor (PF), **(D)** total thermal conductivity ($\kappa_{tot}$) and lattice thermal conductivity ($\kappa_{lat}$), **(E)** Temperature-dependent figure-of-merit (*zT*). The uncertainty in the zT measurement is 20%, and **(F)** Experimentally measured power density from a single-leg thermoelectric module made using Se co-doped AgSb$_{1.01}$Cd$_{0.04}$Te$_{1.86}$Se$_{0.20}$.

We fabricated a single-leg thermoelectric module (Methods) using Se co-doped AgSb$_{1.01}$Cd$_{0.04}$Te$_{1.86}$Se$_{0.20}$. We measured a maximum power density value of ~70 mW cm$^{-2}$ for a temperature difference of 290 K, comparable to the power density reported by Roychowdhury et al.[22]

Hall measurements were performed for undoped AgSb$_{1.05}$Te$_{2.06}$, Cd-doped AgSb$_{1.01}$Cd$_{0.04}$Te$_{2.06}$ and Se co-doped AgSb$_{1.01}$Cd$_{0.04}$Te$_{1.86}$Se$_{0.20}$ (Supplementary Section S5). The negative Hall coefficient ($R_H$) suggests electrons as the majority carriers. This is supported by our first principles calculations, which show that Ag states are deep in the valence band and therefore, theoretically, are not expected to contribute to transport. The p-type states are dominated by Te p-states and Sb p-states near the conduction band (Fig. 35 and S36). However, the positive Seebeck coefficient confirms dominant hole transport possibly due to a high hole carrier density with low mobility originating from heavy valence bands. [42]The non-linearity in Hall resistivity and the presence of positive magnetoresistance suggest that both electrons and holes contribute to charge transport.[22] In contrast, the field-independent Hall resistivity and magnetoresistance observed in Se co-doped AgSb$_{1.01}$Cd$_{0.04}$Te$_{1.86}$Se$_{0.20}$ suggests a reduction in electron-dominated transport, likely due to an increased pseudo-bandgap induced by Se co-doping.[41,43] These results are in-line with the observed increase in the optical band gap with doping (Fig. S7). Further details on the Hall and magnetoresistance measurements are provided in the Supplementary Information S5.

Next, to understand the origin of the low thermal conductivity, we measured thermal conductivity using time-domain thermoreflectance (TDTR) for all Cd-doped samples near room temperature (~ 330 K) and compared it with laser flash analysis (LFA) measurements. Additionally, we also performed TDTR two-dimensional (2D) maps for the undoped AgSb$_{1.05}$Te$_{2.06}$ and Se co-doped AgSb$_{1.01}$Cd$_{0.04}$Te$_{1.86}$Se$_{0.2}$ samples to gain comprehensive insights into local properties of our samples. The TDTR measurements were performed with a large 1/e$^2$ radius spot size of 29 μm and a relative low modulation frequency of 1.8 MHz. With a lower modulation frequency, TDTR can probe deeper into the samples and thus effectively obtain a mean thermal conductivity over a larger volume. For each sample, we measured 4 – 5 different spots and calculated the mean values. Fig. 5A shows the mean lattice thermal conductivity values compared to the LFA measurements near room temperature for Se co-doped samples. Additional TDTR measurements of Cd-doped samples can be found in Supplementary Section S9 (Fig.S43A). We observe that both Cd doping in AgSb$_{1.05}$Te$_{2.06}$ and Se co-doping in AgSb$_{1.01}$Cd$_{0.04}$Te$_{2.06}$ are effective in suppressing thermal conductivity. This also indicates that while low-angle grain boundaries and dislocations scatter heat-carrying phonons effectively, the disappearance of Moiré fringes upon Se co-doping (Fig. S21 and S22) has no influence on the thermal transport.

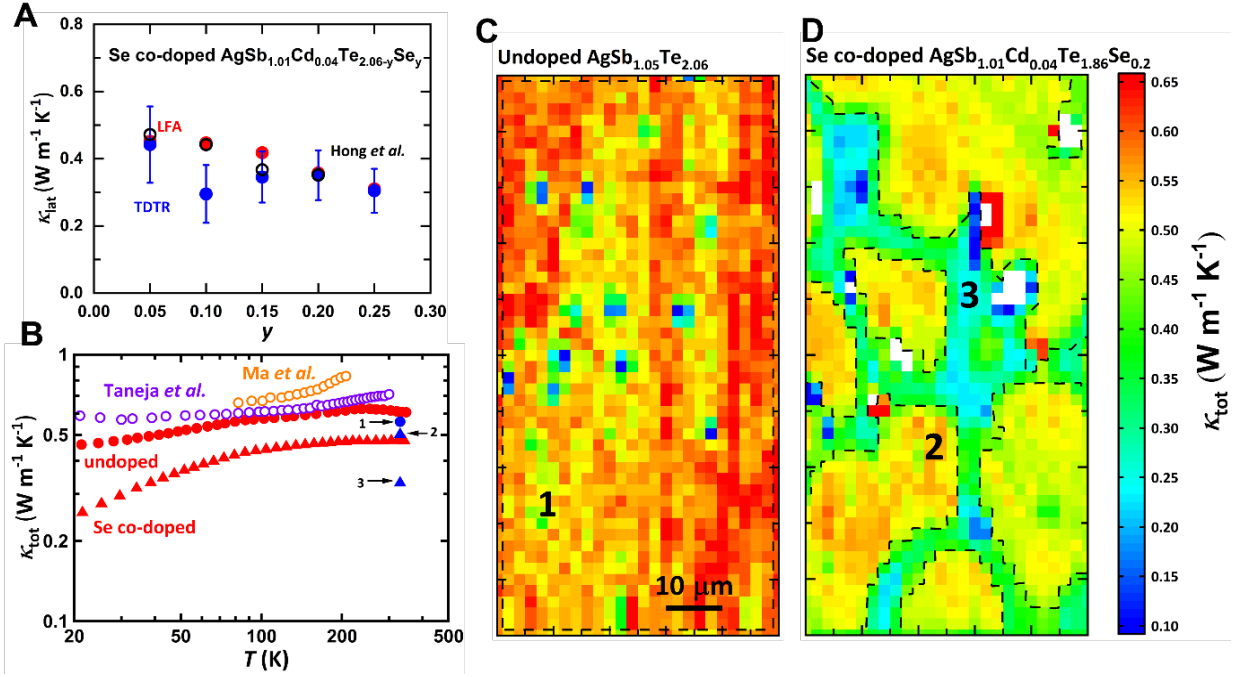

**Fig. 5. Thermal conductivity measurements. (A)** $\kappa_{lat}$ of Se co-doped $AgSb_{1.01}Cd_{0.04}Te_{2.06-y}Se_y$ series measured by TDTR (blue) and LFA (red), compared to data of $AgSbTe_{2-y}Se_y$ from Hong *et al.* (black).[44] The blue circles and the error bars represent the mean values and the standard deviations of 4 – 5 TDTR measurements performed at random locations, respectively. **(B)** Temperature dependence of $\kappa_{lat}$ of undoped $AgSb_{1.05}Te_{2.06}$ (circles) and $AgSb_{1.01}Cd_{0.04}Te_{1.86}Se_{0.20}$ (triangles) measured by LFA (red), compared to data from Ma *et al.*[45] (Orange) and Taneja *et al.*[23] (Violet), as labeled. The blue circles and triangles are the averages of $\kappa_{lat}$ of regions 1, 2, and 3 (as labeled) from the TDTR maps of undoped $AgSb_{1.05}Te_{2.06}$ **(C)** and Se co-doped $AgSb_{1.01}Cd_{0.04}Te_{1.86}Se_{0.2}$ **(D)**, respectively. Measurements with signal amplitude <60% of the maximum, (Fig. S44A) are deemed surface-defective and removed from the maps (appearing white). The dashed lines are the boundaries of the IL-matrix (further details in Supplementary Information S9).

First, we look into the possible $\kappa_{lat}$ reduction due to Rayleigh scattering by point defects, generally silver vacancies and dopant impurities for $AgSbTe_2$[46–48], which are also expected to be present in our DJS synthesized undoped, Cd-doped and Se co-doped samples. Scattering rate by Rayleigh scattering scales with $\omega^4$, where $\omega$ is the phonon frequency, and as a result, one would expect $\kappa_{lat} \sim T$ if phonons are dominantly scattered by Rayleigh scattering.[48] The weak temperature dependence occurs because Rayleigh scattering is particularly ineffective in scattering low-energy, long-wavelength phonons that dominantly carry heat at low temperatures. To test the role of point defects, we plot $\kappa_{tot}$ as a function of temperature in Fig. 5B. Since $\kappa_{tot} = \kappa_{lat} + \kappa_e$ and $\kappa_e$ is negligible below cryogenic temperatures, our observation of further reduction of $\kappa_{tot}$ even below cryogenic temperatures contrasts with the temperature dependence expected from phonon scattering dominated by Rayleigh scattering, negating this possibility.

Next, we explore whether $\kappa_{lat}$ reduction is due to phonon scattering by the ILs in $AgSb_{1.01}Cd_{0.04}Te_{1.86}Se_{0.20}$. To examine the effect of the microstructure and ILs on the thermal conductivity, we employ a TDTR mapping technique and study the spatial dependence of thermal conductivity of both undoped $AgSb_{1.05}Te_{2.06}$ and Se co-doped $AgSb_{1.01}Cd_{0.04}Te_{1.86}Se_{0.20}$ (Fig. 5C and 5D). Details of the mapping are given in the Methods section. Interestingly, we observe clear high-$\kappa_{tot}$ matrix (>10 μm) and low-$\kappa_{tot}$ IL (2-10 μm) regions in the thermal conductivity map of Se co-doped $AgSb_{1.01}Cd_{0.04}Te_{1.86}Se_{0.20}$ (Fig.5D). We plot in Fig. 5B the average $\kappa_{tot}$ of undoped $AgSb_{1.05}Te_{2.06}$ (labeled "1"), and the matrix (labeled "2") and IL (labeled "3") regions of the Se co-doped sample. We find that $\kappa_{tot}$ of the matrix region in the co-doped sample is ~20% smaller than $\kappa_{tot}$ of the undoped $AgSbTe_2$ sample. The lower thermal conductivity, despite similar chemical compositions and structures, could be due to additional scattering of long wavelength phonons with long mean-free-paths by the ILs. Importantly, the $\kappa_{lat}$ derived from the average $\kappa_{tot}$ from the thermal maps in Fig.5D are within 15% from the corresponding mean $\kappa_{lat}$ reported in Fig. 5A measured with a large spot size of 29 μm and a thermal penetration depth of ~300 nm, suggesting that ballistic phonons do not contribute significantly even in the mapping measurements, where the spot size is only 3 μm and the thermal penetration depth is only ~100 nm. When TDTR is performed with a small spot size and/or a high frequency, *i.e.* a short thermal penetration depth, the apparent thermal conductivity measured by TDTR could be significantly reduced because phonons traverse ballistically across the spot size or thermal penetration depth and do not contribute to the heat dissipation from the surface measured by TDTR.[49,50] Thus, the lack of the ballistic effects further supports the posit that long-wavelength phonons with long mean-free-paths are efficiently scattered by the ILs unique to our DJS-grown Se co-doped $AgSbTe_2$ samples. The strong scattering by ILs, where cation-ordered domains are observed, could also explain the temperature dependence we observe in Fig. 5B.

Our results suggest that the ILs are responsible for additionally scattering phonons with long mean-free-paths beyond $\kappa_{lat}$ of $AgSbTe_2$ prepared by other techniques. This opens new opportunities for further $\kappa_{lat}$ reduction through a synergistic combination of cation-ordered domains located at the interfacial layers in the materials.

**Discussion**

We demonstrate a rapid joule-heating based DJS synthesis technique on Ag-Sb-Te-based stoichiometries with controlled doping using Cd and Se to generate fully dense samples with micro- and nanostructures comparable to thermodynamically grown Bridgman $Ag_{50-x}Sb_xSe_{50-y}Te_y$ samples.[33] EBSD and HR-TEM reveal a non-equilibrium solidification induced lamellar microstructures, where an $AgSbTe_2$ matrix ($Ag_{25.5}Sb_{28.2}Te_{46.3}$) co-solidifies with Ag-Te interlayers (ILs) ($Ag_{67.3}Te_{32.7}$), similar to zone-refined $AgSbTe_2$ ingots known to influence electronic and thermal transport.[43] The matrix consists of large grains indexed to rocksalt $AgSbTe_2$ ($Fm\bar{3}m$, PDF 00-015-0540), confirmed by SAED (Fig. 3F, 3M and 3T, Fig. S21).

While matrix structures remain consistent for undoped, Cd-doped and Se co-doped samples, IL composition is shown to vary. Undoped $AgSb_{1.05}Te_{2.06}$ ILs contain $Ag_2Te$ (*P21/c*, PDF 00-034-0142) with Sb nanoparticles (~100 nm) at IL edges. In Cd-doped $AgSb_{1.01}Cd_{0.04}Te_{2.06}$, CdTe nanoparticles are found along with Ag-Te binary compounds in the ILs, most likely formed through solid-solid precipitation during cooling.[51] For Se co-doped $AgSb_{1.01}Cd_{0.04}Te_{1.86}Se_{0.20}$, Se preferentially incorporates into the matrix, enhancing Cd solubility. However, near IL edges in $AgSb_{1.01}Cd_{0.04}Te_{2.06}$, periodic modulation and FFT satellite spots indicate cation-ordered domains in IL (Fig. S23).

In terms of thermoelectric transport, Cd doping enhances power factor and suppresses lattice thermal conductivity, raising *zT* to 1.5 at 573 K. Further, Se co-doping improves power factor and drives lattice thermal conductivity toward the amorphous limit (~0.2 W m$^{-1}$ K$^{-1}$). TDTR measurements confirm that scattering of both long- and short-wavelength phonons by ILs are the primary source of this reduction, consistent with HR-TEM observations of cation-ordered domains acting as phonon scattering centres for wavelengths near their characteristic length scale.[22]

By precisely tuning doping, microstructure and disorder, DJS offers a high-throughput, scalable route to optimize bulk materials, with broader implications for phase-change memory, metal alloys, solid-state batteries, refractory ceramics and more.

# Conclusion

The Direct Joule-heated Solid-State Synthesis (DJS) technique offers a rapid, scalable and cost-effective approach for achieving fully dense polycrystalline samples. Through detailed structural analysis using XRD, EBSD and TEM, we unveil the microstructure at different length scales, including large grain growth and formation of interface layers with nano-structure precipitates, in addition to lamella consistent to eutectic solidification. These structural features play a crucial role in enhancing the thermoelectric performance and achieve a high figure-of-merit value of $zT$~2.3 at 573K for $AgSb_{1.01}Cd_{0.04}Te_{1.86}Se_{0.20}$. This competitive value of $zT$ through our DJS technique, with an acceleration factor of $10^5$ compared to conventional synthesis methods, makes DJS the state-of-the art and generalized technique for bulk inorganic materials synthesis, paving the way for next generation of thermoelectric applications.

# Methods

## Experimental setup for Direct Joule-heating Solid-State Synthesis (DJS)

Ball milled precursor powders are sandwiched between several 8 mm diameter of carbon felt with thickness of 2.5mm (Sigracell KFD 2.5EA), all inside an 8 mm quartz tube (MTI Group-Shen Zhen Kejing Star Technology Company). A graphite rod with 8 mm diameter was used to connect both carbon felt and conductive copper tape. A spring was put on one side of electrode to compensate the volume shrinkage of powder densification process. An eTOMMENS-6020C was used as a DC power source with tuneable current (0-20A) and voltage (0-60V). The temperature inside the tube was measured with an IR-gun in the range of 300-1800°C with an uncertainty of ± 0.5%.

## Materials

Silver powder (Ag, Sigma Aldrich >99.9% trace metals basis, 2-3.5μm), antimony powder (Sb, Sigma Aldrich 99.5% trace metals basis, 100mesh), cadmium telluride powder (CdTe, ANR 99.999%, 60 mesh), tellurium chunks (Te, ANR 99.999%, 3-5mm) and selenium (Se, Sigma Aldrich 99.99% trace metals basis, 100 mesh) were used for ball milling and synthesis without further purification.

## Preparation of pristine and doped $AgSbTe_2$ samples

Ball milling

Stoichiometric amounts of Ag, Sb, Te, CdTe and Se were placed inside tungsten carbide (WC) jars with a ball-to-mass ratio of 5 (10 WC balls of 10 mm diameter) for 15 g sample. Powders were milled in a planetary ball milling instrument (QM04, CHISHUN Tech) at 600 rpm for 15 min clockwise and anticlockwise. In order to keep the temperature low in the jars so as to avoid unwanted side reactions or welding, 10-minute intervals were allowed between rounds. The total time was 400 min for 10 rounds. After milling, powders were dried in a vacuum oven (<1Pa, 50 °C) for 2 hours.

DJS Reaction

About 3 grams of dried ball milled powder was loaded in the quartz tube. All DJS synthesis experiments were conducted in the ambient atmosphere. The passage current and voltage were set for 15A, 30V. Time was recorded by a timer. The synthesis was stopped by visually inspecting when full melting was achieved. The temperature *vs.* time profiles were recorded using IR gun (YCR, D30180AR).

Spark Plasma Sintering

In order to compare the microstructure of a DJS sample with that of conventional sintering techniques, we synthesize undoped $AgSb_{1.05}Te_{2.06}$ using DJS and ground the obtained pellet into powder using pestle and mortar. Subsequently, we densify the ground powder using Spark Plasma Sintering (SPS, Ed-PassIVJ, 6T-3P-30, Japan). Powders were pressed for 5 min at 350 ºC with 50 MPa pressure under vacuum.

**Materials characterization**

Scanning Electron Microscopy (SEM)

The morphology and elemental distribution of the samples were investigated using Scanning electron microscopy (SEM) using a JEOL 7800F field-emission SEM equipped with an Oxford INCA energy dispersive x-ray (EDX) detector. The accelerating voltage was 20kV. Maps were acquired for at least three minutes to investigate elemental homogeneity.

Powder X-ray Diffraction (XRD)

XRD was conducted to investigate the crystal structure of the samples using a Bruker D8 Advanced diffractometer under coupled theta-2theta geometry with a Cu $K_α$ X-ray source operated at 40 kV and 40 mA ($\lambda_{CuKα}$ = 1.54056 Å) and a beam knife for low-background low-angle signal. The following parameters were fixed for all measurements: time per step of 0.8 s/step, slit width of 0.6mm and a rotating speed of 15 rpm. Rietveld refinement was performed using TOPAS V6.

Electron Backscatter Diffraction (EBSD)

*Sample preparation*

Samples were mechanically grinded using sandpapers from P500 to P4000 each for 5 min with a rotational speed of 150 rpm, using Struers RotoPol-15 polishing machine. Samples were further final polished using Buehler's MasterPrep suspension for 30mins with a rotational speed of 40rpm. Samples were subjected to a final vibratory polishing with MasterMet suspension for 8 hours.

*Imaging and data analysis*

Electron backscattered diffraction (EBSD) measurements were performed with electron accelerating voltage of 20kV using JEOL IT500HR field emission scanning electron microscope (FESEM) equipped with Oxford Instruments Symmetry EBSD detector. Composition of samples was measured concurrently with EBSD measurements using Oxford XMax 80 mm$^2$. Post processing analysis of the EBSD data was performed using HKL Channel 5 (Oxford Instruments) software.

Transmission Electron Microscopy (TEM)

*Sample preparation*

a) Sample polishing: All samples for TEM study were prepared using standard tripod polishing on a silicon carbide grinding paper of grit sizes P400-P500. Each sample was first ground to ~20 micrometers thickness and then both surfaces were polished using fine grit till no observable scratch. After polishing, the sample was attached to a Cu ring using glue (3M structural adhesive, DP100 Plus) for further ion milling.

b) Argon Ion Milling: The electron transparency is achieved by low-angle argon ion (Ar$^+$) milling with the precision ion polishing system (PIPS, GATAN). Firstly, the incident Ar$^+$ is applied in the angle of +5° down and -5° up with an acerating voltage of 5 keV. After the specimen exhibiting a small transparent hole, the Argon milling is further applied for half an hour with a reduced angles of +3° down and -3° up and reduced accelerating voltage of 3keV. In both steps the sample was set to continuous rotation mode to avoid any preferential ion milling.

*Imaging and data analysis*

Transmission Electron Microscopy (TEM) characterizations were performed on Titan 80-300 kV TEM (Thermo Fisher Scientific, Waltham, MA USA; formerly manufactured by FEI) equipped with a 4096×4096 pixels$^2$ OneView CMOS camera (Gatan, Inc., Pleasanton, CA, USA). TEM images from many different areas and at magnifications ranging from 3800x to 490,000x were acquired using full camera resolution 4096×4096 pixels$^2$ with 1 sec exposure times. We also acquired selected area electron diffraction (SAED) from several areas using a selected area aperture of 580 nm. To capture SAED of individual grains, we used a smaller selected area aperture of about 150 nm. For all diffractions the camera length was fixed at 1.1 m and exposure time was set to 2 s. Scanning TEM (STEM) images and Energy Dispersive X-ray (EDX) mapping were acquired using Talos 80-200 kV TEM (Thermo Fisher Scientific) equipped with a 4096×4096 pixels$^2$ Thermo Fisher Scientific Ceta camera, Super-X four silicon drift detectors (SDD) for fast EDX mapping. For this study, both TEMs were operated at 200 kV. Images were processed to enhance brightness and contrast using standard microscopy software such as Gatan Digital Micrograph (DM) and open-source ImageJ (National Institutes of Health).

Optical measurements

Bruker VERTEX 80v FTIR spectrometer was used to measure optical band gap measurement. A sample powder (10 mg) mixed with KBr powder (1g) and a pellet was made for measurement. Transmission mode was used to measure the band gap. Wavelength number range from 400 cm$^{-1}$ to 7200 cm$^{-1}$ was used in measurement.

**Electronic transport measurements**

As-synthesized pellets were directly used for measurements. A diamond wire saw (STX-202AQ Lab Precision Compact Diamond Wire Saw with Pneumatic Tension System) was used as a cutting tool. Bulk ingots were cut and polished into parallelepiped shape of typical dimension ~2 mm × 2 mm × 8 mm. Temperature-dependent electrical conductivity and Seebeck coefficient were measured simultaneously under low pressure He atmosphere from room temperature to 600 K using a Cryoall CTA instrument.

The Hall and magnetoresistance measurements were conducted using the Electrical Transport Option (ETO) in a Physical Property Measurement System (PPMS), Quantum Design USA.

**Thermal transport measurements**

Pellets were cut and polished into disk shaped specimens of typical dimension > 6 mm diameter and less than 1 mm thickness. Thermal diffusivity ($D$) was directly measured in the 323 − 600 K temperature range by laser flash diffusivity method using a Netzsch LFA-457. The sample specimens were coated with a thin layer of graphite to reduce emissivity and minimize the consequent error in thermal conductivity measurement. The temperature dependent heat capacity ($C_p$) was derived using a standard sample (Pyroceram 9606) in LFA-457, which is in good agreement with the Dulong-Petit value and previously reported values. The total thermal conductivity ($\kappa_{tot}$) was calculated using the formula, $\kappa_{tot} = D\rho C_p$ where $\rho$ is the sample density. The measured sample densities were equal or greater than 97 % of theoretical density.

**Time-domain thermoreflectance (TDTR)**

We coated a ~ 100 nm Pd (for thermal measurement) or Al layer (for thermal mapping) using an e-beam evaporator onto the samples for TDTR measurements. Laser pulses from an ultrafast laser were split into a pump and a probe beam. The pump beam heated the samples periodically, creating a temperature oscillation. The pump beam periodically heated the samples, creating a temperature oscillation. The probe beam monitored this oscillation at the sample surface through thermoreflectance (*i.e.*, the change in reflectance with temperature). Since the induced temperature oscillation depends on the thermal properties of the sample, the thermal conductivity can be extracted by comparing the measured cooling curves to calculations from a thermal model.[1] In this study, we used a $1/e^2$ laser radius of 29 μm and a modulation frequency of 1.8 MHz for thermal conductivity measurements, and 2.9 μm and 9.7 MHz for thermal mapping to improve spatial resolution. For thermal conductivity measurements, the pump power is 15 mW, and the probe power is 9 mW. The uncertainty of thermal measurement was estimated to be 12%.

The thermal mappings were conducted on a 100 μm × 50 μm area on the samples. The step length is 2 μm, with a total of 1,326 points measured per scan. the total The dwell time per point was 7 seconds. The delay time was fixed at 1 ns. The pump power is 9 mW, and the probe power is 6 mW. the uncertainty of the thermal map was estimated to be ~20%.

**<u>Electronic thermal conductivity and Lorentz number calculation</u>**

Electronic thermal conductivity ($\kappa_e$) of samples is estimated by Wiedemann-Franz law, $\kappa_e = L\sigma T$, where $L$ is the Lorenz number and $\sigma$ is the electrical conductivity at temperature T.[2] Calculated the value of an effective Lorentz number ($L_{eff}$) by solving the general equation for Seebeck (Equation S1), electrical conductivity (Equation S3) and Lorentz number (Equation S4) derived from the Boltzmann Transport Equation (BTE), under scattering time approximation (STA).[3] Notably, STA is the only assumption, which means that these equations can be used for materials with any band structure regardless of complexity.[2]

The Seebeck coefficient (S) can be expressed as:

$$S = \frac{k_B}{e}\left[\frac{(s+1)F_s(\eta)}{sF_{s-1}(\eta)} - \eta\right] \quad (S1)$$

where $k_B$ is the Boltzmann constant ($1.38 \times 10^{-23}$ J·K$^{-1}$), e is the electron charge ($1.602 \times 10^{-19}$ C) s is is the energy-dependent scattering parameter, $\eta = \frac{E_F}{k_B T}$ is the reduced chemical potential, which depends on the energy of the Fermi level ($E_F$), and $F_i(\eta)$ is the *i-th* order Fermi integral (Equation S2), which can be evaluated numerically:

$$F_i(\eta) = \int_0^\infty \frac{x^i}{e^{x-\eta}+1} \tag{S2}$$

The total conductivity can be expressed as:

$$\sigma = \sigma_{E_0}(T) \times sF_{s-1}(\eta) \tag{S3}$$

where $\sigma_{E0}$ is an energy-independent transport parameter.

The Lorentz number is expressed as:

$$L = \left(\frac{k_B}{e}\right)^2 \left(\frac{(s+2)F_{s+1}(\eta)}{sF_{s-1}(\eta)} - \left[\frac{(s+1)F_s(\eta)}{sF_{s-1}(\eta)}\right]^2\right) \tag{S4}$$

Values of s are considered for the three main scattering mechanism affecting electronic transport in thermoelectric materials: for acoustic phonon scattering (APS) s takes unity value (s=1), for acoustic phonon scattering (APS) s=1, for polar optical phonon scattering (POP) s = 2 and for ionised impurity scattering (IIS) s =3.[4] Measured values of Seebeck coefficient and electrical conductivity values are used to interpolate the value of $\eta$, which in turn is used to find the effective Lorentz number for our case. This is made possible because for a fixed scattering mechanism, $L$ only depends on $\eta$. The values of Lorentz number used to extract $\kappa_e$ are $1.53 \times 10^{-8}$ $V^2$ $K^{-2}$, $1.56 \times 10^{-8}$ $V^2$ $K^{-2}$ and $1.57 \times 10^{-8}$ $V^2$ $K^{-2}$ for undoped $AgSb_{1.05}Te_{2.06}$, Cd-doped $AgSb_{1.01}Cd_{0.04}Te_{2.06}$ and Se co-doped $AgSb_{1.01}Cd_{0.04}Te_{1.86}Se_{0.20}$, respectively.

**Single leg device measurement**

A single-leg thermoelectric generator was assembled using co-doped $AgSb_{1.01}Cd_{0.04}Te_{1.86}Se_{0.20}$. The powder was pressed together with Cu/Fe end layers using spark plasma sintering (SPS) for 5min at 350 °C with 50 MPa pressure under vacuum. The dimension of the pressed sample was 10.21 mm diameter and 7.38 mm height. Thermoelectric output power of the fabricated single-leg thermoelement was estimated using mini-PEM module testing system (Advance Riko). The temperature difference across the device, ΔT, was defined as the difference between the hot-side temperature ($T_H$) and the cold-side temperature ($T_C$)

**First principles calculations**

Density functional theory (DFT) calculations were performed using the Vienna *Ab initio* Simulation Package (VASP).[5–8] VASP employs the projector augmented-wave (PAW) pseudopotential approach to treat core-valence electron interactions.[9,10]

For the $Fd\bar{3}m$ structure, calculations employed a plane-wave energy cutoff of 350 eV and a Γ-centred k-point mesh of 7 x 7 x 7. For the Fm-3m structure, an energy cutoff of 350 eV and a 10 x 5 x 5 Γ-centred k-point mesh were used. Total energies were converged to within 1 meV atom$^{-1}$. Structural optimisations of primitive unit cells for both space groups were conducted with a force convergence criterion of 0.01 eV Å$^{-1}$, employing an increased energy cutoff of 455 eV to account for Pulay stress.

For optimisation, electronic and vibrational property calculations of the $Fd\bar{3}m$ structure, the Generalised Gradient Approximation (GGA)[11,12] Perdew-Burke-Ernzerhof functional revised for solids, PBEsol,[13] was used with an effective Hubbard U parameter of 5.17 applied to the Ag atoms.[14,15] Following initial optimisation, the structure was further relaxed to a tighter convergence criterion of 0.0001 eV Å$^{-1}$. Subsequently, a 3 x 3 x 3 supercell containing 432 atoms was generated for phonon calculations using the Phonopy code.[16,17]

The disordered $Fm\bar{3}m$ structure was modelled by a special quasirandom structure (SQS) generated using the icet package.[18,19] Structural relaxation of this cell employed the Heyd-Scuseria-Ernzerhof hybrid (HSE06) functional[20], and the unfolded band structure was computed using the easyunfold package.[21]

Spin-orbit coupling effects were incorporated into density of states and band structure calculations for both systems. Results were visualised using the sumo code.[22]

The AMSET code was used to model the electronic transport properties by solving the linearised Boltzmann transport equation.[23] Both hybrid and GGA functionals were used for these calculations.

**Inductively coupled plasma optical emission spectroscopy (ICP-OES)**

The concentration ratios of Ag, Sb, Cd, Te and Se in the synthesized pellets were determined using inductively coupled plasma optical emission spectroscopy (ICP-OES, Avio 550Max, PerkinElmer). Prior to the ICP-OES analysis, the solid pellets were digested overnight in aqua regia, prepared with a 1:3 ratio of concentrated nitric acid to concentrated hydrochloric acid. The resulting solution was subsequently diluted more than tenfold with Milli-Q (MQ) water to ensure that the concentration of each element in the prepared solution was within the appropriate range. Ag standard (1000 mg/L, Sigma-Aldrich), Sb standard (1000 mg/L, Sigma-Aldrich), Te standard (1000 mg/L, Sigma-Aldrich), Cd standard (1000 mg/L, Sigma-Aldrich) and Se standard (1000 mg/L, Sigma-Aldrich) were used for ICP composition determination.

**Data and materials availability:**

All data are available in the main text or the supplementary materials.

**Acknowledgements**.

K.H. acknowledges funding from the Singapore National Research Foundation (NRF) Fellowship No. NRF-NRFF132021-0011. D.V.M.R, P.K and K.H. acknowledge funding from the MAT-GDT Program at A*STAR via the AME Programmatic Fund by the Agency for Science, Technology and Research (A*STAR) under Grant No. M24N4b0034. K.B. acknowledges Swarnajayanti Fellowship, India. Y.K.K. and Y.J. acknowledges Singapore MOE Tier Grant for funding. Z. A. acknowledges support from the Singapore National Research Foundation's Competitive Research Program (NRF-CRP23-2019-0001). C.Z. thanks Dr. Jing Lu for assistance conducing surface roughness measurements. J. R-G. thanks Dr. Samuel A. Morris for helpful discussions of the powder diffraction data. C.Z. thanks Dr. Jinfeng Dong for precision ion polishing sample preparation and discussion. Z. C and J. R-G. thank T. N-W. for facilitating data analysis.

**Contributions:**

Conceptualization: CZ, JRG, KH.

Methodology: CZ, JRG, ZA, YJ, SYT, CJJL, AS, HD, WWT, FW and DVMR.

Investigation: CZ, JRG, ZA, YJ, LJ, HL, YQ, SH, VT, PK, ZL and DVMR.

Funding acquisition: ZA, QS, DS, KB, YYK and KH.

Supervision: QS, DS, KB, YYK and KH.

Writing – original draft: CZ, JRG, YJ, DVMR and KH.

Writing – review & editing: all authors.

**Corresponding author**

Correspondence to kedar@ntu.edu.sg, joserecatala.gomez@ntu.edu.sg, repakadvm@imre.a-star.edu.sg, kohyeekan@nus.edu.sg

**Competing interests:**

Authors declare that they have no competing interests.

# Supplementary Information for

## Direct Joule-Heated Synthesis Enables High Performing Bulk Thermoelectrics

Chenguang Zhang[1]*, Jose Recatala-Gomez[1]*#, Zainul Aabdin[2]*, Yi Jiang[3]*, Luyang Jiang[1], Sze Yu Tan[2], Hong Liu[4,5], Yuting Qian[4,5,6], Coryl Jing Jun Lee[2], Sabrine Hachmioune[7], Vaishali Taneja[8], Anqi Sng[2], Pawan Kumar[2], Haiwen Dai[1], Zhiqian Lin[1], Weng Weei Tjiu[2], Fengxia Wei[2], Qianhong She[4,5], D. V. Maheswar Repaka[2#], David Scanlon[9], Kanishka Biswas[8], Yee Kan Koh[3#], Kedar Hippalgaonkar[1,2#]

*These authors contributed equally to this work

#Corresponding author. Email: kedar@ntu.edu.sg, joserecatala.gomez@ntu.edu.sg, repakadvm@imre.a-star.edu.sg, kohyeekan@nus.edu.sg

The file includes:

List of compounds prepared in this work

Extended benchmarking of the thermoelectric performance

Supplementary Section S1: Further details of DJS reactions

Supplementary Section S2: Elemental compensation of pristine $AgSbTe_2$

Supplementary Section S3: Detailed analysis of XRD and EBSD of $AgSbTe_2$ samples

Supplementary Section S4: Detailed analysis of TEM results

Supplementary Section S5: Detailed analysis of the thermoelectric transport

Supplementary Section S6: Repeatability of the thermoelectric properties

Supplementary Section S7: Complementary structural and microstructure analysis for $Bi_{0.5}Sb_{1.5}Te_3$.

Supplementary Section S8: First principles electron bandstructure and phonon calculations and detailed analysis of electronic transport

Supplementary Section S9: Further TDTR Measurements and Boundary determination of thermal mapping.

Figures S1 to S44

Tables S1 to S8

Supplementary Movie Legends for Movie S1

**List of compositions prepared in this work**

The density of the samples was measured using the Archimedes method. Theoretical density of $AgSbTe_2$ is 7.16 g cm$^{-3}$.

Table S1. Measured density of synthesized samples.

| Composition | Density (g cm$^{-3}$) | Relative Percentage |
|---|---|---|
| $AgSbTe_2$ | 7.08 | 98.9 |
| $AgSb_{1.03}Te_{2.03}$ | 6.96 | 97.2 |
| $AgSb_{1.05}Te_{2.05}$ | 7.05 | 98.6 |
| $AgSb_{1.05}Te_{2.06}$ | 7.05 | 98.6 |
| $AgSb_{1.04}Cd_{0.01}Te_{2.06}$ | 7.02 | 98.1 |
| $AgSb_{1.03}Cd_{0.02}Te_{2.06}$ | 7.03 | 98.3 |
| $AgSb_{1.01}Cd_{0.04}Te_{2.06}$ | 7.02 | 98.3 |
| $AgSb_{0.99}Cd_{0.06}Te_{2.06}$ | 7.01 | 98.2 |
| $AgSb_{1.01}Cd_{0.04}Te_{2.01}Se_{0.05}$ | 6.95 | 97.6 |
| $AgSb_{1.01}Cd_{0.04}Te_{1.96}Se_{0.10}$ | 6.98 | 98.2 |
| $AgSb_{1.01}Cd_{0.04}Te_{1.91}Se_{0.15}$ | 6.90 | 97.0 |
| $AgSb_{1.01}Cd_{0.04}Te_{1.86}Se_{0.20}$ | 6.88 | 97.3 |
| $AgSb_{1.01}Cd_{0.04}Te_{1.81}Se_{0.25}$ | 6.78 | 96.0 |

**Extended benchmarking of the thermoelectric performance**

Table S2. Comparison of synthesis time and thermoelectric performance of DJS samples with respect to other $AgSbTe_2$ samples.

| Reference | Nominal Composition | Synthesis | Max $zT$ |
|---|---|---|---|
| (5) | $AgSb_{0.94}Cd_{0.06}Te_2$ | Furnace synthesis~2760 min | ~2.6 @ 573K |
| (6) | $AgSb_{0.96}Yb_{0.04}Te_2$ | Furnace synthesis~2760 min | ~2.4 @ 573K |
| (7) | $AgSb_{0.96}Hg_{0.04}Te_2$ | Furnace synthesis ~2760 min | ~2.4 @ 570K |
| (8) | $AgSb_{0.97}Mg_{0.03}Te_{1.95}S_{0.05}$ | Furnace synthesis ~3480 min | ~1.96 @ 600K |
| (9) | $(AgSbTe_2)_{0.98}(AgAlSe_2)_{0.02}$ | Furnace synthesis ~2970 min + hot press 50 min | ~1.9 @ 623K |
| (10) | $AgSbTe_{1.85}Se_{0.1}S_{0.05}$ | Furnace synthesis ~1320 min + SPS 2 min | ~2.3 @ 673K |
| (11) | $AgSbTe_{1.85}Se_{0.15}$ | Furnace synthesis ~5280 min + BM 10mins + SPS 5 min | ~2.0 @ 573K |
| (12) | AgSb$_{0.98}$Ce$_{0.02}$Te$_2$ | Furnace synthesis ~600 min + Hot Press 20 min | ~1.59 @ 673K |
| (13) | AgSb$_{0.96}$Mg$_{0.02}$Ti$_{0.02}$Te$_2$ | Furnace synthesis ~2760 min + SPS 5 min | 1.45 @ 523K |
| (14) | AgSb$_{0.97}$Ca$_{0.03}$Te$_2$ | Furnace synthesis ~6300 min | 1.17 @ 623K |
| (15) | AgSb$_{0.93}$In$_{0.07}$Te$_2$ | Furnace synthesis ~9840 min | 1.35 @ 650K |
| (16) | AgMnSbTe$_3$ | Furnace synthesis ~1440 min | ~1.46 @ 823K |
| (17) | AgSbTe$_2$ | Furnace synthesis ~1216 min +SPS 5 mins | 1.15 @ 623K |
| (18) | AgSb$_{0.94}$Sn$_{0.06}$Te$_2$ | Furnace synthesis 1320 min + SPS 2 min | 2.5 @ 673 K |
| This work | Se co-doped AgSb$_{1.01}$Cd$_{0.04}$Te$_{1.86}$Se$_{0.20}$ | BM 400 mins + DJS synthesis ~ 2 mins | ~2.3 @ 573K |

**Supplementary Section S1: Further details of DJS reactions**

The preparation of a DJS sample is shown in Fig. S1. A puncher is typically used to ensure a good size match between the carbon felt disks and the internal diameter of the tube. Through trial-and-error, we found that for 3 grams of undoped or doped AgSbTe$_2$ precursor powder, 8 carbon felt disks per side results in a smooth reaction when an 8 mm diameter quartz tube is used. There are four distinct steps in a typical DJS synthesis, shown in Fig. S2. The synthesis proceeds as such. At the start, the current flows through the carbon felt and passes directly through the powder. Consequently, Joule heating is generated due to the contact resistance arising between the powder and the carbon felt disk directly in contact with it (Fig. S2A). As time increases, there is an increase in resistance with constant current (Fig. S2B). As time increases, the temperature of the sample increases and melt the loaded powder (Fig. S2B). Once the melt has been completed (Fig. S2C), the current is stopped, and the system is left to cool down to ambient temperature (Fig. S2D).

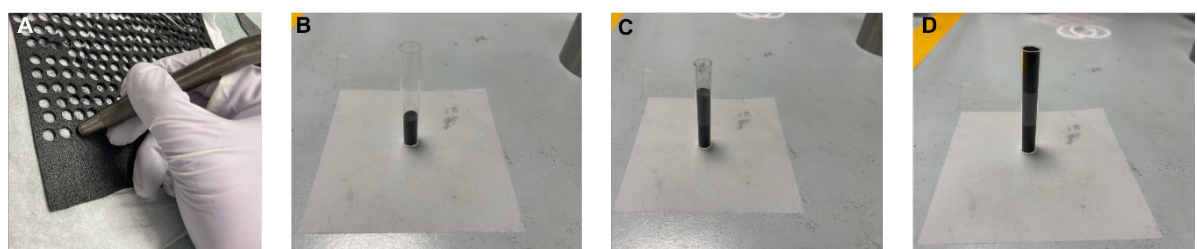

**Fig. S1. Stepwise procedure for preparing DJS samples. (A)** Carbon felt are cut using a puncher, thus ensuring a perfect size match between the carbon felt disks and the quartz tube.

Then, they are introduced in the quartz tube forming a "carbon felt column" **(B)**. Afterward, the powder that will undergo DJS reaction is introduced **(C)**. Finally, the rest of the carbon felt disks are introduced **(D)**. To ensure a smooth reaction, the whole tube must be filled with the carbon felt column – powder – carbon felt column.

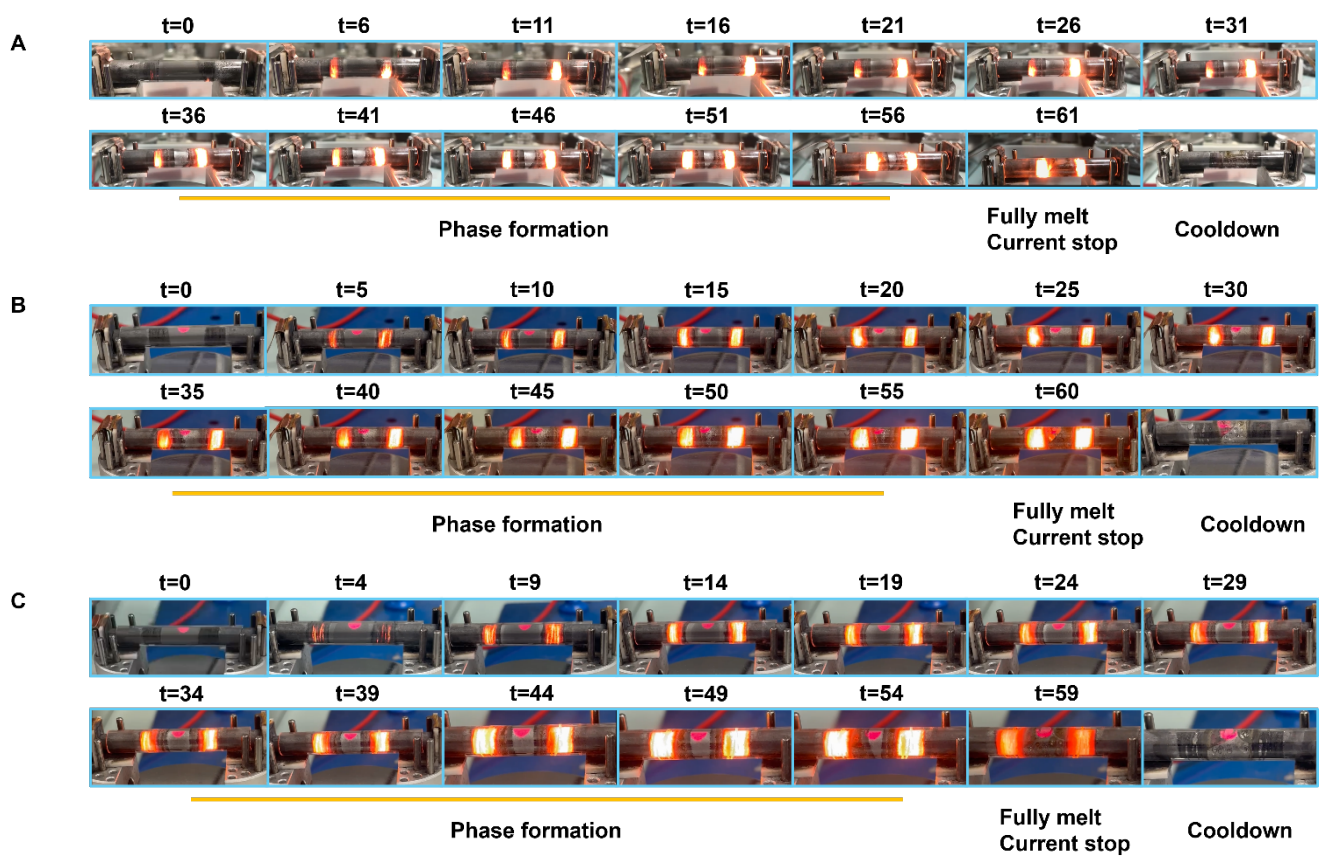

Fig S2. Repeatability of DJS synthesis of AgSbTe$_2$ compounds.

**Supplementary Section S2: Elemental compensation of pristine AgSbTe$_2$**

Compensation for samples was conducted to account for losses during DJS reaction. Initially, we conducted ICP-OES to accurately determine the concentration of Ag, Sb, Cd, Te and Se in the as-synthesized pellets. Numbers in brackets indicate the error. The standard deviation of ICP-OES is ±0.002. All numbers were normalized to Ag.

**Table S3.** Determination of the concentration of Ag and its ratios with respect to Sb, Te, Cd and Se. Ratios were calculated with respect to Ag because its concentration was normalized to 1.

| Composition | Ag | Sb/Ag | Te/Ag | Te/Sb | Cd/Ag | Se/Ag |
|---|---|---|---|---|---|---|
| AgSbTe$_2$ | 1 | 0.98(2) | 1.88(6) | 1.92(0) | | |
| AgSb$_{1.03}$Te$_{2.03}$ | 1 | 1.01(1) | 1.94(6) | 1.92(5) | - | - |
| AgSb$_{1.05}$Te$_{2.05}$ | 1 | 1.08(2) | 2.03(9) | 1.88(4) | - | - |
| AgSb$_{1.05}$Te$_{2.06}$ | 1 | 1.06(1) | 2.01(0) | 1.89(4) | - | - |
| AgSb$_{1.04}$Cd$_{0.01}$Te$_{2.06}$ | 1 | 1.00(7) | 1.94(7) | 1.93(3) | 0.011(2) | - |
| AgSb$_{1.03}$Cd$_{0.02}$Te$_{2.06}$ | 1 | 1.01(3) | 1.96(9) | 1.94(3) | 0.021(7) | - |
| AgSb$_{1.01}$Cd$_{0.04}$Te$_{2.06}$ | 1 | 1.00(2) | 1.96(7) | 1.96(1) | 0.047(7) | - |
| AgSb$_{0.99}$Cd$_{0.06}$Te$_{2.06}$ | 1 | 1.00(4) | 1.98(8) | 1.98(0) | 0.065(0) | - |
| AgSb$_{1.01}$Cd$_{0.04}$Te$_{2.01}$Se$_{0.05}$ | 1 | 0.98(9) | 1.98(5) | 2.00(7) | 0.035(1) | 0.043(1) |
| AgSb$_{1.01}$Cd$_{0.04}$Te$_{1.96}$Se$_{0.1}$ | 1 | 1.02(2) | 1.99(8) | 1.95(5) | 0.046(7) | 0.116(4) |
| AgSb$_{1.01}$Cd$_{0.04}$Te$_{1.91}$Se$_{0.15}$ | 1 | 1.03(0) | 1.98(8) | 1.92(9) | 0.045(3) | 0.170(6) |
| AgSb$_{1.01}$Cd$_{0.04}$Te$_{1.86}$Se$_{0.2}$ | 1 | 1.08(2) | 2.02(8) | 1.87(4) | 0.040(7) | 0.202(7) |

**Supplementary Section S3: Detailed analysis of XRD and EBSD of AgSbTe$_2$ samples**

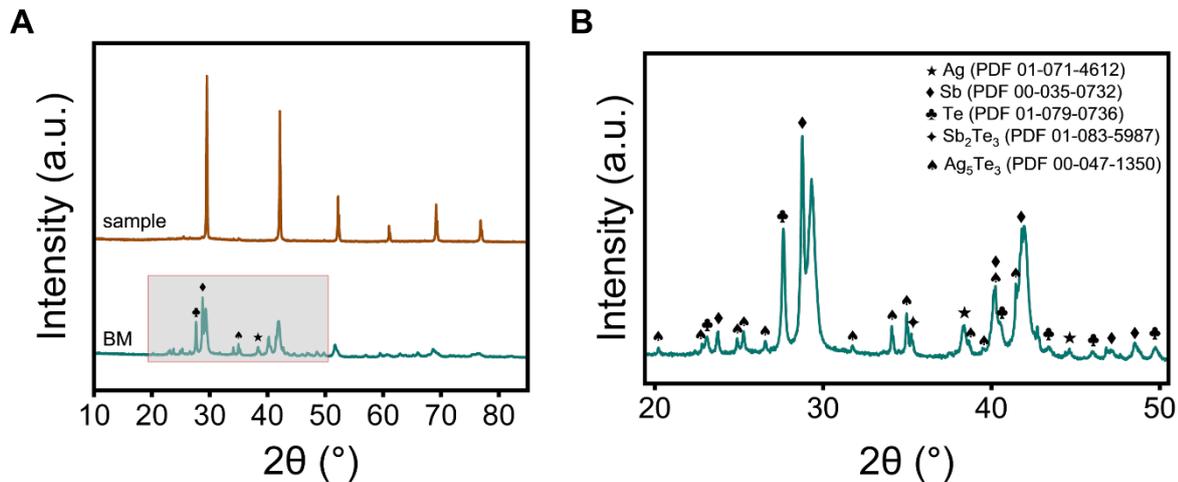

**Fig. S3. Structural evolution of AgSbTe$_2$ in DJS. Unindexed peaks represent cubic phase AgSbTe$_2$. (A)** Phase evolution over time in DJS; **(B)** Indexed impurity of ball milling precursor in zoom-in figure from 2θ of 20°- 50°.

The structure of Cd-doped AgSb$_{1.01}$Cd$_{0.04}$Te$_{2.06}$ samples is investigated using XRD. For concentrations of CdTe below 4 mol %, we observe a linear decrease in the lattice parameter, indicating that Cd occupies lattice sites. The lattice parameter remains constant for values equal or greater than 4 mol % CdTe, which indicates that Cd no longer occupies lattice sites and precipitates out. This is clearly observed in both TEM analysis (Fig. 3) and SEM-EDX analysis (Fig. S9) where CdTe precipitates are observed.

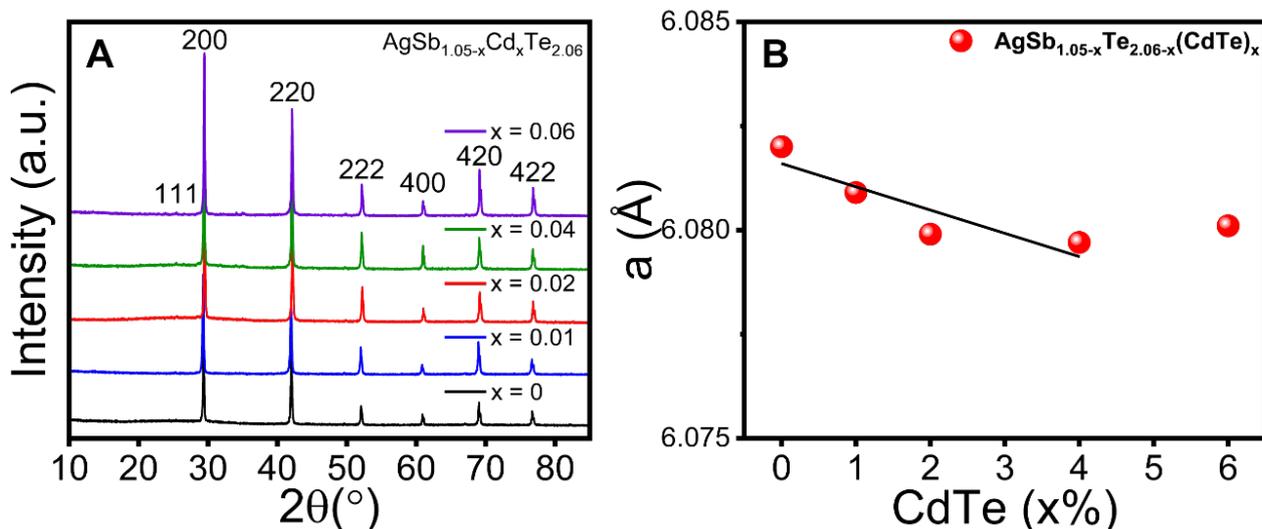

**Fig S4. (A)** Room temperature powder X-ray diffraction pattern for Cd-doped $AgSb_{1.01}Cd_{0.04}Te_{2.06}$ and **(B)** Rietveld refined lattice parameter vs. mol % dopant.

The structure of Se co-doped $AgSb_{1.01}Cd_{0.04}Te_{2.06-y}Se_y$ samples is investigated using XRD and the results are shown in Fig. S5. As Se content increases, we observe a monotonic displacement of the diffraction peaks towards higher 2θ (Fig. S6A), corresponding to a shrinkage of the unit cell, indicating that Se (198 picometers) is replacing Te (221 picometers). This is further corroborated by plotting the refined lattice parameter vs. the Se mol % (Fig. S5B). We observe a linear decrease in the lattice parameter as the Se mol % increases, following Vegard's Law, and indicating that indeed all Se occupies Te lattice sites in rocksalt $AgSbTe_2$.

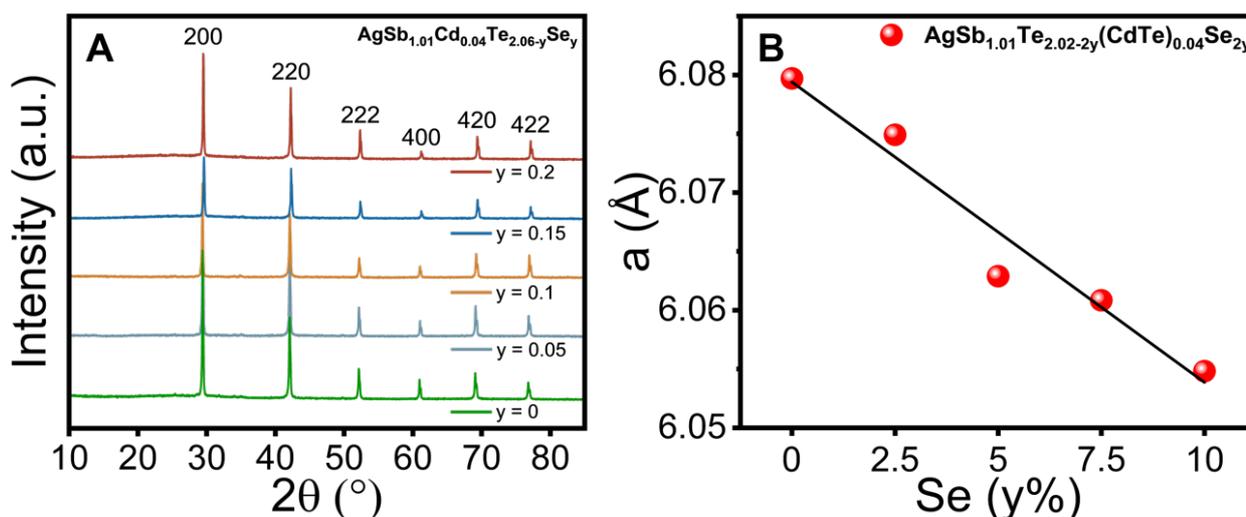

**Fig. S5. (A)** Room temperature powder X-ray diffraction patterns for Se co-doped $AgSb_{1.01}Cd_{0.04}Te_{2.06-y}Se_y$ (y=0.05, 0.1, 0.15, 0.2). **(B)** Rietveld refined lattice parameter vs. mol % dopant.

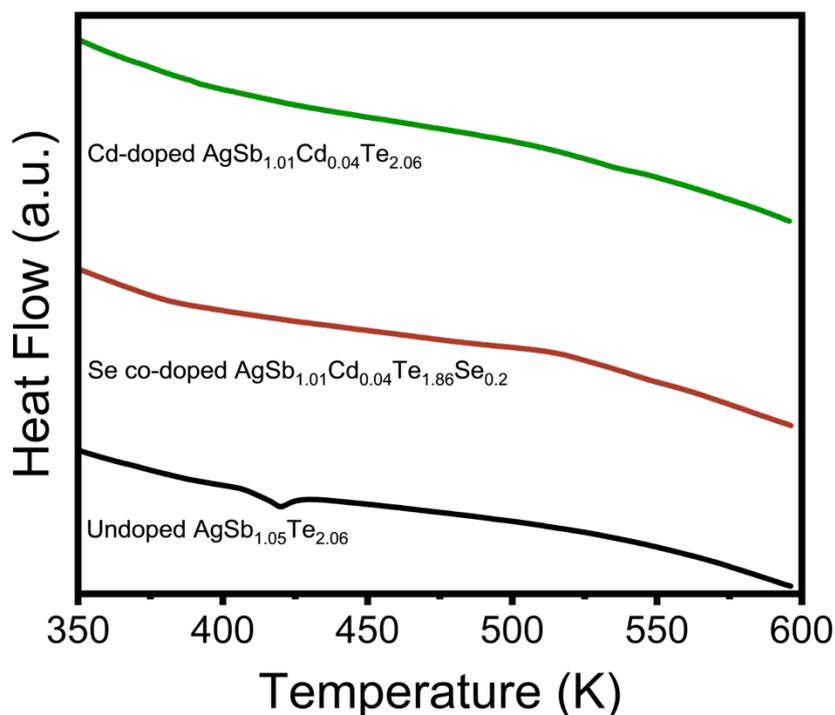

**Fig. S6.** Differential scanning calorimetry (DSC) measurements of undoped $AgSb_{1.05}Te_{2.06}$, Cd-doped $AgSb_{1.01}Cd_{0.04}Te_{2.06}$ and Se co-doped $AgSb_{1.01}Cd_{0.04}Te_{1.86}Se_{0.20}$.

Fig. S6 shows the DSC curves for undoped $AgSb_{1.05}Te_{2.06}$, Cd-doped $AgSb_{1.01}Cd_{0.04}Te_{2.06}$ and Se co-doped $AgSb_{1.01}Cd_{0.04}Te_{1.86}Se_{0.20}$. The small peak at ~420 K in the DSC curve for undoped $AgSb_{1.05}Te_{2.06}$ is attributed to the $Ag_2Te$ phase transition from monoclinic to cubic. No peak was observed in other samples, indicating that Ag-Te binary precipitates were suppressed.

Optical measurements show a bandgap increase for Cd-doped $AgSb_{1.01}Cd_{0.04}Te_{2.06}$ and Se co-doped $AgSb_{1.01}Cd_{0.04}Te_{1.86}Se_{0.2}$ from 0.13 eV to 0.30 eV (Fig. S7). Surprisingly, the bandgap almost does not change from Cd-doped $AgSb_{1.01}Cd_{0.04}Te_{2.06}$ to Se co-doped $AgSb_{1.01}Cd_{0.04}Te_{1.86}Se_{0.20}$. A possible explanation for bandgap saturation after doping 4 mol % CdTe is Fermi level pinning due to dopant states. At high doping concentration, the formation of inactive clusters traps free carriers, leading to Fermi level pinning in n-type Si.(*19*) In Cu-doped CdTe, the Fermi energy is eventually pinned when equal amounts of substitutional Cu and interstitial Cu are formed.(*20*) SEM-EDX maps (Fig. S9) show formation of clusters in Cd-doped samples, which could cause Fermi level pinning in our samples.

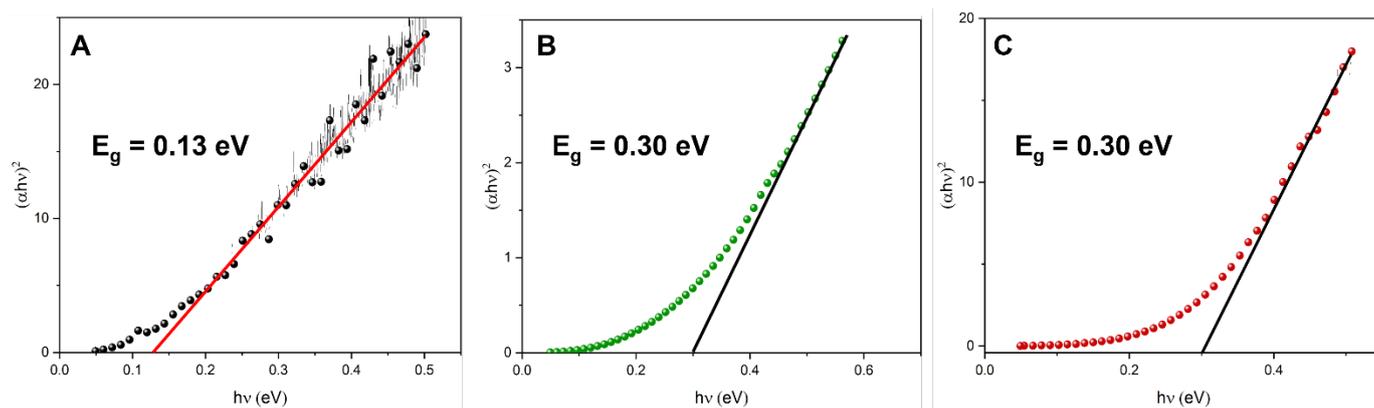

**Fig. S7.** Bandgap ($E_g$) estimation from Tauc plots of (**A**) undoped AgSb$_{1.05}$Te$_{2.06}$ (**B**) Cd-doped AgSb$_{1.01}$Cd$_{0.04}$Te$_{2.06}$ and (**C**) Se co-doped AgSb$_{1.01}$Cd$_{0.04}$Te$_{1.86}$Se$_{0.20}$. Values of the gap are given as labels within the panels.

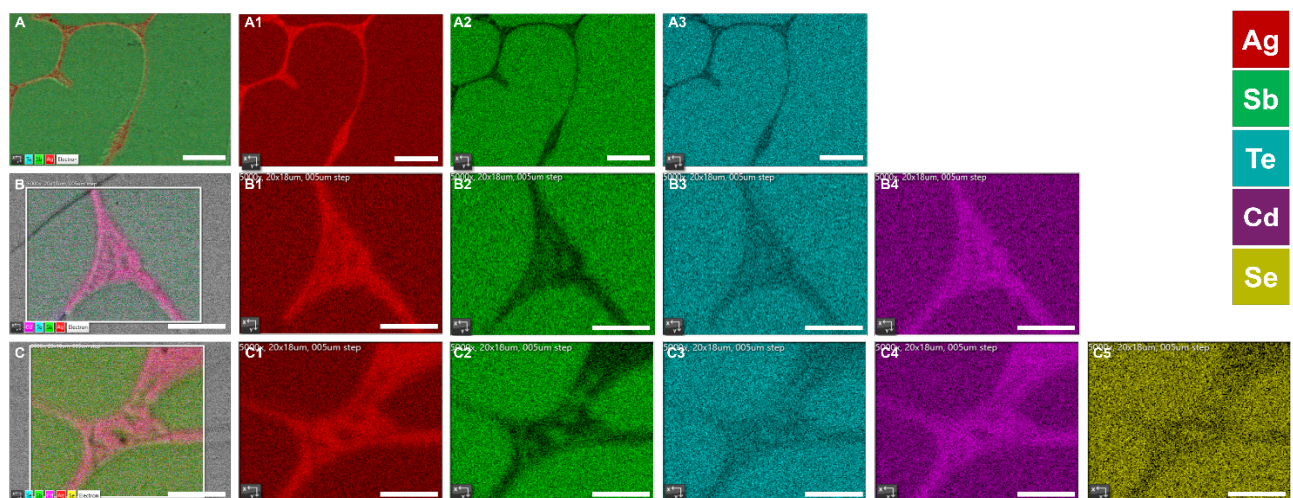

**Fig. S8.** EDX mappings corresponding to the EBSD data in the main text (**A**) undoped AgSb$_{1.05}$Te$_{2.06}$, (**B**) Cd-doped AgSb$_{1.01}$Cd$_{0.04}$Te$_{2.06}$, (**C**) Se co-doped AgSb$_{1.01}$Cd$_{0.04}$Te$_{1.86}$Se$_{0.2}$. Scale bar is 10μm.

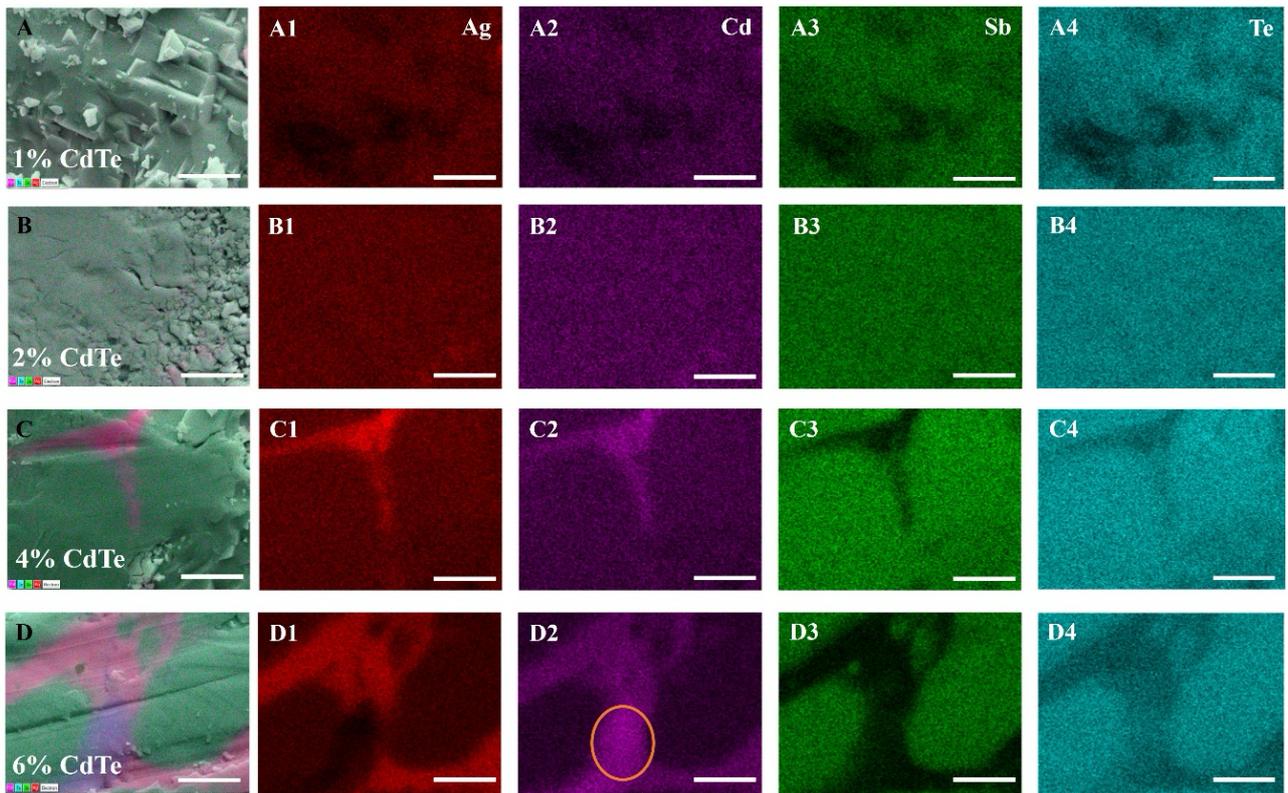

**Fig.S9**. **SEM-EDX mapping of Cd-doped AgSb$_{1.05-x}$Cd$_x$Te$_{2.06}$.** **(A)** 1 mol % CdTe. **(B)** 2 mol % CdTe. **(C)** 4 mol % CdTe. **(D)** 6 mol % CdTe. The orange circle in D1 indicates an area in which a precipitated of CdTe is found. The sale bar is 5μm.

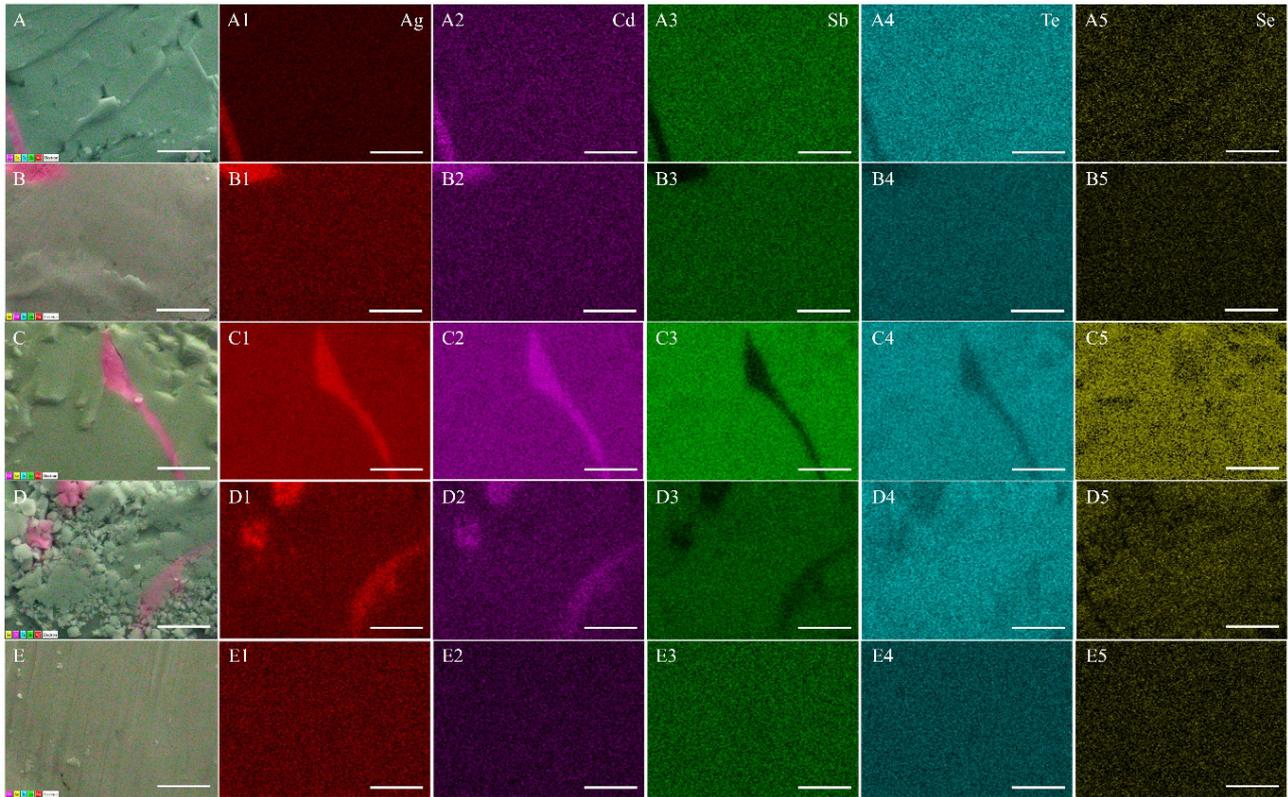

**Fig. S10**. **SEM EDX mapping of Se co-doped AgSb$_{1.01}$Cd$_{0.04}$Te$_{2.06-y}$Se$_y$. A to E** represents y=0.05, 0.1, 0.15, 0.2, 0.25, respectively. The scale bar is 5 μm.

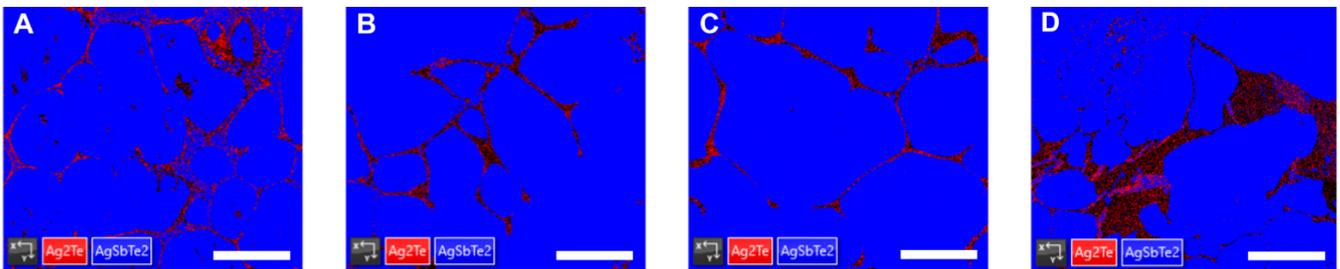

**Fig. S11.** EBSD images used to calculate phase fractions. **A - D** represents undoped AgSb$_{1.05}$Te$_{2.06}$, Cd-doped AgSb$_{1.01}$Cd$_{0.04}$Te$_{2.06}$, Se co-doped AgSb$_{1.01}$Cd$_{0.04}$Te$_{1.86}$Se$_{0.2}$ and undoped AgSb$_{1.05}$Te$_{2.06}$ (SPS), respectively. The scale bar is 25μm.

**Table S4. EBSD phase fractions for selected samples.** The phase fractions have been calculated from the images in Fig. S4 considering AgSbTe$_2$ ($Fm\bar{3}m$) and Ag$_2$Te ($P2_1/c$) as present phases.

| Composition | AgSbTe$_2$ Phase Fraction (%) | Ag$_2$Te Phase Fraction (%) | Zero solution (%) |
|---|---|---|---|
| Undoped AgSb$_{1.05}$Te$_{2.06}$ | 90.75 | 5.24 | 4.02 |
| Cd-doped AgSb$_{1.01}$Cd$_{0.04}$Te$_{2.06}$ | 94.28 | 2.36 | 3.37 |
| Se co-doped AgSb$_{1.01}$Cd$_{0.04}$Te$_{1.86}$Se$_{0.2}$ | 94.88 | 2.32 | 2.80 |
| Undoped AgSb$_{1.05}$Te$_{2.06}$ (SPS) | 83.31 | 5.53 | 11.16 |

**Table S5. EBSD phase fractions for selected samples.** The phase fractions have been calculated from the images in Fig. S4 considering AgSbTe$_2$ ($Fm\bar{3}m$), Ag$_2$Te ($P2_1/c$) and Ag$_{4.96}$Te$_3$ ($P\bar{6}2m$) as present phases. The phase fractions of Ag-Te binary precipitates corresponding to Ag$_2$Te ($P2_1/c$) and Ag$_{4.96}$Te$_3$ ($P\bar{6}2m$) through EBSD are not possible because of the similarity between phases and presence of Ag-Te binary compounds with compositions in between Ag$_2$Te and Ag$_{4.96}$Te$_3$.

| Composition | AgSbTe$_2$ Phase Fraction (%) | Ag$_2$Te Phase Fraction (%) | Ag$_{4.96}$Te$_3$ Phase Fraction (%) | Zero solution (%) |
|---|---|---|---|---|
| Undoped AgSb$_{1.05}$Te$_{2.06}$ | 78.90 | 19.90 | 0.10 | 1.81 |
| Cd-doped AgSb$_{1.01}$Cd$_{0.04}$Te$_{2.06}$ | 92.2 | 2.57 | 0.09 | 5.22 |
| Se co-doped AgSb$_{1.01}$Cd$_{0.04}$Te$_{1.86}$Se$_{0.2}$ | 94.7 | 2.62 | 1.01 | 1.67 |
| Undoped AgSb$_{1.05}$Te$_{2.06}$ (SPS) | 89.7 | 6.33 | 0.99 | 2.98 |

**Note 1:** XRD results indicate that Ag$_{4.96}$Te$_3$ is only present in Cd-doped AgSb$_{1.01}$Cd$_{0.04}$Te$_{2.06}$, Se co-doped AgSb$_{1.01}$Cd$_{0.04}$Te$_{1.86}$Se$_{0.2}$ and Undoped AgSb$_{1.05}$Te$_{2.06}$ (SPS).
**Note 2:** XRD results indicate that no Ag$_2$Te is present in Se co-doped AgSb$_{1.01}$Cd$_{0.04}$Te$_{1.86}$Se$_{0.2}$ and Undoped AgSb$_{1.05}$Te$_{2.06}$ (SPS).

**Table S6. Phase fraction for selected samples from Rietveld refinement.** Data calculated from diffraction patterns shown in Fig. 2, Column 1.

| Composition | AgSbTe$_2$ Phase Fraction (%) | Ag$_2$Te Phase Fraction (%) | Ag$_{4.96}$Te$_3$ Phase Fraction (%) | Sb$_2$Te$_3$ Phase Fraction (%) |
|---|---|---|---|---|
| Undoped AgSb$_{1.05}$Te$_{2.06}$ | 91.40 | 8.60 | - | - |
| Cd-doped AgSb$_{1.01}$Cd$_{0.04}$Te$_{2.06}$ | 92.19 | 2.92 | 4.89 | - |
| Se co-doped AgSb$_{1.01}$Cd$_{0.04}$Te$_{1.86}$Se$_{0.2}$ | 94.29 | - | 5.71 | - |
| Undoped AgSb$_{1.05}$Te$_{2.06}$ (SPS) | 92.80 | - | 6.20 | 0.99 |

**Supplementary Section S4: Detailed analysis of TEM results**

TEM/STEM imaging, selected area electron diffraction (SAED) and STEM-EDX maps for undoped AgSb$_{1.05}$Te$_{2.06}$, Cd-doped AgSb$_{1.01}$Cd$_{0.04}$Te$_{2.06}$ and Se co-doped AgSb$_{1.01}$Cd$_{0.04}$Te$_{1.86}$Se$_{0.2}$ are conducted to for a complete picture of microstructures and defects, phase analysis and elemental distribution homogeneity in the DJS samples. The following table summarizes the series of analysis performed for each set of samples. HR-TEM images (S12-S14 and S21-S23) show information about the lattice and microstructure/defects/cation ordering, while STEM-HAADF images (S15, S17, S19) provide details about the large grains and ILs. The STEM-EDX images (S16, S18, S20) provide information about elemental homogeneity.

To confirm the crystallinity of the large grains, we acquired TEM images (Fig. 3D, 3K, 3R) and SAED (Fig. 3F, 3M, 3T) patterns. SAED patterns match rocksalt AgSbTe$_2$ structure for all three samples. High-resolution TEM images and corresponding SAEDs (Fig. 3E, 3L, 3S; Fig. S21) confirm single crystalline nature of large grains with diffraction indexing and lattice fringe spacing matching well with face-centered-cubic (FCC) AgSbTe$_2$ ($Fm\bar{3}m$, PDF 00-015-0540). Most of the reflections from the matrix match (FCC) AgSbTe$_2$, with some minority phases matching monoclinic Ag$_2$Te (*P21/c*, PDF 00-034-0142) in undoped AgSb$_{1.05}$Te$_{2.06}$ and Cd-doped AgSb$_{1.01}$Cd$_{0.04}$Te$_{2.06}$.

To study in detail the phase/precipitate formation and obtain elemental quantification of the dopants, we performed STEM-EDX chemical mapping for all three samples (Fig. 3G, 3N, 3U). EDX maps of undoped AgSb$_{1.01}$Te$_{2.06}$ (Fig. S16) show that the IL is mainly comprised by Ag$_2$Te precipitates, with Sb-rich precipitates at the IL-bulk interfaces. Surprisingly, the matrix of undoped AgSb$_{1.05}$Te$_{2.06}$ is full of Ag$_2$Te nanoprecipitates (Fig. 3G, Fig. S16).

These Ag-Te binary precipitates likely introduce stress and defects in the large grains that explains the observed texture in TEM images (Fig. 3D), resembling polycrystallinity. However, SAED measurements confirm that the large grains are, in fact, monocrystalline (Fig. 3F). For the Cd-doped $AgSb_{1.01}Cd_{0.04}Te_{2.06}$, EDX mapping of the matrix (Fig. 3N, Fig. S17) shows a Cd-deficient (~1 at %) matrix, uniquely indexed to (FCC) $AgSbTe_2$. The IL of this sample is also comprised by $Ag_2Te$, although this time there is a minority phase corresponding to CdTe precipitates embedded in the $Ag_2Te$ layer.

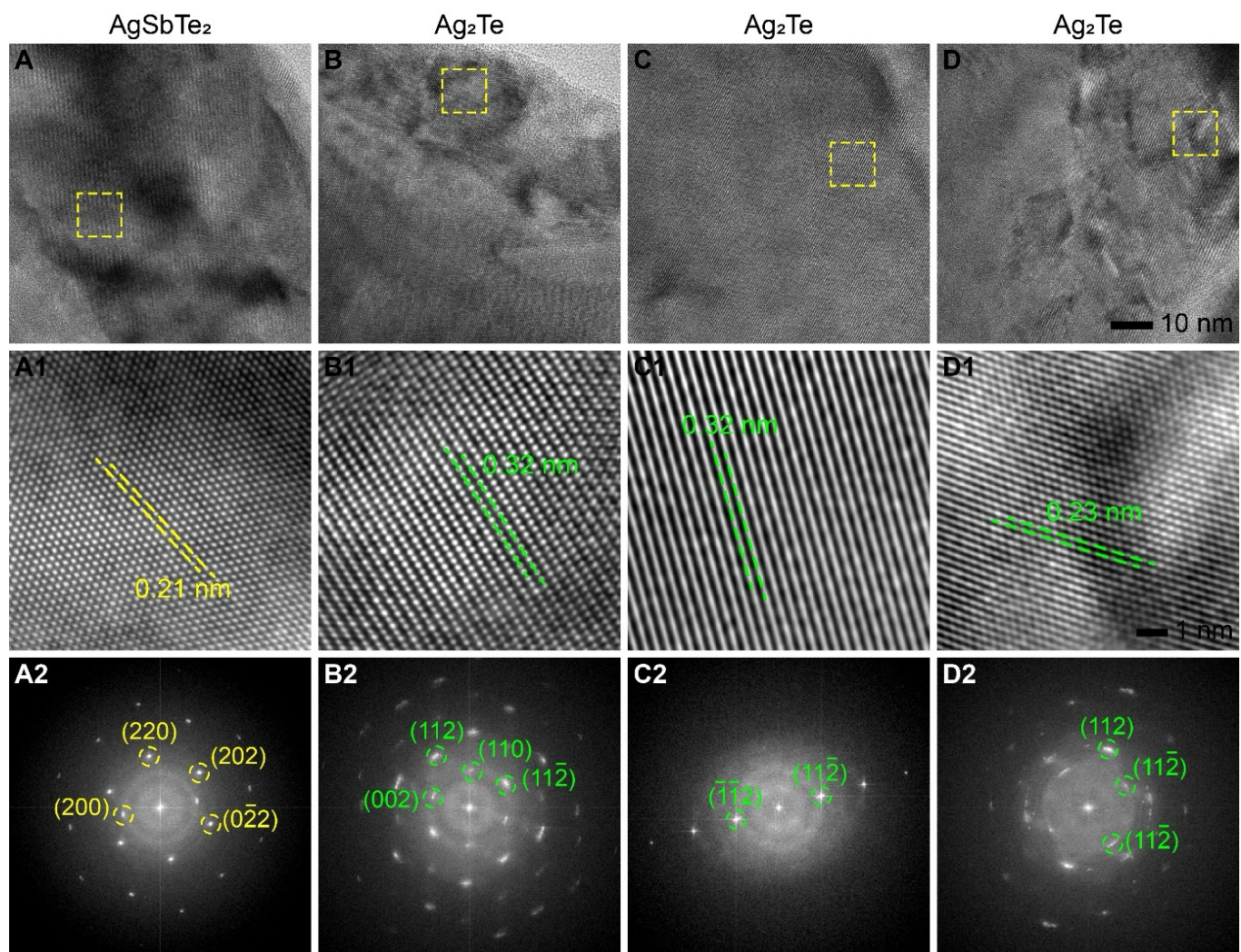

Fig. S12. High-resolution TEM (HR-TEM) imaging and phase identification for undoped $AgSb_{1.05}Te_{2.06}$ sample. (A-D) HR-TEM images from four different areas. (A1-D1) Corresponding zoomed-in sections of the areas marked with dotted yellow rectangles in (A-D), showing lattice fringes and measured lattice fringe spacing indicated by dotted lines. (A2-D2) Corresponding Fast Fourier Transform (FFT) images. The indexed FFT confirms the presence of two phases: $AgSbTe_2$ and $Ag_2Te$. Detailed STEM imaging and STEM-EDX chemical mapping are provided in **SI Fig. S15** and **S16**, respectively.

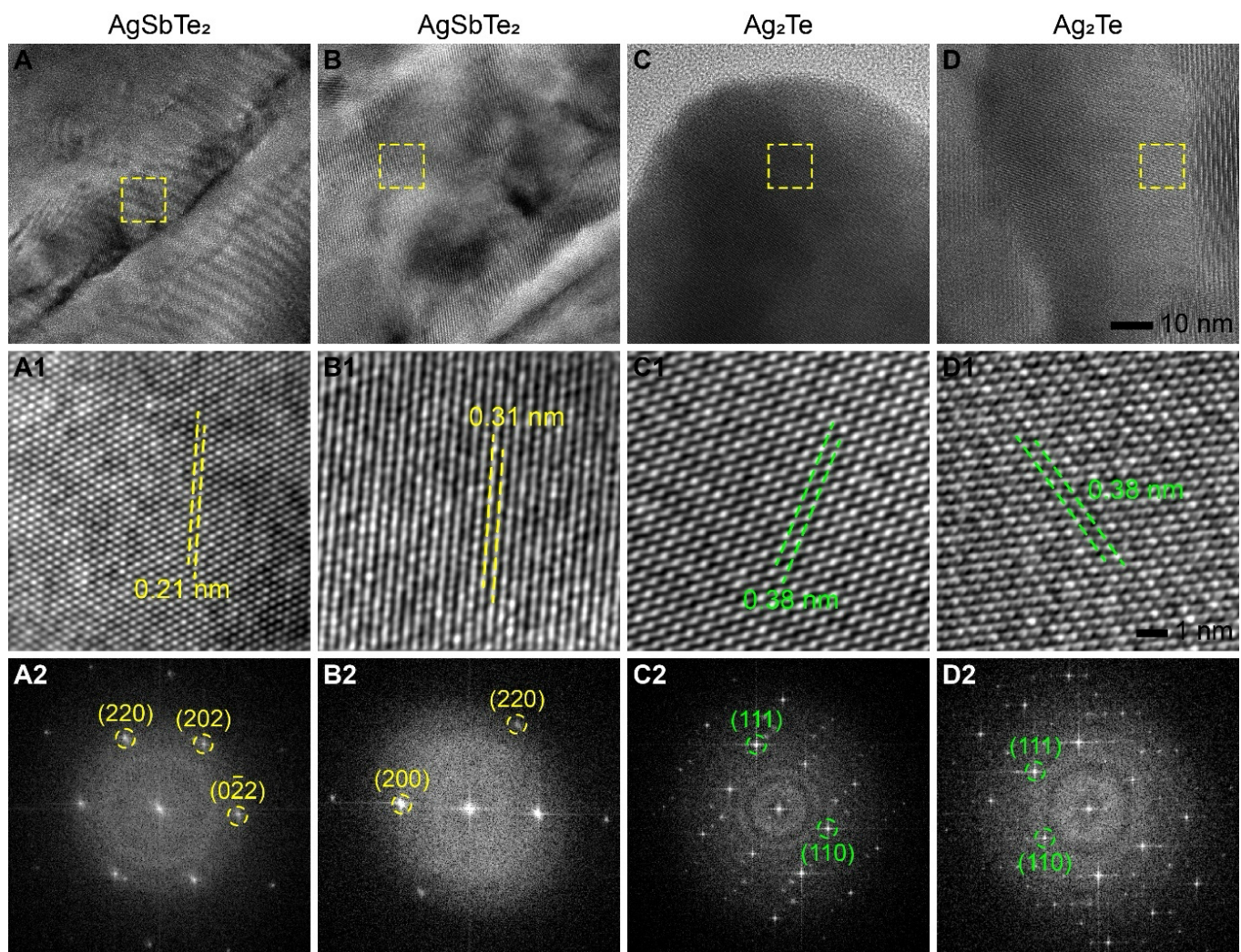

**Fig. S13. High-resolution TEM (HR-TEM) imaging and phase identification for Cd-doped AgSb$_{1.01}$Cd$_{0.04}$Te$_{2.06}$ sample.** (**A-D**) HR-TEM images from four different areas. (**A1-D1**) Corresponding zoomed-in sections of the area marked with dotted yellow rectangles in (A-D), showing lattice fringes and measured lattice fringe spacing indicated by dotted lines. (**A2-D2**) Corresponding Fast Fourier Transform (FFT) images. The indexed FFT confirms the presence of two phases: AgSbTe$_2$ and Ag$_2$Te. Detailed STEM imaging and STEM-EDX chemical mapping are provided in **SI Fig. S17** and **S18**, respectively.

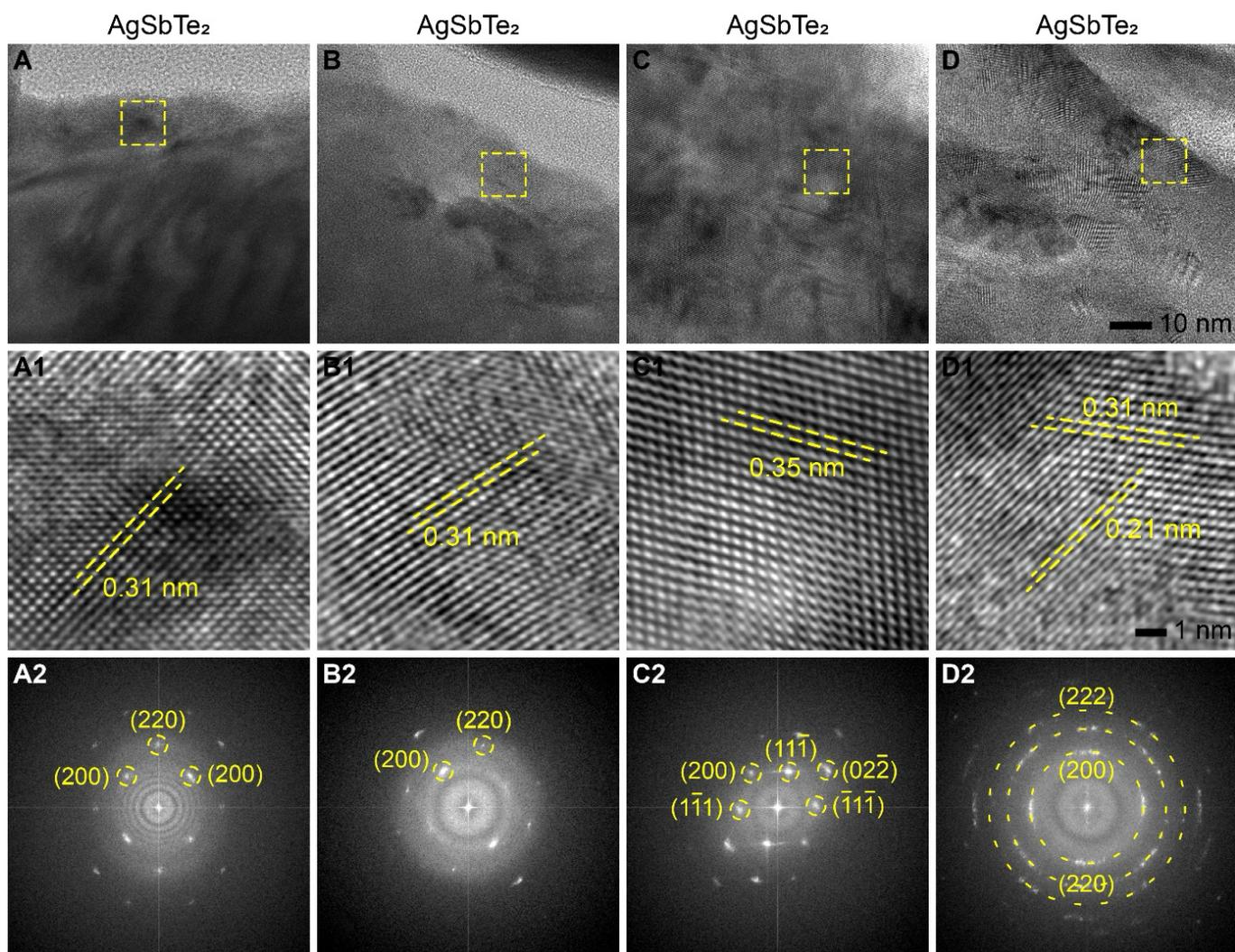

**Fig. S14. High-resolution TEM (HR-TEM) imaging and phase identification for Se co-doped AgSb$_{1.01}$Cd$_{0.04}$Te$_{1.86}$Se$_{0.2}$ sample.** (**A-D**) HR-TEM images from four different areas. (**A1-D1**) Corresponding zoomed-in sections of the area marked with dotted yellow rectangles in (A-D), showing lattice fringes and measured lattice fringe spacing indicated by dotted lines. (**A2-D2**) Corresponding Fast Fourier Transform (FFT) of images. In this case, only the AgSbTe$_2$ phase was observed, with no evidence of the Ag$_2$Te phase. Detailed STEM imaging and STEM-EDX chemical mapping are provided in **SI Fig. S19** and **S20**, respectively.

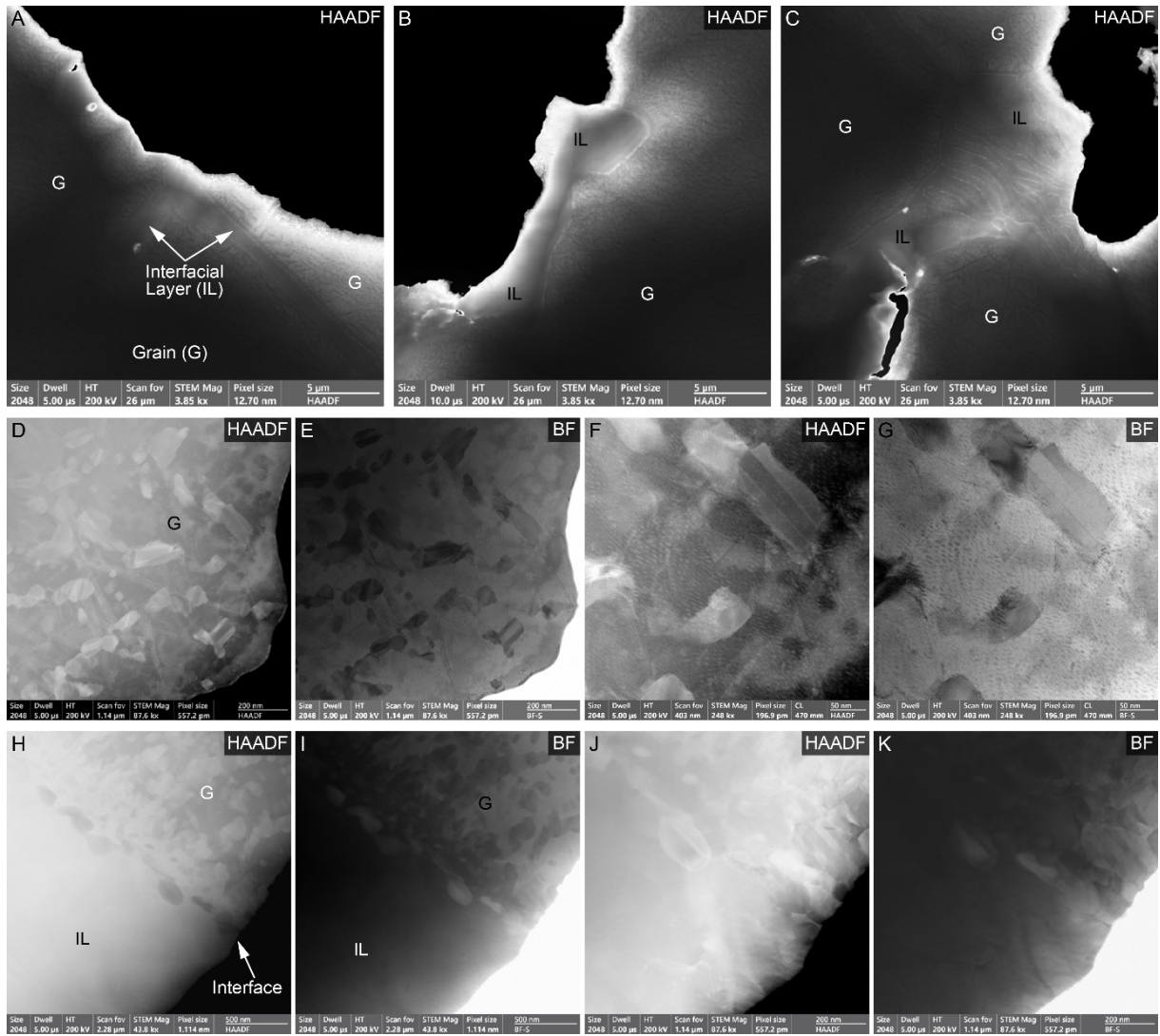

**Fig. S15. STEM imaging of undoped AgSb$_{1.05}$Te$_{2.06}$ sample, revealing the interfacial layer (IL) and its microstructure.** (**A-C**) Low-magnification STEM-HAADF images from three different areas showing the microstructure of grains and IL High-magnification STEM-HAADF and STEM-BF images of (**D-G**) a large grains and (**H-K**) A detailed microstructural characterization and chemical analysis are presented in **SI Fig. S12** and **S16**, respectively.

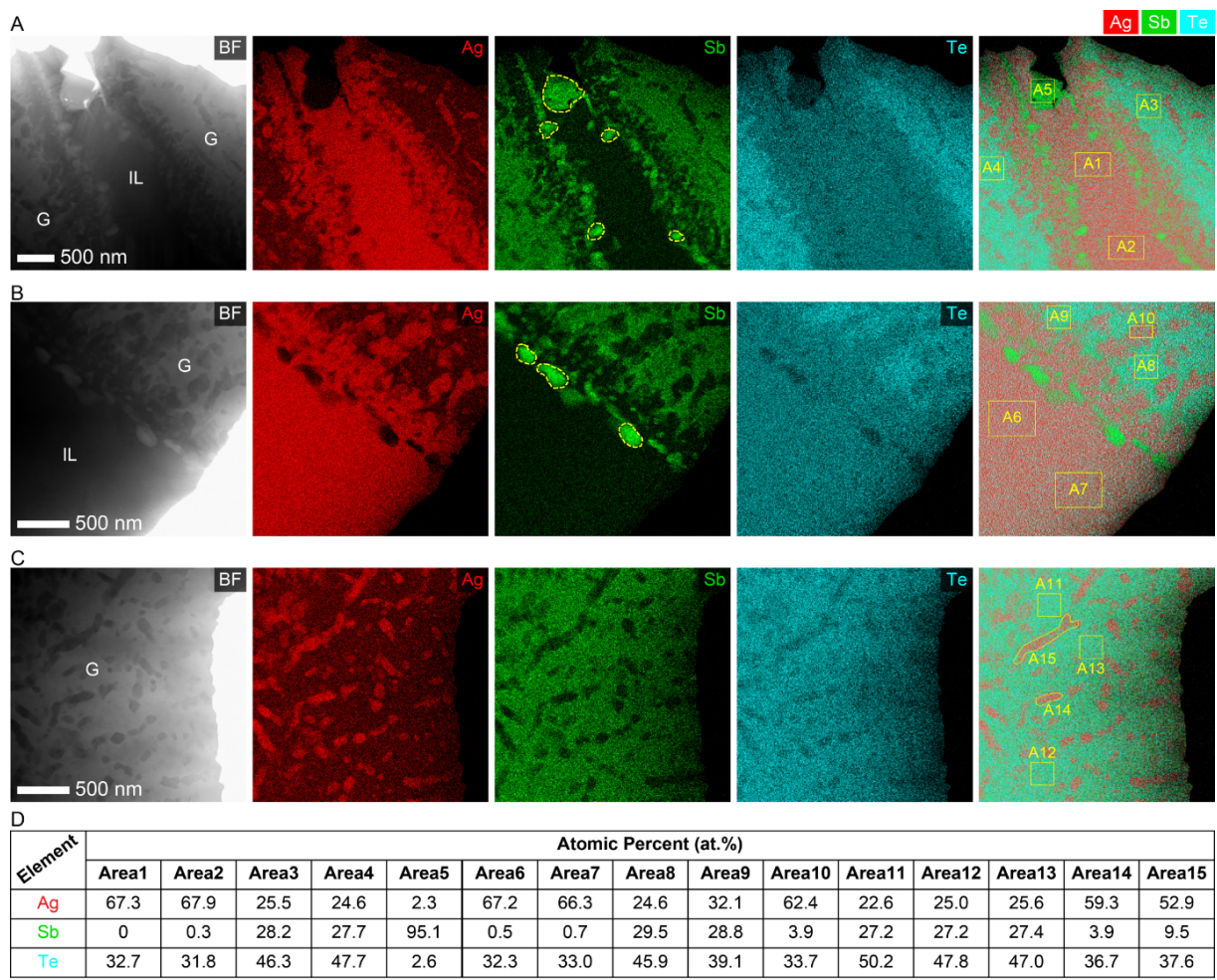

**Fig. S16. STEM-EDX elemental mapping of undoped AgSb$_{1.05}$Te$_{2.06}$.** (**A-C**) STEM-EDX elemental maps from three different regions, showing the chemical composition of the IL and grains. Each row shows STEM-BF, elemental maps of Ag (*red*), Sb (*green*), Te (*cyan*), and overlay (Ag+Sb+Te) image from left to right. (**D**) Chemical quantification of the regions marked with yellow boxes in the overlay images reveal the chemical composition of the grain (A3, A4, A8, A11, A12, A13) matches week with AgSbTe$_2$, while the IL (A1, A2, A6, A7) is mostly Ag$_2$Te with Sb precipitates decoration at the IL edges (A5). The precipitates within the grains (A10, A14, A15) are also mostly Ag$_2$Te.

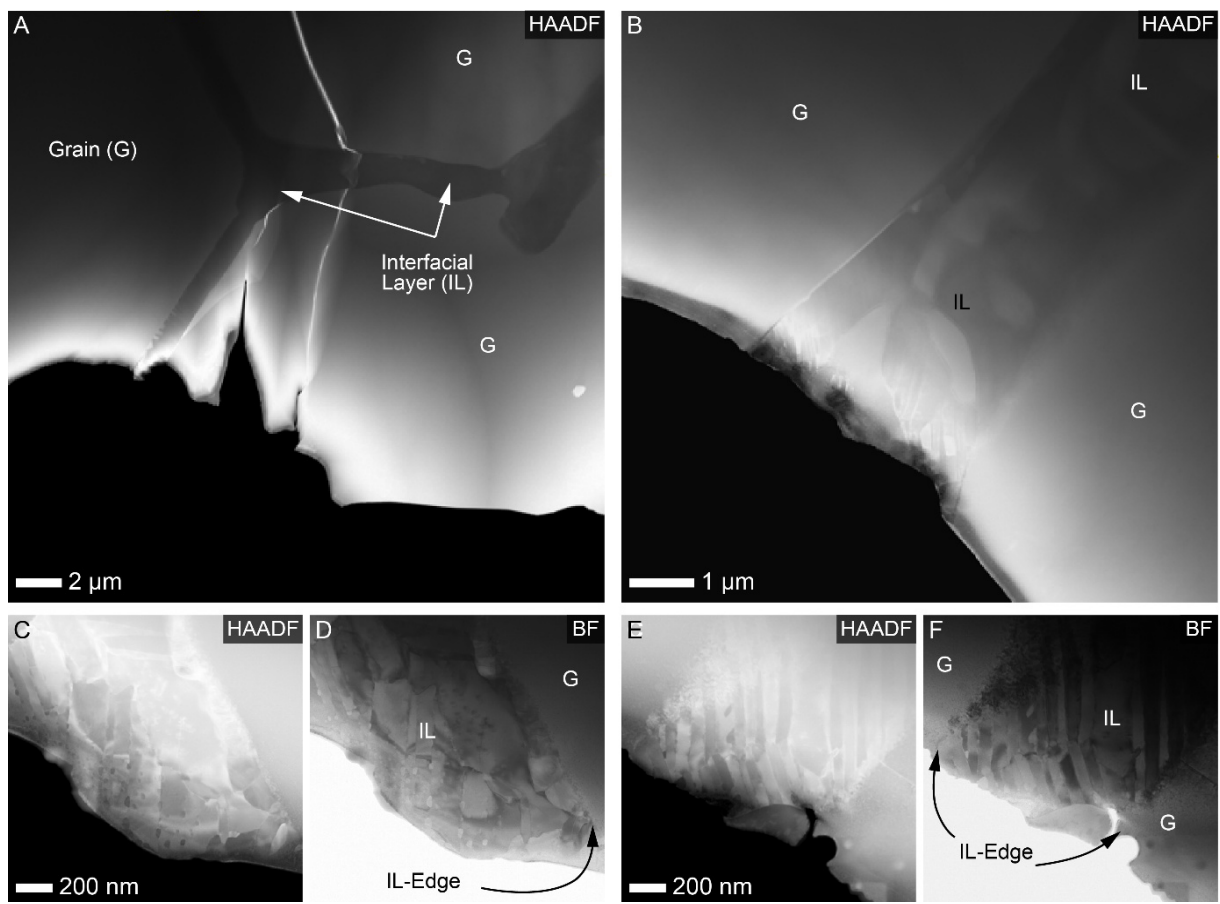

**Fig. S17. STEM imaging of Cd-doped (AgSb$_{1.01}$Cd$_{0.04}$Te$_{2.06}$) sample, revealing the interfacial layer (IL) and its microstructure.** (**A-B**) Low-magnification STEM-HAADF images from two different areas, showing the microstructure of grains and IL. (**C-F**) High-magnification STEM-HAADF and STEM-BF images of the IL. Interestingly, in this case we see no precipitates within the large grains, instead, the IL consists of lamellar Ag$_2$Te precipitates. This observation is in contrast to that we have seen in the undoped sample. A detailed microstructural characterization and chemical analysis are presented in **SI Fig. S13** and **S18**, respectively.

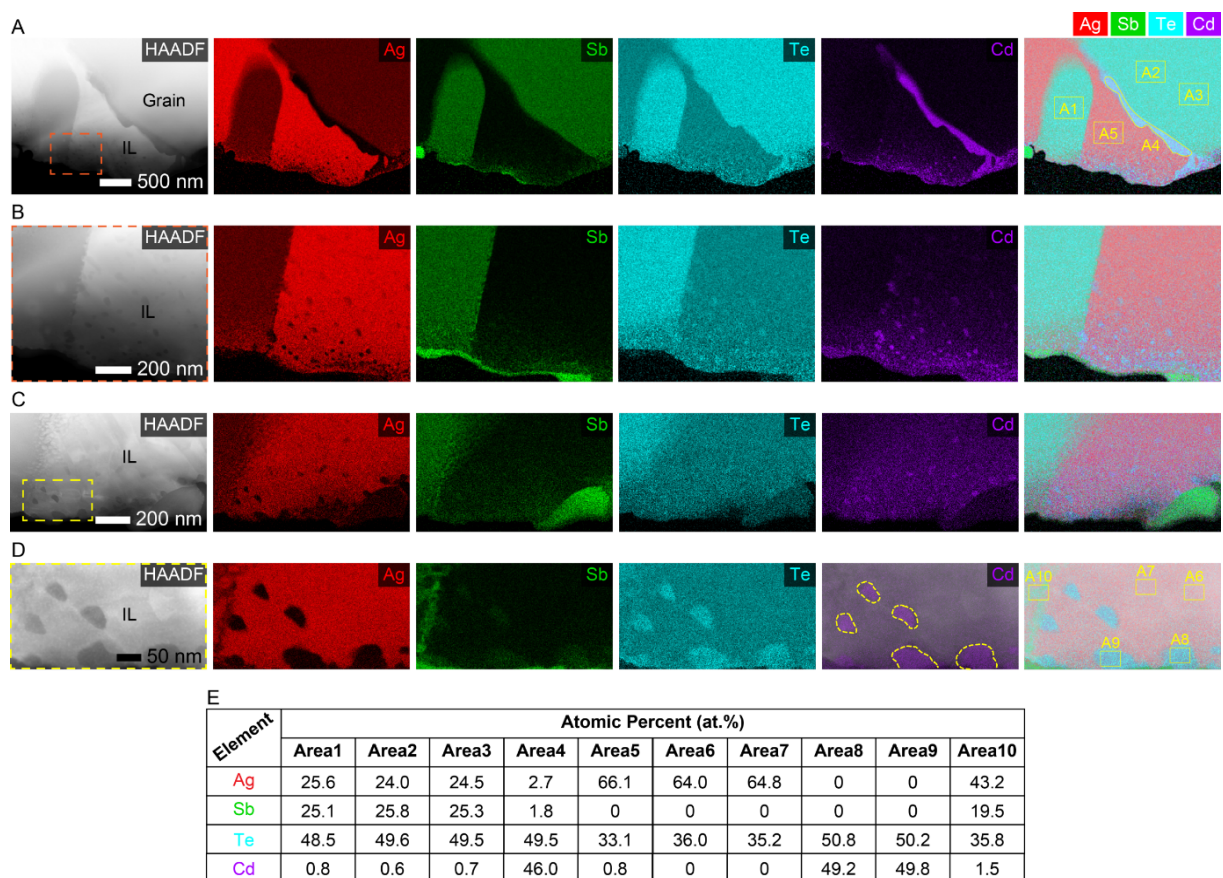

| Element | Atomic Percent (at.%) | | | | | | | | | |
|---|---|---|---|---|---|---|---|---|---|---|
| | Area1 | Area2 | Area3 | Area4 | Area5 | Area6 | Area7 | Area8 | Area9 | Area10 |
| Ag | 25.6 | 24.0 | 24.5 | 2.7 | 66.1 | 64.0 | 64.8 | 0 | 0 | 43.2 |
| Sb | 25.1 | 25.8 | 25.3 | 1.8 | 0 | 0 | 0 | 0 | 0 | 19.5 |
| Te | 48.5 | 49.6 | 49.5 | 49.5 | 33.1 | 36.0 | 35.2 | 50.8 | 50.2 | 35.8 |
| Cd | 0.8 | 0.6 | 0.7 | 46.0 | 0.8 | 0 | 0 | 49.2 | 49.8 | 1.5 |

**Fig. S18. STEM-EDX elemental mapping of Cd-doped ($AgSb_{1.01}Cd_{0.04}Te_{2.06}$) sample.** (**A**) and (**C**) STEM-EDX elemental mapping of Cd-doped ($AgSb_{1.01}Cd_{0.04}Te_{2.06}$) sample from two different areas showing the chemical composition of the IL and grain. Each row shows STEM-HAADF, elemental maps of Ag (*red*), Sb (*green*), Te (*cyan*), Cd (*purple*) and overlay (Ag+Sb+Te+Cd) images from left to right. (**B**) and (**D**) are chemical mapping at higher magnification to resolve the precipitates within the IL. (**E**) Chemical quantification of the regions marked with yellow boxes in the overlay images reveals the chemical composition of the grain (A1, A2, A3) matches well with $AgSbTe_2$, while the IL is mostly $Ag_2Te$ with embedded CdTe precipitate (A5, A6, A7). These results also show that the Cd composition is much lower (~ 1 at.%) in the grain than the expected doping level (4 at.%). and although the large grains are free from any precipitates, nanoscale Cd-rich precipitates (A8, A9) are observed in the IL.

The presence of $Ag_5Te_3$ in Cd-doped $AgSb_{1.01}Cd_{0.04}Te_{2.06}$ and Se co-doped $AgSb_{1.01}Cd_{0.04}Te_{1.86}Se_{0.20}$ could not be fully confirmed through TEM (contrary to XRD). This is because their d-spacing is too close to be able to differentiate them through DP or HR-TEM (d-spacing for most intense peaks of $Ag_2Te$ and $Ag_5Te_3$ is 0.2299 nm and 0.2118 nm, respectively). Additionally, even though there are compositional fluctuations, our EDX quantitative analysis shows better matching to $Ag_2Te$ (Table S7 and S8).

**Table S7.** STEM-EDX composition calculations for Cd dopped AST (Fig. S18).

| | IL | | | Grain | | | |
|---|---|---|---|---|---|---|---|
| **Elements** | Area 5 | Area 6 | Area 7 | | | | |
| Ag | 66.1 | 64 | 64.8 | | | | |
| Sb | 0 | 0 | 0 | | | | |
| Te | 33.1 | 36 | 35.2 | | | | |
| Cd | 0.8 | 0 | 0 | | | | |
| Ratio Ag/Te | 1.996979 | 1.777778 | 1.840909 | No precipitate found in the grain | | | |

**Note 1:** Ag/Te = 2 if it is $Ag_2Te$; Ag/Te = 1.67 if it is $Ag_5Te_3$.

**Note 2:** Area 5 and 7 are certainly $Ag_2Te$. Area 6 is quantitively closer to $Ag_5Te_3$, however, image A6 and 7 are very similar. Therefore, this is not a different compound but rather a fluctuation in composition.

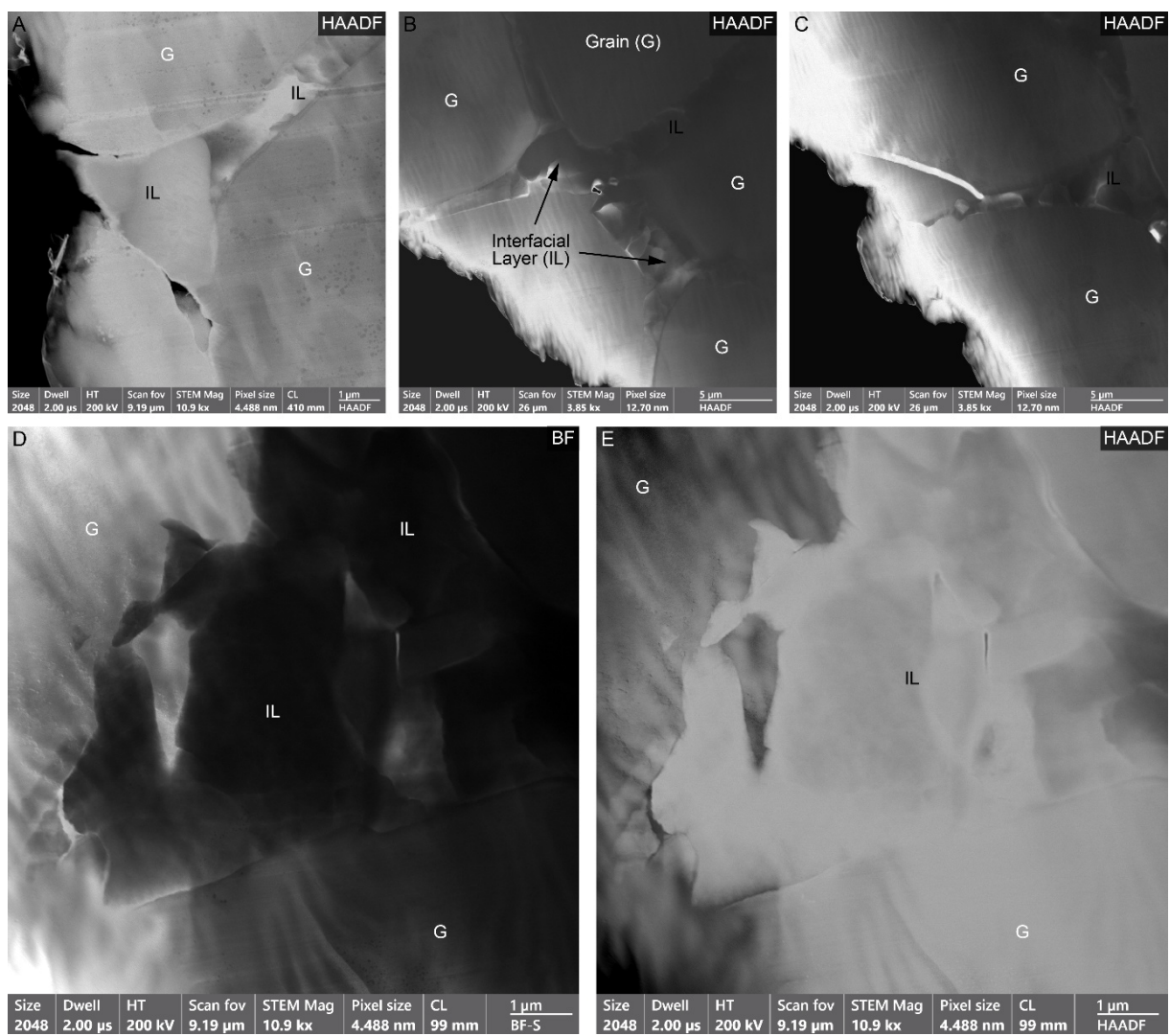

**Fig. S19. STEM imaging of Se co-doped (AgSb$_{1.01}$Cd$_{0.04}$Te$_{1.86}$Se$_{0.2}$) sample, revealing the interfacial layer (IL) and its microstructure.** (**A-C**) Low-magnification STEM-HAADF images from three different areas, showing the microstructure of grains and IL. High-magnification (**D**) STEM-HAADF and (**E**) STEM-BF images of the IL. In this case again the large grains has no precipitates while the IL consists Ag$_2$Te precipitates. This observation is similar to Cd-doped sample. A detailed microstructural characterization and chemical analysis are presented in SI Fig. S14 and S20, respectively.

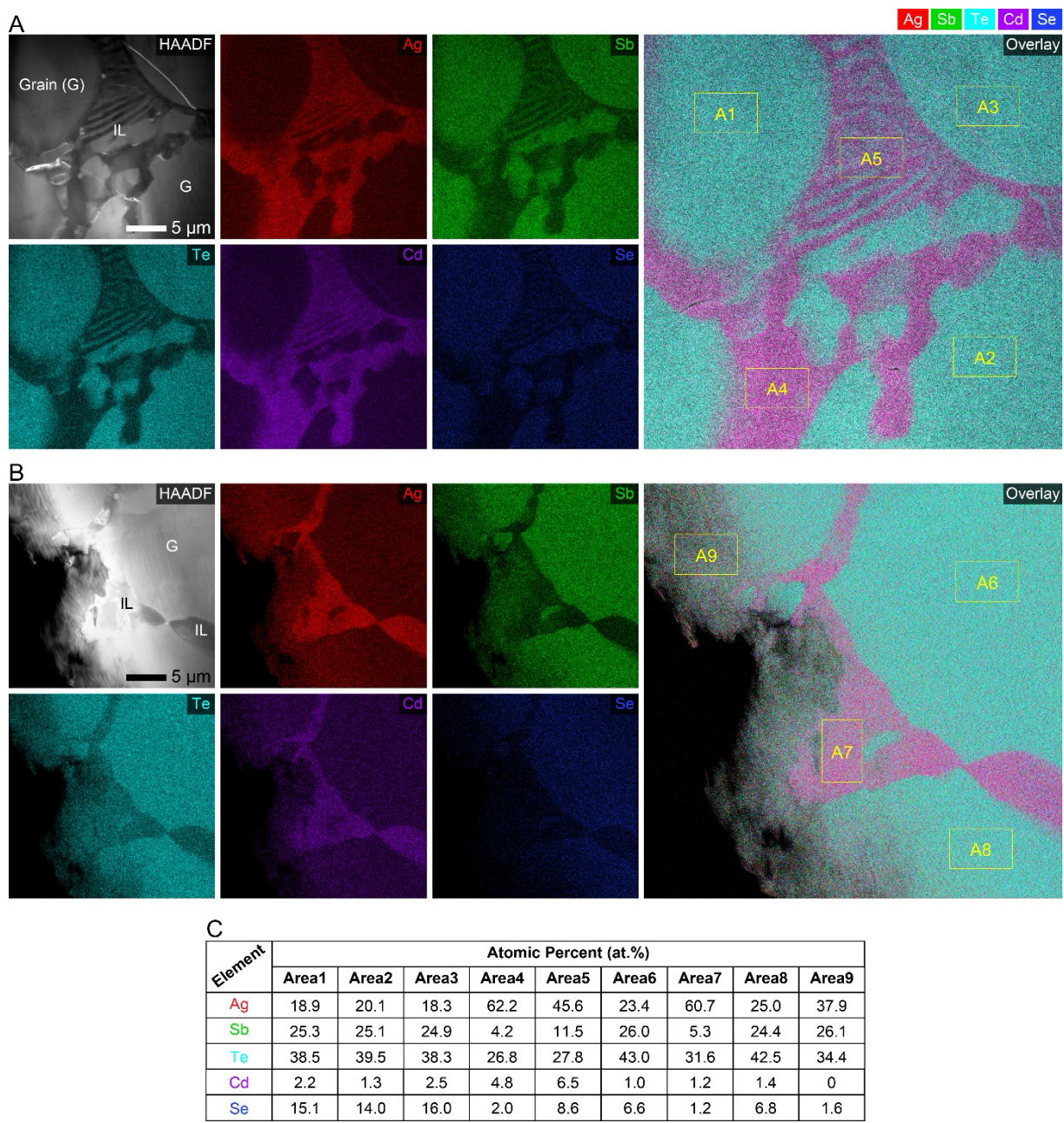

**Fig. S20. STEM-EDX elemental mapping of Se co-doped (AgSb$_{1.01}$Cd$_{0.04}$Te$_{1.86}$Se$_{0.2}$) sample.** (**A**) and (**B**) STEM-EDX elemental mapping from two different areas, showing the chemical composition of the IL and grain. Each row shows HAADF-STEM, elemental maps of Ag (*red*), Sb (*green*), Te (*cyan*), Cd (*purple*), Se (*blue*) and overlay (Ag+Sb+Te+Cd+Se) images. (**C**) Chemical quantification of the regions marked with yellow boxes in the overlay images reveals the chemical composition of the grain (A1, A2, A3, A6, A8) and the IL (A4, A5, A7). The grain composition is well matching with AgSbTe$_2$, whereas the ILs mostly show Ag$_2$Te precipitate with varying 1-4 at. % doping of Cd and Se. Within the grains Cd doping is lower while Se is higher than the expected doping.

Table S8. STEM-EDX composition calculations for Se co-dopped AST (Fig. S20).

| Elements | IL | | | Grain |
|---|---|---|---|---|
| | Area4 | Area5 | Area7 | |
| Ag | 62.2 | 45.6 | 60.7 | |
| Sb | 4.2 | 11.5 | 5.3 | |
| Te | 26.8 | 27.8 | 31.6 | |
| Cd | 4.8 | 6.5 | 1.2 | |
| Se | 2 | 8.6 | 1.2 | |
| Ratio Ag/Te | 2.320896 | 1.640288 | 1.920886 | No precipitate found in the grain |

**Note 1:** Ag/Te = 2 if it is $Ag_2Te$; Ag/Te = 1.67 if it is $Ag_5Te_3$.

**Note 2:** Area 4 and 7 are certainly $Ag_2Te$. Area 5 cannot be considered as it includes clearly other area than just precipitate (see EDX map).

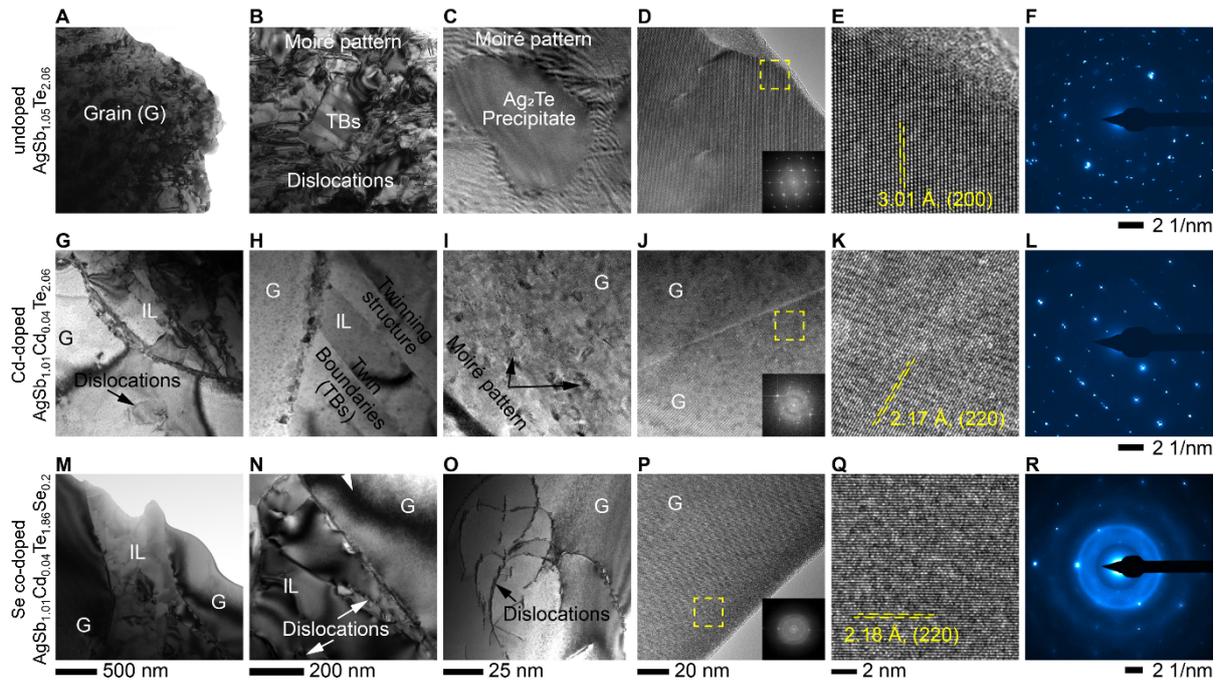

**Fig. S21. Detailed microstructural characterization and comparison of undoped, Cd-doped and Se co-doped samples using TEM imaging and diffraction techniques.** (**A**) Low-magnification, (**B**) high-magnification, (**C, D**) high-resolution TEM image of undoped sample. (**E**) A zoomed-in view of the rectangular region in (D), showing lattice fringes that match well with the pristine $AgSbTe_2$ phase. (F) SAED of the grain near the IL. The second row (**G-L**) and third row (**M-R**) shows exactly the same analysis under similar conditions for Cd-doped $AgSb_{1.01}Cd_{0.04}Te_{2.06}$ and Se co-doped $AgSb_{1.01}Cd_{0.04}Te_{1.86}Se_{0.2}$ samples, respectively. Both samples show similar microstructure with polycrystalline interfacial layer (IL) and monocrystalline large grains. In contrast, the undoped sample shows polycrystalline phase within the large grains.

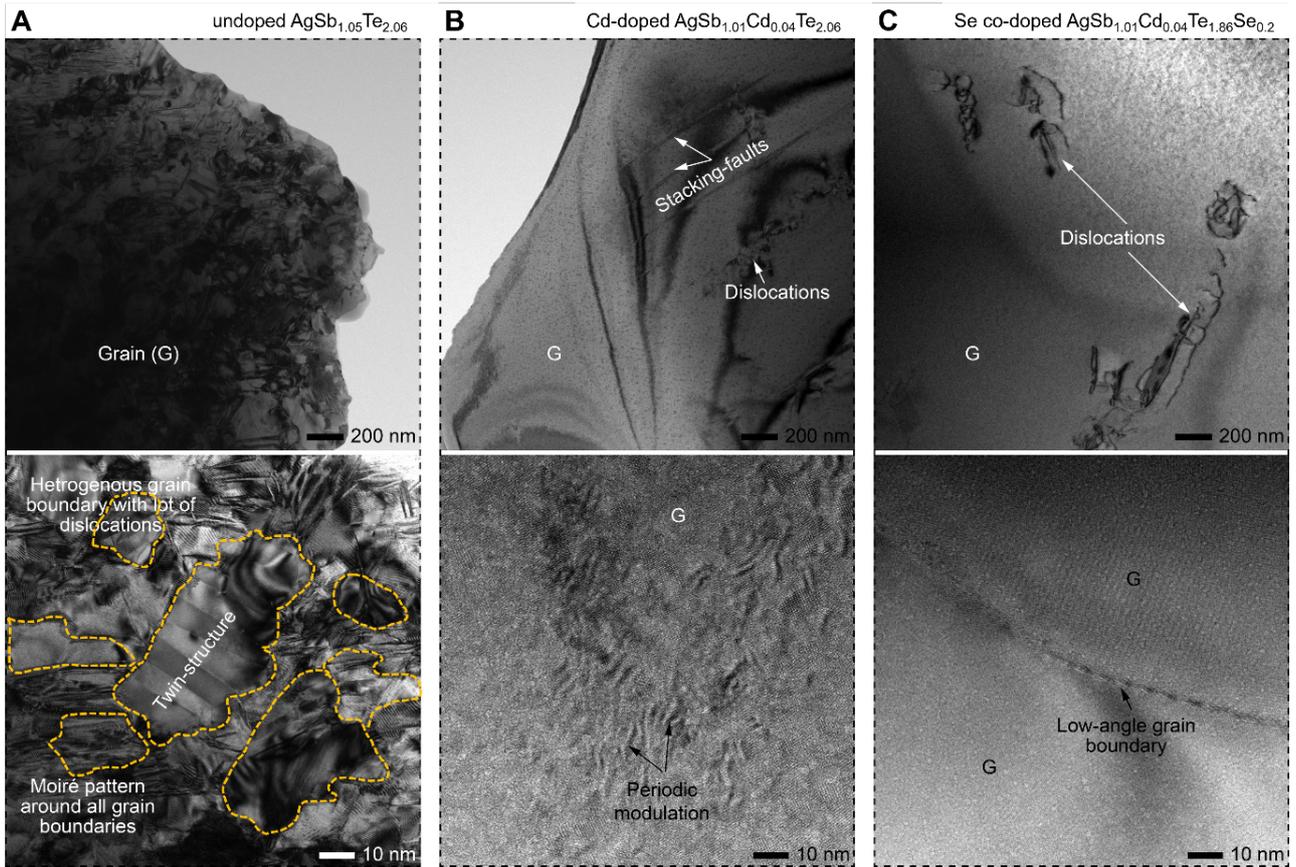

**Fig. S22. Highlight of the major microstructural differences between undoped, Cd-doped and Se co-doped samples.** Low-magnification *(top row)* and high-magnification (*bottom row*) TEM images of (**A**) undoped $AgSb_{1.05}Te_{2.06}$, (**B**) Cd-doped $AgSb_{1.01}Cd_{0.04}Te_{2.06}$ and (**C**) Se co-doped $AgSb_{1.01}Cd_{0.04}Te_{1.86}Se_{0.2}$ samples, showing different types of crystallographic defect formation due to secondary phase precipitation in the large grains in case of undoped sample; contrary, Cd-doped and Se co-dopped samples are free from secondary phase precipitation and large grains are mostly monocrystalline. While, Cd-doped sample shows periodic modulation, especially near the IL-edge, indicating cationic ordering, no such ordering observed for Se co-dopes sample except dislocations network randomly distributed within the large grains.

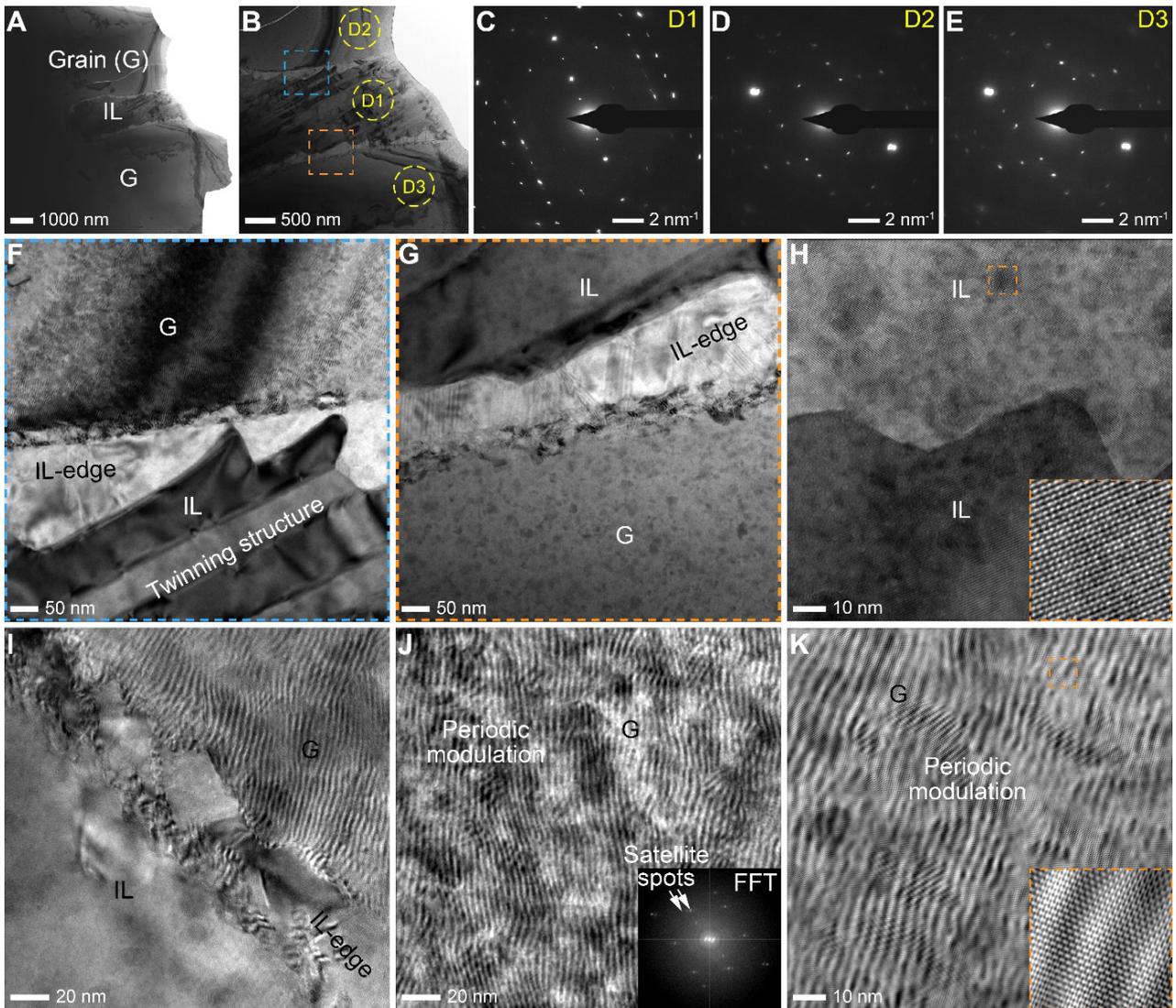

**Fig. S23. Evidence of cation ordering in Cd-doped AgSb$_{1.01}$Cd$_{0.04}$Te$_{2.06}$.** (**A**, **B**) Low-magnification TEM images, showing IL and grain. SAEDs of (**C**) IL, and (**D**, **E**) grains near the IL edge, confirming observed polycrystalline laminar structure of IL and monocrystalline structure of grains in the (**F**, **G**) high-magnification TEM images of interface regions. High-resolution TEM images of (**H**) center of IL, (**I**) IL-edge, and (**J**, **K**) grain near the IL-edge showing periodic modulation in the grains near the IL-edges. Insets in (H) and (K) show a zoomed-in view lattice fringes and in (J) FFT from the periodic modulation, showing satellite spots—an indication of cationic ordering.

**Supplementary Section S5: Detailed analysis of thermoelectric transport**

- **Detailed analysis of the thermoelectric properties of compensated AgSbTe$_2$**

Near room-temperature, we report a Seebeck coefficient of 257.57 µV K$^{-1}$ (Fig. S24A) and an electrical conductivity of 128.35 S cm$^{-1}$ (Fig. S24B) for undoped AgSb$_{1.05}$Te$_{2.06}$, with a maximum power factor of 8.51 µW cm$^{-1}$ K$^{-2}$ (Fig. 24C). The positive value of the Seebeck coefficient indicates that the sample is p-type, in very good agreement with previous reports of traditionally synthesized AgSbTe$_2$. These values generally increase when the temperature is increased, achieving maximum values at 600 K. At this temperature, we report electrical conductivity, Seebeck coefficient and power factor of 155.57 S cm$^{-1}$, 271.49 µV K$^{-1}$ and 11.47 µW cm$^{-1}$ K$^{-2}$ respectively. Despite a general increasing trend of electrical conductivity with respect to temperature, it is not clear enough so that semiconducting behaviour can be ascribed. We measured the temperature-dependent total thermal conductivity ($\kappa_{tot}$) of undoped AgSb$_{1.05}$Te$_{2.06}$ (Fig. S24D) and observe a gradual decrease in $\kappa_{tot}$ from 0.58 W m$^{-1}$ K$^{-1}$ at near-ambient temperature to 0.55 W m$^{-1}$ K$^{-1}$ at 375 K. This value further decreases to ~0.47 W m$^{-1}$ K$^{-1}$ and remains constant until 525 K, temperature at which $\kappa_{tot}$ increases again until 0.55 W m$^{-1}$ K$^{-1}$ at 600 K. Lattice thermal conductivity values were obtained using the Wiedemann-Franz Law ($\kappa_e = L\sigma T$). The Lorentz number was calculated by numerically solving its Boltzmann Transport Equation (Methods). The temperature-dependent lattice thermal conductivity for undoped AgSb$_{1.05}$Te$_{2.06}$ is shown in Fig. S24E. We observe the usual T$^{-1}$ dependence of $\kappa_{lat}$, indicating that phonon scattering dominates thermal transport, with a minimum value of $\kappa_{lat}$ ~ 0.41 W m$^{-1}$ K$^{-1}$ at 573 K. Considering this, we attribute the increase in total thermal conductivity to increased electronic thermal conductivity ($\kappa_e$) due to bipolar conduction. Finally, we report a maximum zT value of ~1.2 at 600 K, a steep increase from a near-ambient value of ~0.4 (Fig. S24F). Compared to previously reported pristine AgSbTe$_2$ the samples in this work are beyond the optimal carrier concentration value *i.e.* overdoped. This is evidenced by several transport coefficients and temperature-dependences of our data. We report higher near-ambient values of electrical conductivity and similar Seebeck coefficient relative to the values reported by Roychowdhury *et al.* ($\sigma$ ~ 121 S cm$^{-1}$ and S ~279 µV K$^{-1}$), Bhui *et al.* ($\sigma$ ~ 104 S cm$^{-1}$ and S ~273 µV K$^{-1}$), Hu *et al.* ($\sigma$ ~ 120 S cm$^{-1}$ and S ~260 µV K$^{-1}$) and Pathak *et al.* ($\sigma$ ~ 100 S cm$^{-1}$ and S ~275 µV K$^{-1}$). These values lead to a much higher power factor values for our undoped AgSb$_{1.05}$Te$_{2.06}$ than for those in the literature (Roychowdhury *et al.* ~9.5 µW cm$^{-1}$ K$^{-2}$, Bhui *et al.* ~7.8 µW cm$^{-1}$ K$^{-2}$). Thus, the *zT* values obtained for undoped AgSb$_{1.05}$Te$_{2.06}$ in this work are much higher than those reported in the literature: ~0.6 and ~0.9 for Roychowdhury *et al.* and Bhui *et al.*, respectively.

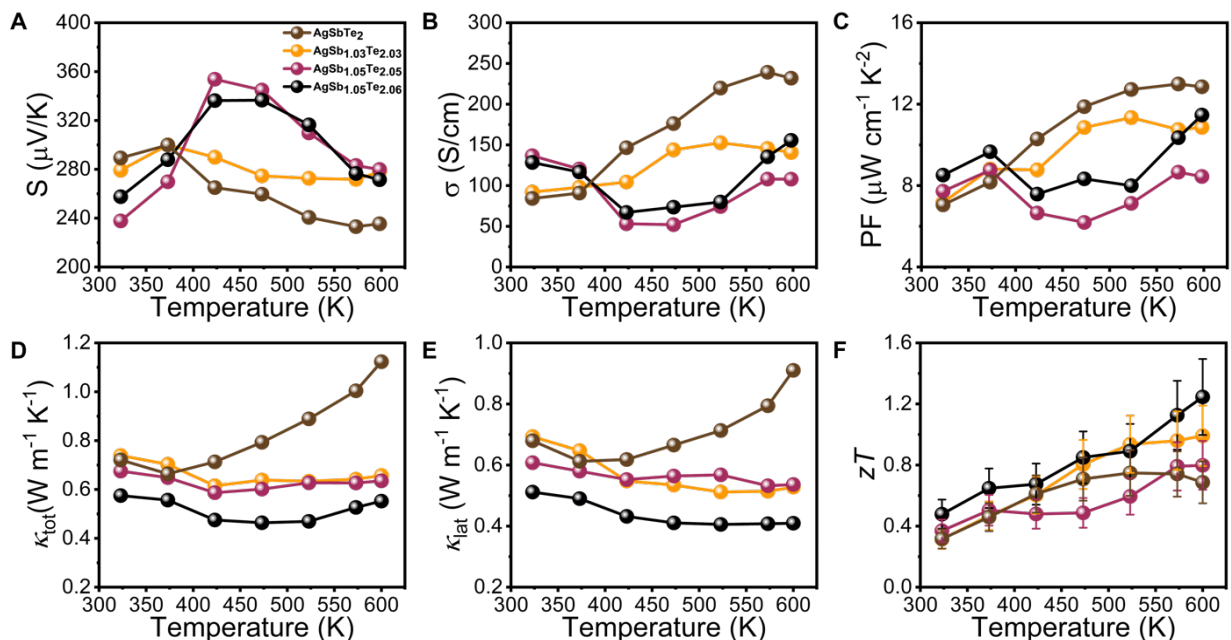

**Figure. S24.** Thermoelectric properties of polycrystalline nominal AgSbTe$_2$ and compensated AgSbTe$_2$ with different amounts of volatile elements (Sb and Te). The uncertainty in the $zT$ measurement is 20%. ICP-OES and thermoelectric transport determine that after elemental loss during DJS reaction, undoped AgSb$_{1.05}$Te$_{2.06}$ matches nominal AgSbTe$_2$. The compensated AgSb1.05Te2.06 achieved power factor of 11.47 µW cm$^{-1}$ K$^{-2}$ and $zT$ of 1.2 at 600K.

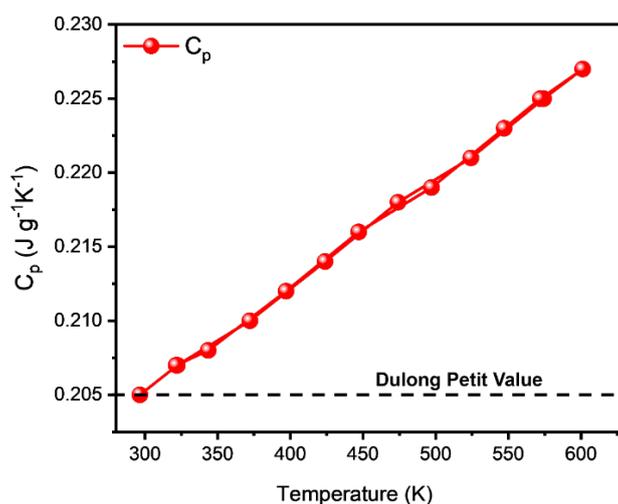

**Figure. S25.** Temperature-dependent heat capacity measurement of undoped AgSb$_{1.05}$Te$_{2.06}$.

- **Detailed analysis of the thermoelectric properties of Cd-doped AgSb$_{1.05-x}$Cd$_x$Te$_{2.06}$**

We measure the temperature-depending transport properties of the series (Fig. S26) to determine the best performing and select it for further experiments. Doping CdTe in AgSb$_{1.05}$Te$_{2.06}$ results in a decrease in Seebeck coefficient (Fig. S26A) for the measured temperature range. Near-room temperature, S decreases from 257.57 µV K$^{-1}$ for AgSb$_{1.05}$Te$_{2.06}$ to 241.30 µV K$^{-1}$ in 6 mol % CdTe–doped AgSb$_{1.05}$Te$_{2.06}$. The reduction in Seebeck coefficient is accompanied by an increase in electrical conductivity (Fig. S26B), from 128.35 S cm$^{-1}$ in the baseline material to 159.97 128.35 S cm$^{-1}$ in AgSb$_{0.99}$Cd$_{0.06}$Te$_{2.06}$. For all CdTe-doped samples, the electrical conductivity clearly increases with temperature, indicating semiconducting behaviour, in stark contrast with the baseline compound. The increased σ while retaining large values of S result in an overall large enhancement of the power factor throughout the measured temperature range, reaching a maximum value of 15.63 µW cm$^{-1}$ K$^{-2}$ at 573 K Fig. S26C). The total thermal conductivity for the doped samples decreases for all temperature values (Fig. S26D). We extracted $\kappa_{lat}$ values using a calculated Lorentz number (Methods) and we observe a decrease in $\kappa_{lat}$ with increasing mol % CdTe, from ~0.51 W m$^{-1}$ K$^{-1}$ at near-ambient temperature for the pristine sample to ~0.45 W m$^{-1}$ K$^{-1}$ for 6 mol % CdTe. As temperature increases, we observe a gradual decrease in $\kappa_{lat}$, with a T$^{-1}$ dependence. The minimum value of $\kappa_{lat}$ is achieved at 573 K for 4 % mol CdTe: 0.33 W m$^{-1}$ K$^{-1}$. Consequently, the largest *zT* value for Cd-doped AgSb$_{1.05-x}$Cd$_x$Te$_{2.06}$ is achieved when x = 0.04, corresponding to Cd-doped AgSb$_{1.01}$Cd$_{0.04}$Te$_{2.06}$, which has a zT ~1.5 at 573 K (Fig. S26F). These values fall short from those reported by Roychowdhury *et al.* for AgSb$_{0.94}$Cd$_{0.06}$Te$_2$ (zT~2.6 at 573 K), Bhui *et al.* for AgSb$_{0.96}$Hg$_{0.04}$Te$_2$ (zT ~2.4 at ~570 K) and Taneja *et al.* for AgSb$_{0.96}$Yb$_{0.04}$Te$_2$ (zT ≈ 2.4 at 573 K) but are higher than any other single cation-site doped AgSbTe$_2$ (Table S3).

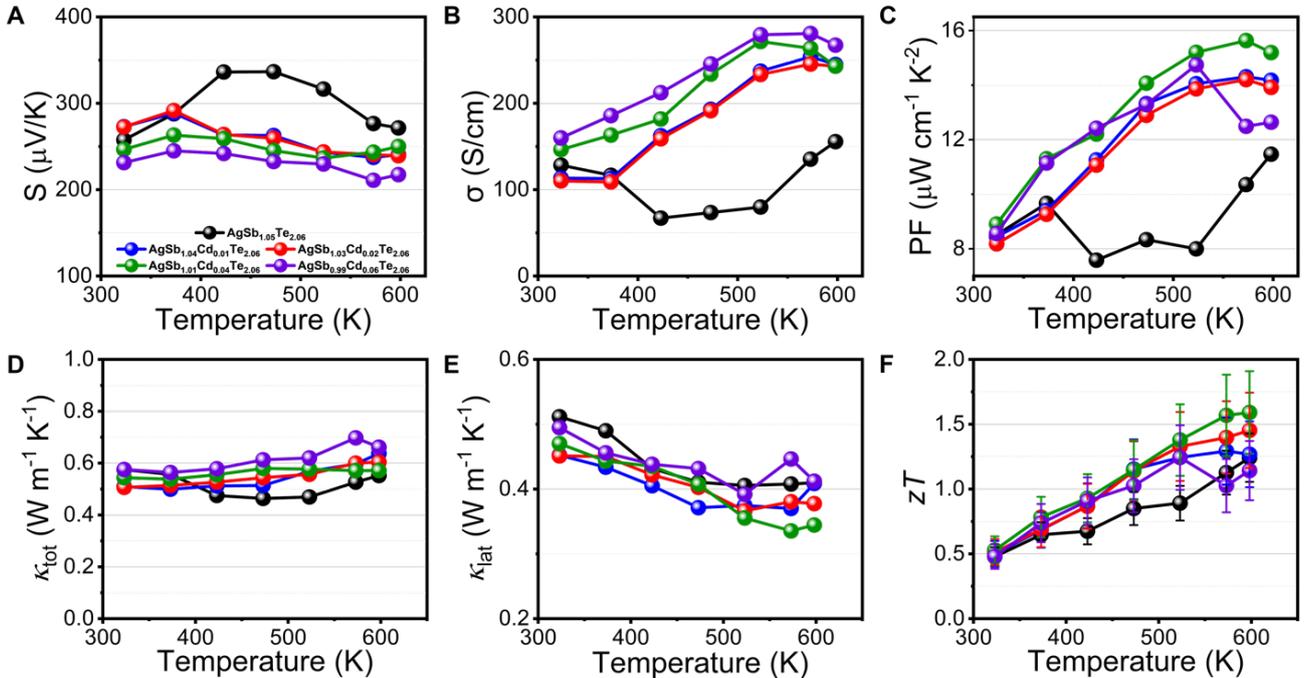

**Fig. S26 Thermoelectric properties of Cd-doped AgSb$_{1.05-x}$Cd$_x$Te$_{2.06}$ (x = 0 - 0.06).** Temperature dependent **(A)** Seebeck coefficient (*S*), **(B)** electrical conductivity (σ), **(C)** power factor (σ*S*$^2$), **(D)** total

thermal conductivity ($\kappa_{tot}$), **(E)** lattice thermal conductivity ($\kappa_{lat}$), **(F)** $zT$. The uncertainty of thermoelectric transport measurement is about 20%.

- **Hall measurements for selected samples**

Figure S27A shows the change in Hall resistivity with magnetic field and S27B shows magneto resistance for undoped AgSb$_{1.05}$Te$_{2.06}$ at 300 K. The inset S27A shows the linear fit for the smaller magnetic fields (< 1T) to obtain the Hall coefficient, R$_H$ (-1/nq), which is negative and equal to -0.926 m³/C. The non-linearity of Hall resistivity with applied magnetic field suggests that two carriers transport contribute to transport. And Fig. S26B shows a positive magnetoresistance, expected for bipolar conduction. The negative Hall coefficient at first glance indicates that the dominant carrier are electrons. However, the Seebeck coefficient is positive throughout the measured temperature range for undoped AgSb$_{1.05}$Te$_{2.06}$ (Fig. S24), which confirms p-type conduction.

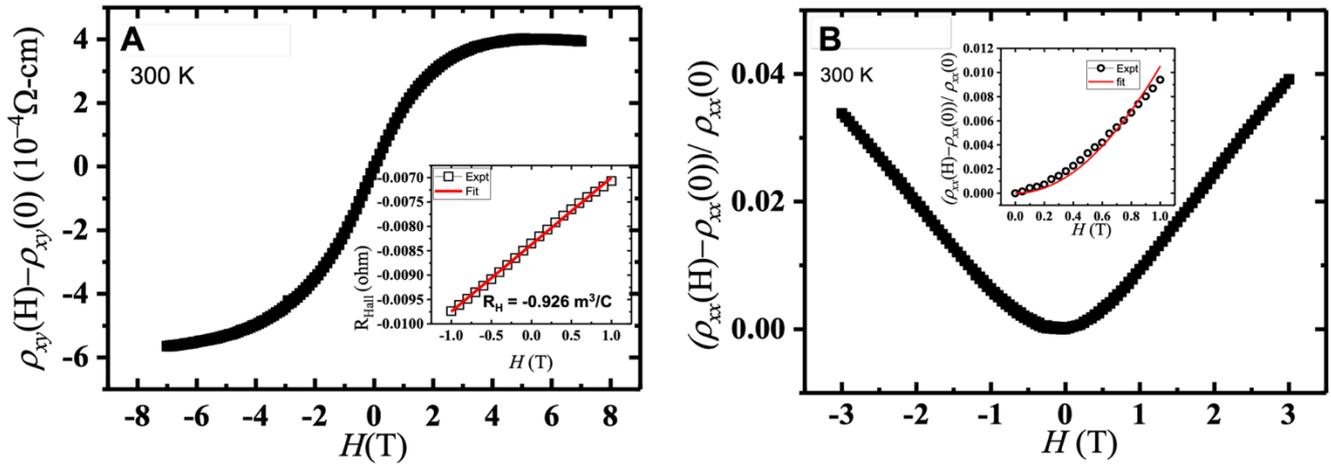

**Figure. S27. (A)** Change in the Hall resistivity and **(B)** change in magnetoresistance resistivity for the undoped AgSb$_{1.05}$Te$_{2.06}$ at 300 K. The inset **(A)** is a fit to Hall resistance to obtain the Hall coefficient **(B)** is the 2 carrier model data fit data for low magnetic fields < 1T.

The negative Hall coefficient in AgSbTe$_2$ has been attributed to the dominance of high-mobility, low-concentration narrow-band electrons, which outweigh the contribution of holes in Hall transport due to their superior mobility. The positive Seebeck coefficient indicates hole-dominated transport, where the high hole concentration arises from multiple degenerate valence bands, leading to a large effective mass, enhanced Seebeck coefficient due to band degeneracy, and low hole mobility.[24,25]

Using the two-carrier model, the Hall resistivity $\rho_{xy}(H)$ and the longitudinal resistivity $\rho_{xx}(H)$ can be expressed as:

$$\frac{\Delta \rho}{\rho} = \frac{\rho_{xx}(H)}{\rho_{xx}(0)} - 1 = \frac{(\mu_h + \mu_e)^2 np \mu_h \mu_e}{(n\mu_e + p\mu_h)^2} H^2 - \frac{(\mu_h + \mu_e)^2 np(n-p)^2 \mu_h^3 \mu_e^3}{(n\mu_e + p\mu_h)^4} H^4 \quad \text{(S5)}$$

$$\rho_{xy}(H) = \frac{1}{e}\left[\frac{n\mu_e^2(1+\mu_h^2 H^2) - p\mu_h^2(1+\mu_e^2 H^2)}{n^2\mu_e^2(1+\mu_h^2 H^2) + p^2\mu_h^2(1+\mu_e^2 H^2) + 2pn\mu_h\mu_e(1-\mu_h\mu_e H^2)} H\right] \quad \text{(S6)}$$

The inset of Fig. S27B shows the magnetoresistance fit for low magnetic fields, < 1T, to obtain the carrier transport parameters. We obtain the hole and electron carrier concentration, $p = 4.48 \times 10^{19}$ cm$^{-3}$ and $n = 9.00 \times 10^{17}$ cm$^{-3}$ with corresponding hole and electron mobilities of $\mu_p = 18$ cm$^2$ V$^{-1}$ s$^{-1}$ and $\mu_n = 2516$ cm$^2$ V$^{-1}$ at 300 K for AgSb$_{1.05}$Te$_{2.06}$.

Figure S28A shows the Hall resistivity as a function of magnetic field, while Figure S26B shows the magnetoresistance at 300 K for both AgSb$_{1.01}$Cd$_{0.04}$Te$_{2.06}$ and AgSb$_{1.01}$Cd$_{0.04}$Te$_{1.86}$Se$_{0.2}$ at 300 K. The inset in Figure S28A shows a linear fit of the Hall resistivity at low magnetic fields (B < 1 T) used to extract the Hall coefficient R$_H$ (-1/nq), which is negative and equal to -0.183 m$^3$/C, indicating electron-dominated Hall transport. Despite this, both doped samples exhibit a positive Seebeck coefficient.

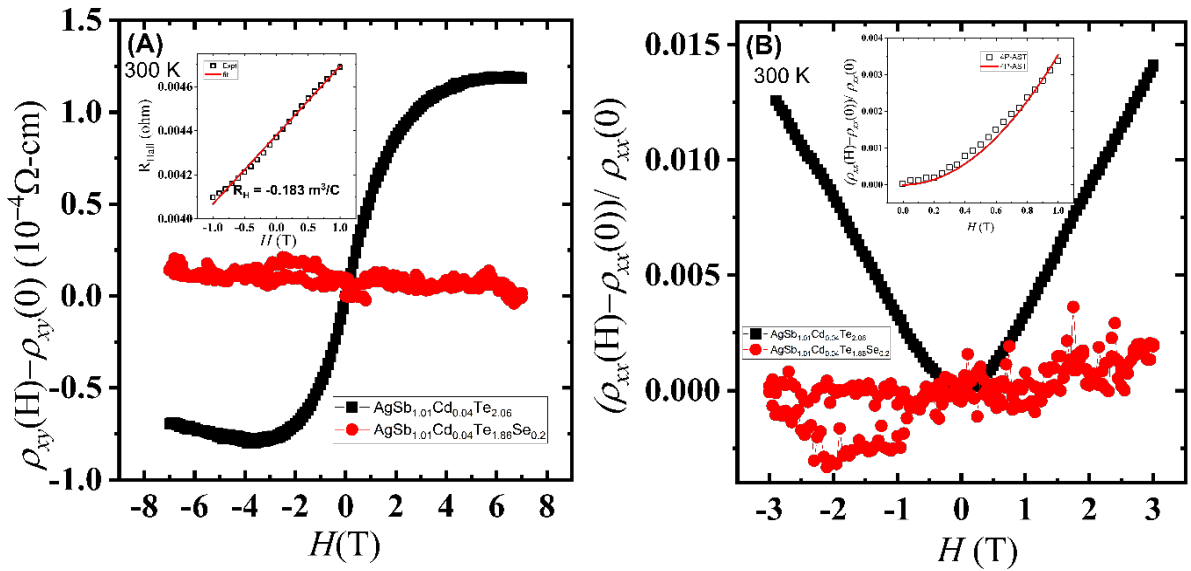

**Fig. S28.** **(A)** Change in the Hall resistivity and **(B)** change in magnetoresistance for both AgSb$_{1.01}$Cd$_{0.04}$Te$_{2.06}$ and AgSb$_{1.01}$Cd$_{0.04}$Te$_{1.86}$Se$_{0.2}$ at 300 K. The inset in **(A)** is a fit to Hall resistance to obtain the Hall coefficient for AgSb$_{1.01}$Cd$_{0.04}$Te$_{2.06}$. The inset in **(B)** is the 2 carrier model data fit data for low magnetic fields < 1T for AgSb$_{1.01}$Cd$_{0.04}$Te$_{2.06}$.

Fig. S28B shows a positive magnetoresistance for AgSb$_{1.01}$Cd$_{0.04}$Te$_{2.06}$. The change in magnetoresistance decreases in the Cd-doped sample compared to undoped AgSb$_{1.05}$Te$_{2.06}$. The inset of fig. S28B shows the magnetoresistance fit for low magnetic fields (< 1T), from which we obtain the hole and electron carrier concentration, $p = 2.37 \times 10^{19}$ cm$^{-3}$ and $n = 1.00 \times 10^{17}$ cm$^{-3}$ with corresponding hole and electron mobilities of $\mu_p = 20$ cm$^2$ V$^{-1}$ s$^{-1}$ and $\mu_n = 1410$ cm$^2$ V$^{-1}$ s$^{-1}$ at 300 K for AgSb$_{1.01}$Cd$_{0.04}$Te$_{2.06}$. Non-linearity in Hall resistivity and magnetoresistance decreased with Se doping. This suggests further suppression of electron carrier contributions to transport, accompanied by a reduction in magnetoresistance, indicative of decreased carrier–carrier scattering in the samples. The field-independent behaviour observed in both Hall resistivity and magnetoresistance may be attributed to an increase of pseudo-bandgap, reflecting lower contributions from narrow-band electrons and a transition from a two-carrier transport regime to a single (hole) carrier dominated transport upon Se substitution.[26,27] Furthermore, Cd and Se doping, suppresses the formation of n-type Ag$_2$Te precipitates as evidenced from our spectroscopic studies.[28]

**Supplementary Section S6: Repeatability of the thermoelectric properties**

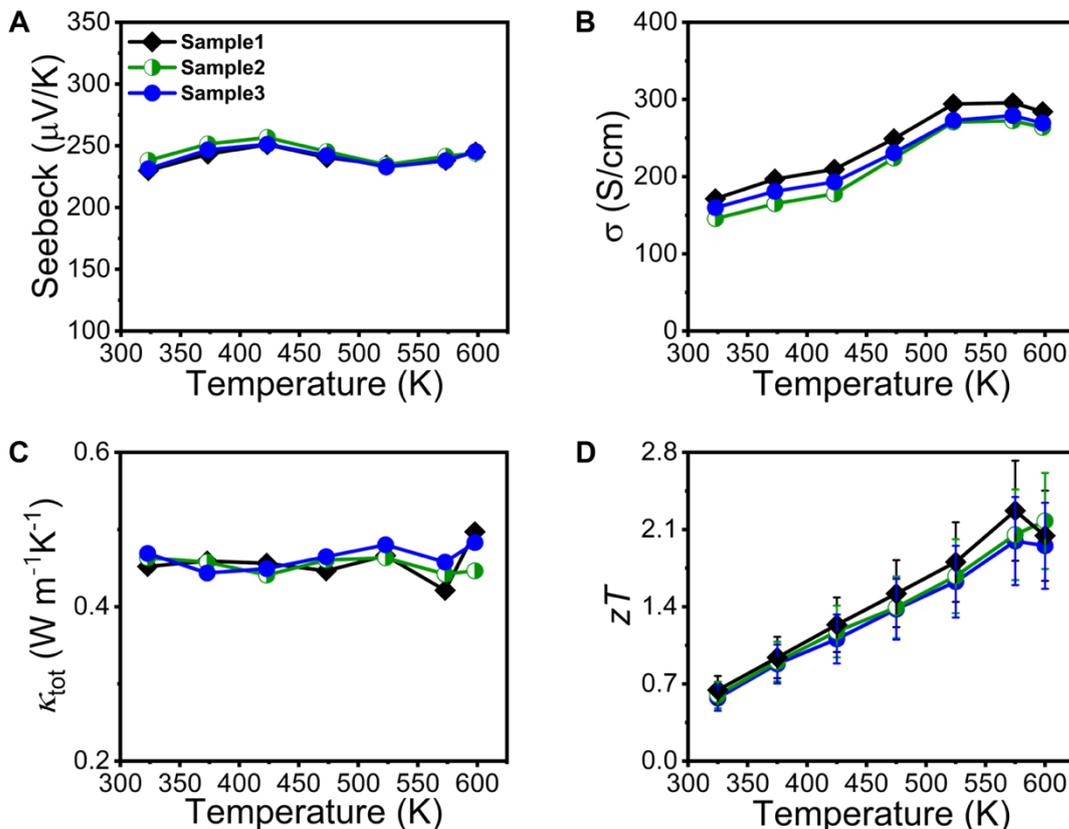

**Figure. S29.** Reproducibility of temperature-dependent thermoelectric properties Se co-doped AgSb$_{1.01}$Cd$_{0.04}$Te$_{1.86}$Se$_{0.2}$. (A) Seebeck coefficient, (B) electrical conductivity, and (C) thermal conductivity, (D) $zT$. A total of three different samples were measured.

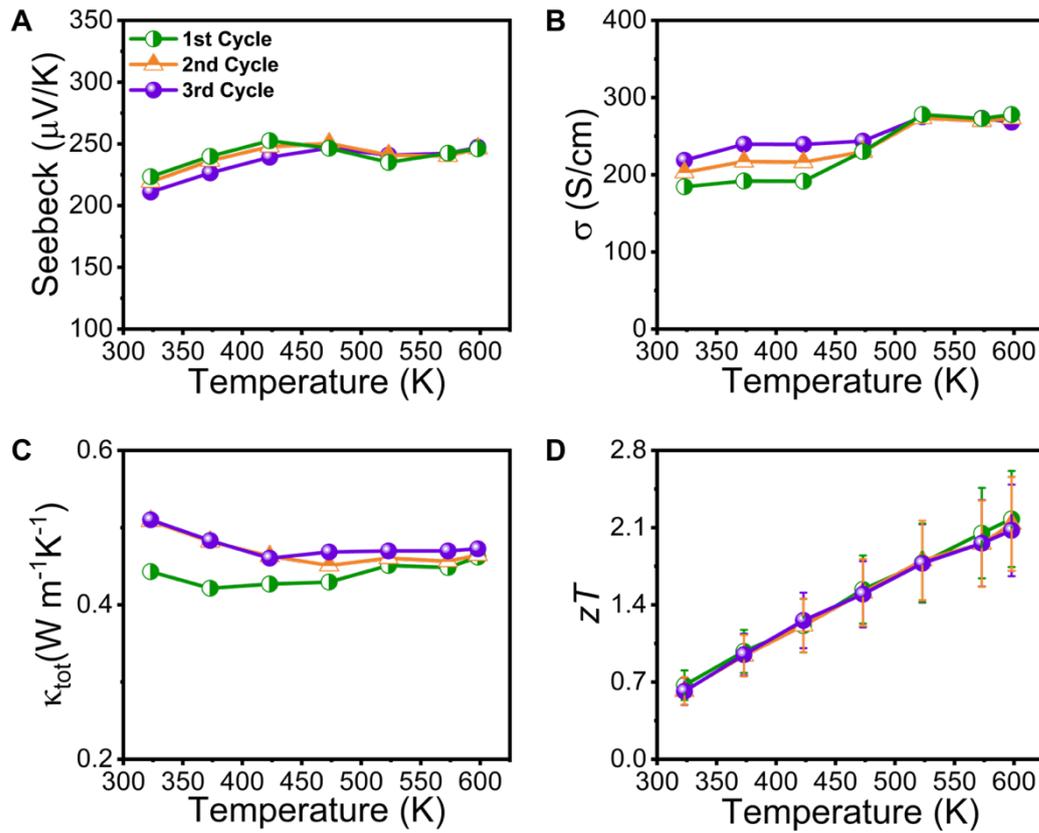

**Figure. S30.** Thermoelectric properties of polycrystalline Se co-doped AgSb$_{1.01}$Cd$_{0.04}$Te$_{1.86}$Se$_{0.2}$ sample with several heating-cooling cycles.

**Supplementary Section S7: Complementary structural and microstructure analysis for $Bi_{0.5}Sb_{1.5}Te_3$.**

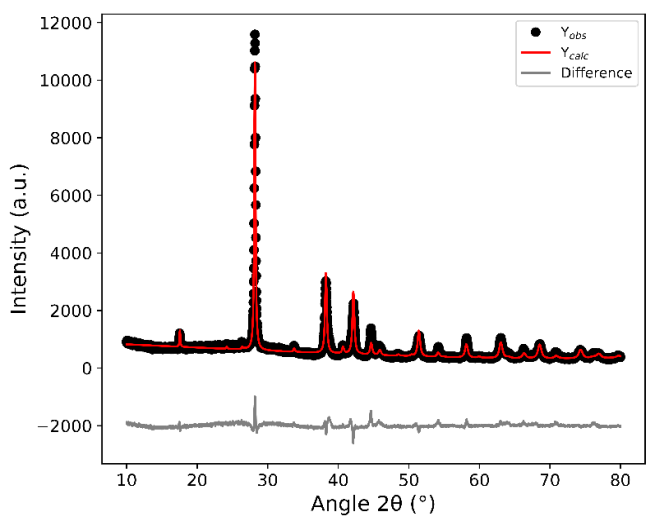

**Figure. S31.** Rietveld refined diffraction pattern for $Bi_{0.5}Sb_{1.5}Te_3$. The final $R_{wp}$ of the refinement was 3.18. Peaks were refined against $Bi_{0.5}Sb_{1.5}Te_3$ (ICSD 184246). Minority impurities matched to $Sb_2Te_3$ (PDF 01-083-5987) were also found.

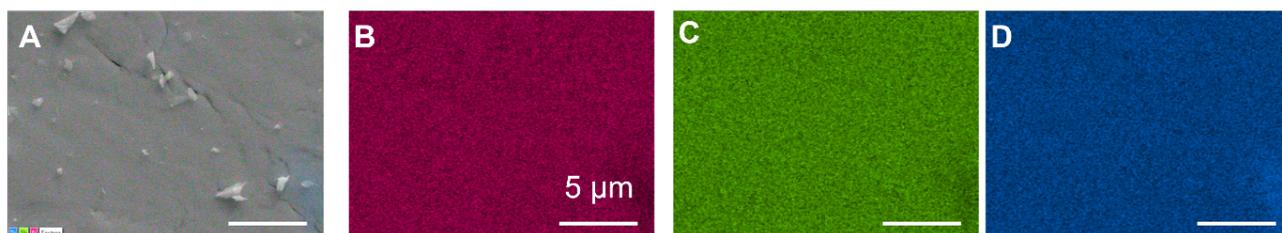

**Figure. S32.** SEM-EDX for $Bi_{0.5}Sb_{1.5}Te_3$. **(A)** Layered image. **(B)** EDX map for Bi. **(C)** EDX map for Sb. **(D)** EDX map for Te.

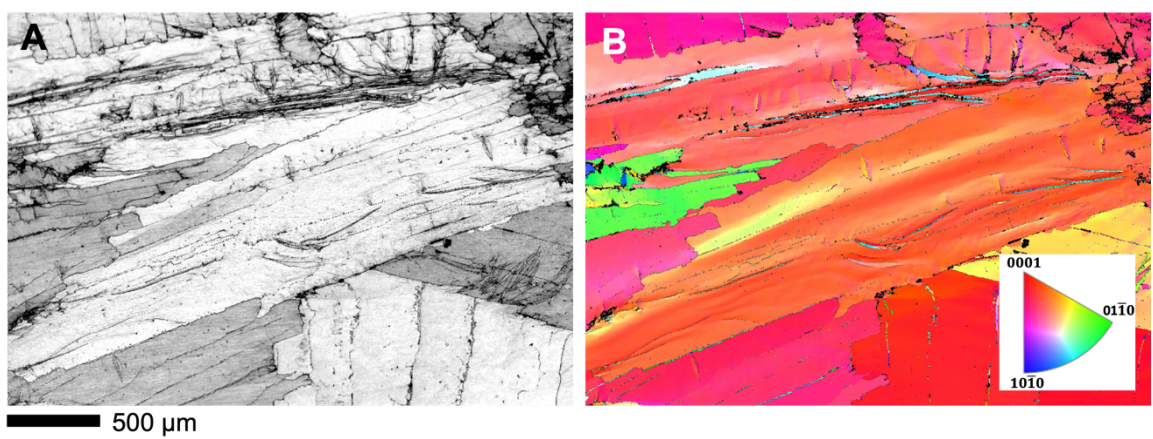

**Fig. S33.** EBSD for $Bi_{0.5}Sb_{1.5}Te_3$. **(A)** Low magnification band contrast image. **(B)** Inverse pole figure (IPF) map corresponding to (A).

**Supplementary Section S8: First principles electron band structure and phonon calculations and detailed analysis of electronic transport**

- **First principles calculations**

Fig. S34 shows the calculated phonon dispersion curves for ordered AgSbTe$_2$ in the $Fd\bar{3}m$ space group. Phonon frequencies were computed using the PBEsol+U functional, showing no imaginary (negative) frequencies across the Brillouin zone, indicating dynamical stability of the $Fd\bar{3}m$ structure. The phonon dispersions show avoided crossings between acoustic and optical modes, providing channels for phonon scattering that reduce lattice thermal conductivity. Additionally, the presence of soft, densely packed optical modes suggests strong phonon-phonon interactions and pronounced anharmonicity in the crystal lattice, which further suppress heat transport.

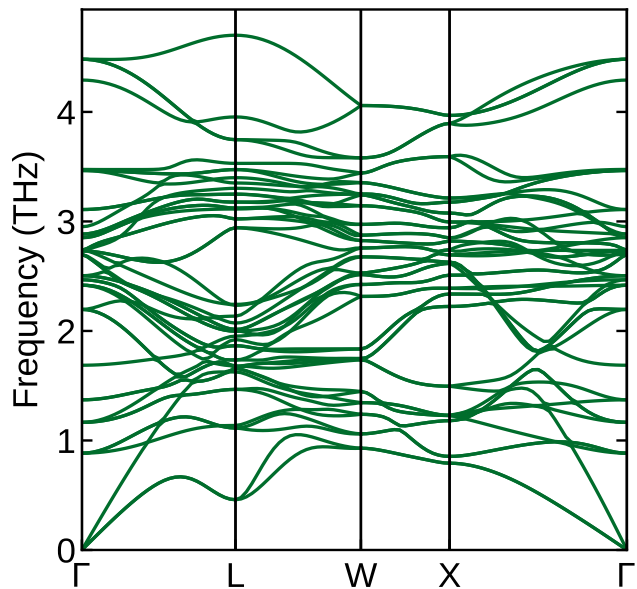

**Fig. S34.** Phonon dispersions of ordered AgSbTe$_2$ with $Fd\bar{3}m$ space group in a 3 x 3 x 3 supercell.

The electronic density of states (DOS) of ordered $Fd\bar{3}m$ AgSbTe$_2$ reveals that the valence band is predominantly composed of Te p-states, with a secondary contribution from Sb p-states and Ag d-states (Fig. S35). The conduction band minimum is mainly derived from Sb and Te p-states, with a minor presence of Ag s- and p-states. The total DOS exhibits a sharp onset at the conduction band edge and a broad peak in the valence band just below the Fermi level, indicative of a high density of available states for hole transport. This distribution suggests that electronic transport in the valence band is likely dominated by Te p-character, while the conduction band transport is governed by mixed Sb–Te p-character. The contribution from Ag is primarily localised and deep in the valence region, consistent with a limited role in conduction but potentially relevant for bonding and structural stability.

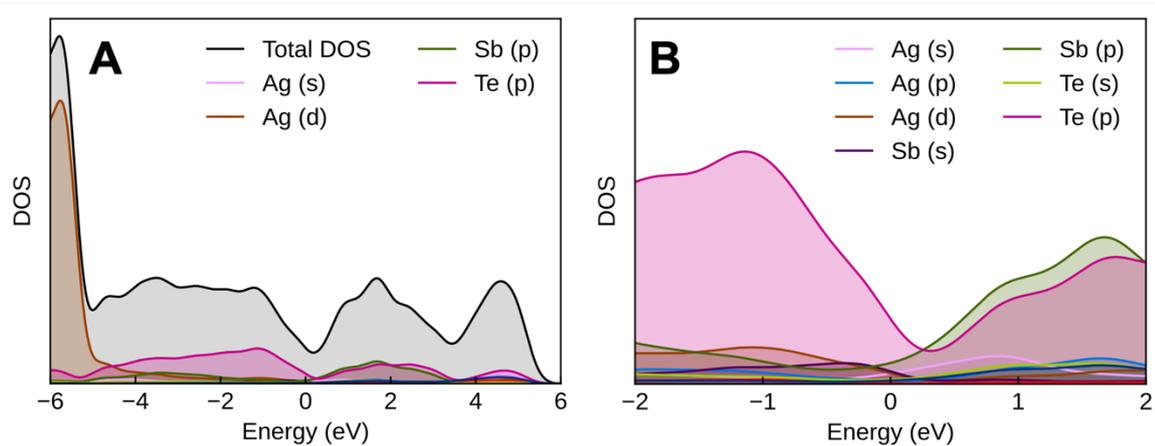

**Fig. S35. (A)** Electronic density of states of ordered $Fd\bar{3}m$ AgSbTe$_2$ calculated using PBEsol + U with spin-orbit coupling (SOC). **(B)** Zoomed-in region showing the DOS around the Fermi level.

Compared to the GGA-based results (Fig. S35), the DOS of ordered $Fd\bar{3}m$ AgSbTe$_2$ calculated using the hybrid functional show the formation of a bandgap (Fig. S35). The conduction band edge is shifted to higher energies, resulting in a clearer separation between the valence and conduction bands. This correction is consistent with the known underestimation of bandgaps by semi-local DFT and supports the use of hybrid functionals for a more accurate description of the electronic structure relevant to transport properties.

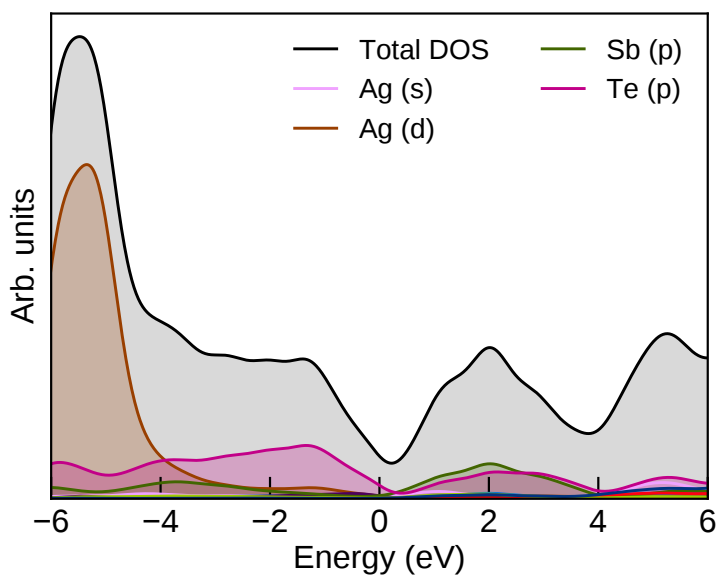

**Fig. S36.** Electronic density of states of ordered $Fd\bar{3}m$ AgSbTe$_2$ using hybrid functional and spin-orbit coupling (SOC).

Whilst a bandgap can be inferred from Fig. S36, it cannot be definitively confirmed from the DOS plot alone. This is because DOS plots have finite smearing (e.g. Gaussian broadening), which can artificially populate the gap region, making it difficult to determine whether the DOS at the Fermi level is truly zero. The apparent gap in the DOS could still correspond to a very narrow pseudo-gap. To definitively confirm the presence and magnitude of a bandgap, the band structure must be examined directly, however, has yet to be achieved for this structure.

The overall shape of the DOS in Fig. S36 remains qualitatively similar to the results in Fig. S35, with Te p-states dominating the valence band and Sb p-states contributing near the conduction band edge. However, the reduced overlap between valence and conduction states may influence carrier excitation and transport, particularly at elevated temperatures. The Ag d-states remain localised well below the Fermi level, and their contribution does not change significantly with the use of the hybrid functional, reinforcing their limited role in charge transport.

Based on the band structure, the valence and conduction bands overlap at the Fermi level (0 eV) and there is no clear gap between occupied and unoccupied states across the Brillouin zone (Fig. S37). This confirms that ordered $Fd\bar{3}m$ AgSbTe$_2$ is metallic at the PBEsol level. The zoomed-in panel shows that the conduction band dips below the Fermi level near X, while the valence band remains high between Γ and L indicating a band overlap rather than a bandgap.

In contrast to the hybrid functional result (Fig. S36), PBEsol fails to open a gap, characterising the system as metallic. This discrepancy reinforces that semi-local functionals like PBEsol are insufficient for predicting the correct electronic ground state in this material, and hybrid DFT is necessary to capture the semiconducting nature of the ordered phase.

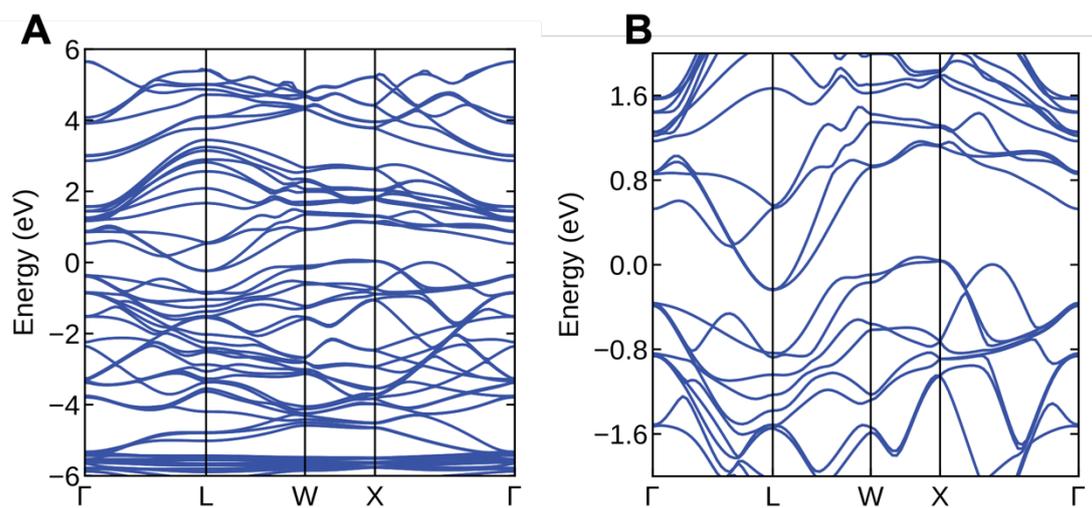

**Fig. S37. (A)** Electronic band structure of ordered AgSbTe$_2$ in $Fd\bar{3}m$ structure calculated using PBEsol + U with spin-orbit coupling (SOC). **(B)** Enlarged section of (A).

The unfolded hybrid-functional band structure of disordered $Fm\bar{3}m$ AgSbTe$_2$ shows that the material appears to be semi-metallic (Fig. S38). There is no clear band gap is visible at the Fermi level (0 eV). Instead, the valence and conduction bands overlap in energy across different k-points. The valence states (mostly Te-derived) cross or touch the Fermi level near Γ–L, and the conduction states (mainly Sb-derived) are also present at or just below the Fermi level, particularly around X. The spectral weight is non-zero at multiple k-points at E=0, indicating a finite density of states at the Fermi level despite the band smearing from configurational disorder.

This strongly suggests that the disordered $Fm\bar{3}m$ phase is semi-metallic at the hybrid DFT level. In contrast to the ordered $Fd\bar{3}m$ phase, which may develop an indirect band gap with hybrid functionals, disorder in $Fm\bar{3}m$ appears to close or significantly reduce the gap, leading to band overlap. This distinction has important implications for transport: carrier concentrations in the disordered phase would be intrinsic and metallic-like rather than thermally activated across a gap.

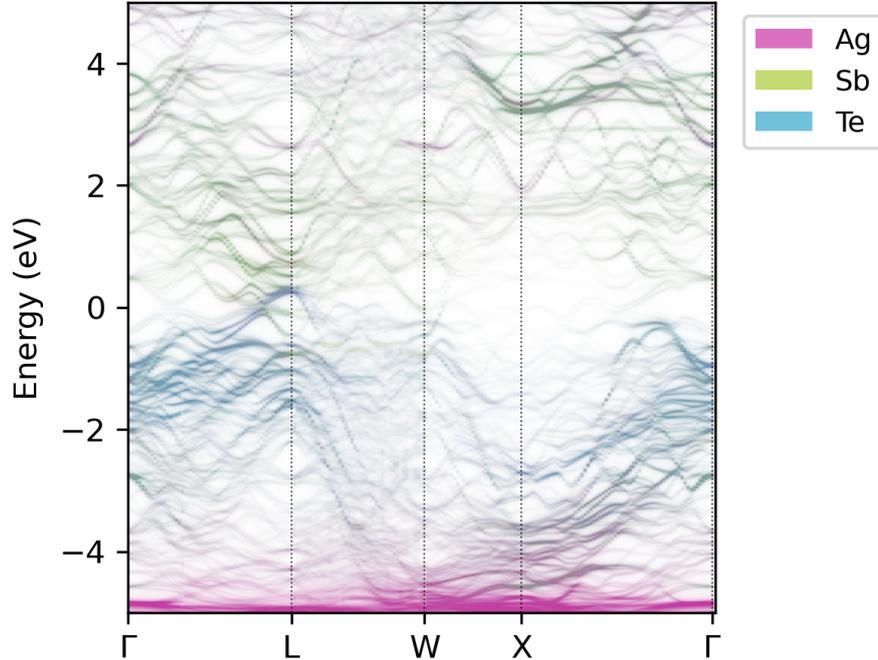

**Fig. S38.** Electronic band structure of disordered $Fd\bar{3}m$ AgSbTe$_2$, calculated using the HSE06 functional and SOC. The spectral weights are colour-resolved by atomic species: Ag (magenta), Sb (green), and Te (blue), showing the projected contribution of each element to the electronic states.

We proceed to use the most stable structure (at DFT level) to perform first principles calculations of the electronic transport properties. A *C2/c* unit cell was used as input for calculation using the *Ab initio* Scattering and Transport (AMSET) method, and the preliminary data is shown in Fig. S39. We observe that our experimentally measured data (in purple, Fig. S39) approaches a carrier concentration between $10^{19}$ and $10^{20}$ cm$^{-3}$, which is in good agreement with our Hall analysis.

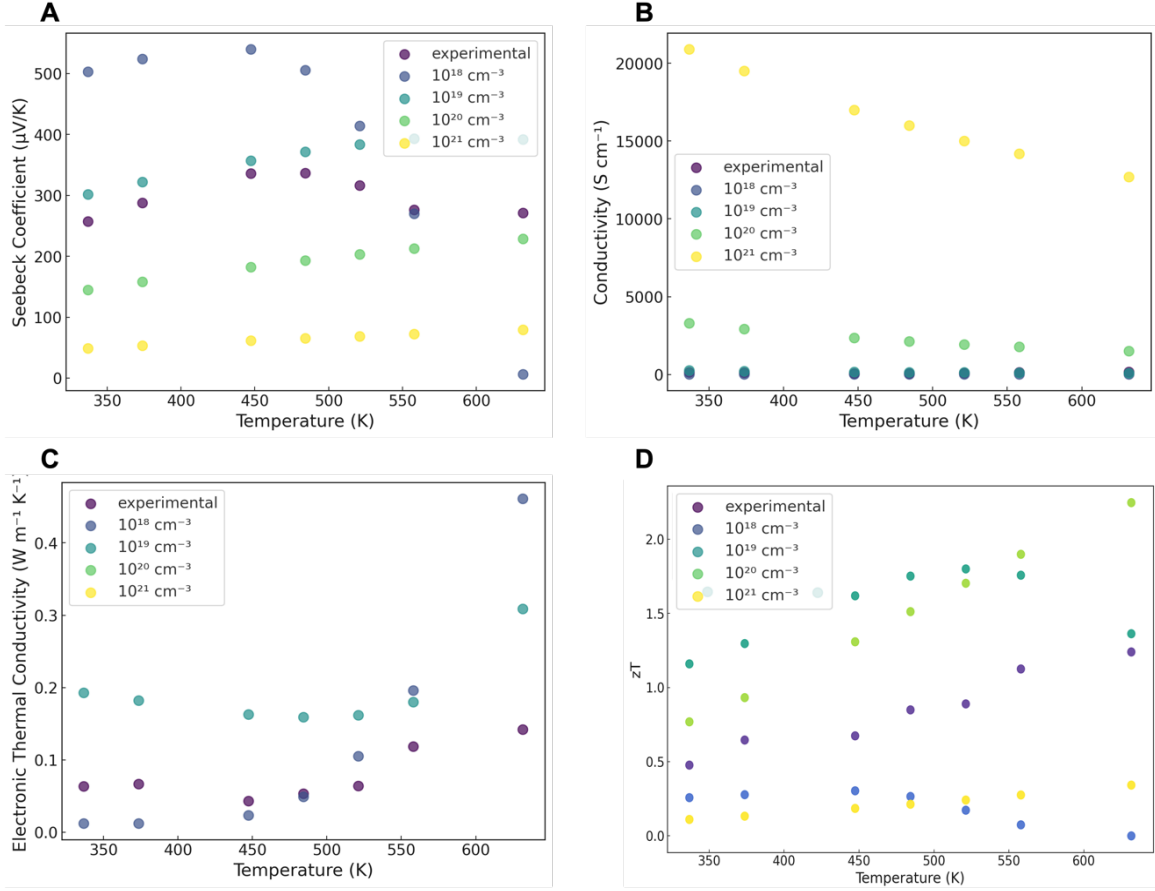

**Figure S39. Comparison between experimental results and theoretical calculations for ordered AgSbTe$_2$.** **(A)** Seebeck coefficient. **(B)** Electrical conductivity. **(C)** Electronic thermal conductivity. **(D)** $zT$. The experimental points correspond to data collected for undoped AgSb$_{1.05}$Te$_{2.06}$.

In order to obtain Jonker's plot as shown in Figure S40 (A-D), the reduced chemical potential, $\eta$ was extracted from equation S1:

$$S = \frac{k_B}{e}\left[\frac{\left(r+\frac{5}{2}\right)F_{r+1.5}(\eta)}{\left(r+\frac{3}{2}\right)F_{r+0.5}(\eta)} - \eta\right] \tag{S1}$$

where F is the Fermi integral. Theoretical Seebeck coefficient value was calculated using $\eta$ and different *r* values such as for acoustic phonon scattering (APS, $r$ = -0.5), polar optical scattering (POP, r=+0.5) and ionized impurity scattering (IIS, $r$ = 1.5). Subsequently, the corresponding *η and r values* can be used to obtain theoretical electrical conductivities $\sigma$ for different transport coefficient values using Eq. S3.

$$\sigma = \sigma_{E_0}(T)\,(r+1.5)\,F_{r+0.5}(\eta) \tag{S3}$$

The theoretical curves of Seebeck coefficient versus electrical conductivity with different scattering parameters were plotted as shown in Figure S40 (A-D). $\sigma_{E_0}$ can be estimated by looking the position and trend of the experimental data from each sample.

Figure S40(E-F) shows the theoretical $zT$ values as a function of $\eta$ for different B values. $zT(B, \eta)$ can be written as:

$$zT(B,\eta) = \left[ \frac{S^2(\eta)}{\left( \frac{\left(\frac{k_B}{e}\right)^2}{B(r+1.5)F_{r+0.5}(\eta)} + L(\eta) \right)} \right] \tag{S7}$$

where $L(\eta)$ is Lorentz number given by Eq. S4. Theoretical curve of $zT$ versus $\eta$ plotted at different $B$ values. Experimental data of $zT$ for $AgSb_{1.05-x}Cd_xTe_{2.06}$ samples and $AgSb_{1.01}Cd_{0.04}Te_{2.06-y}Se_y$ samples at 573 K were placed on those theoretical curves. As seen for a particular B value, experimental $zT$ value is close to optimized $\eta$.

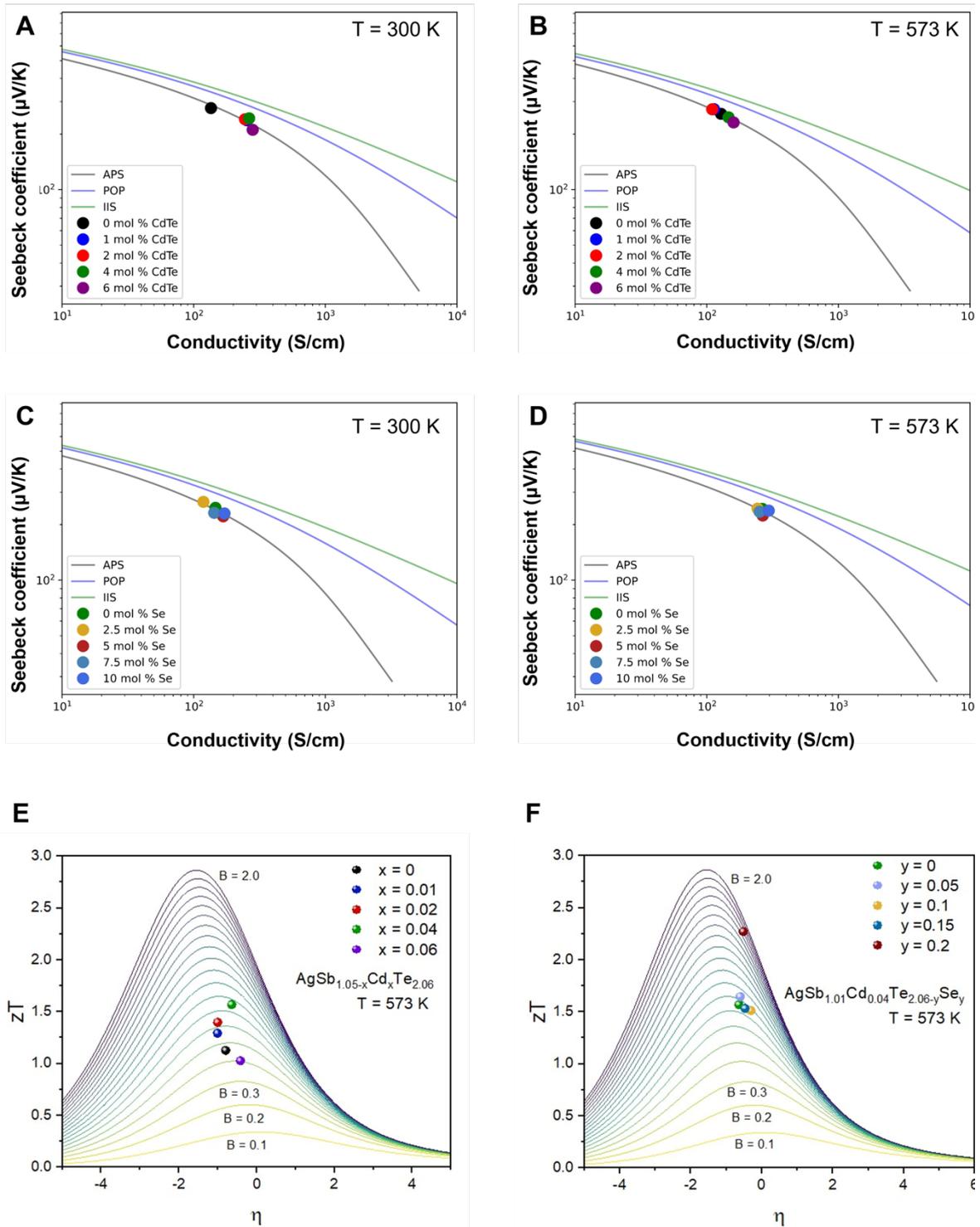

**Figure S40. Electronic transport analysis.** Jonker plots for Cd-doped AgSb$_{1.05-x}$Cd$_x$Te$_{2.06}$ samples at (A) 300 K, (B) at 573 K and for Se co-doped AgSb$_{1.01}$Cd$_{0.04}$Te$_{2.06-y}$Se$_y$ samples (C) 300 K and (D) 573 K. zT as a function of $\eta$ at different B values for (E) Cd-doped AgSb$_{1.05-x}$Cd$_x$Te$_{2.06}$ samples and (F) for Se co-doped AgSb$_{1.01}$Cd$_{0.04}$Te$_{2.06-y}$Se$_y$ samples.

For both Cd-doped $AgSb_{1.05-x}Cd_xTe_{2.06}$ samples and Se co-doped $AgSb_{1.01}Cd_{0.04}Te_{2.06-y}Se_y$ it is clear from Fig S40 A-D that APS dominates at room temperature and above.

The values of $\sigma_{E_0}$ for selected samples are benchmarked with the literature in Fig. S42. $\sigma_{E_0}$ describes the conductive "quality" of charge carriers in the material (magnitude of conductivity for a given value of h). As temperature increases, $\sigma_{E_0}$ increases, which indicates good material quality. Compared to the state-of-the-art material, $AgSb_{0.94}Cd_{0.06}Te_2$, we observe that both at room temperature and 573 K our values are lower. This could be attributed to lower carrier mobility and would also explain why our overall thermoelectric performance is lower.

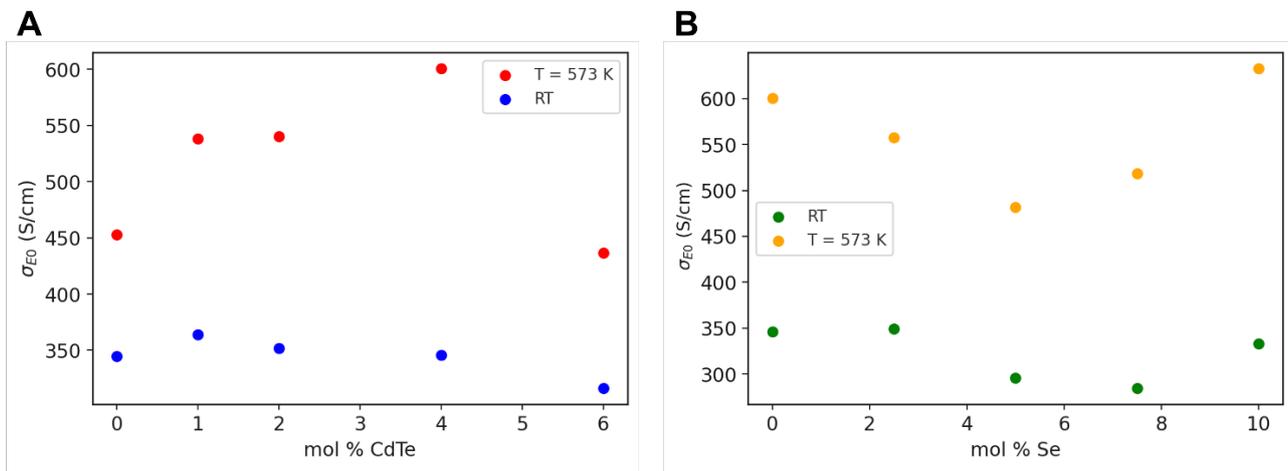

**Figure. S41.** Energy-independent transport parameter ($\sigma_{E_0}$) as a function of doping for room temperature (RT, 300 K) and 573 K. **(A)** Cd-doped $AgSb_{1.05-x}Cd_xTe_{2.06}$ samples at room temperature and 573 K. **(B)** Se co-doped $AgSb_{1.01}Cd_{0.04}Te_{2.06-y}Se_y$ samples at room temperature and 573 K.

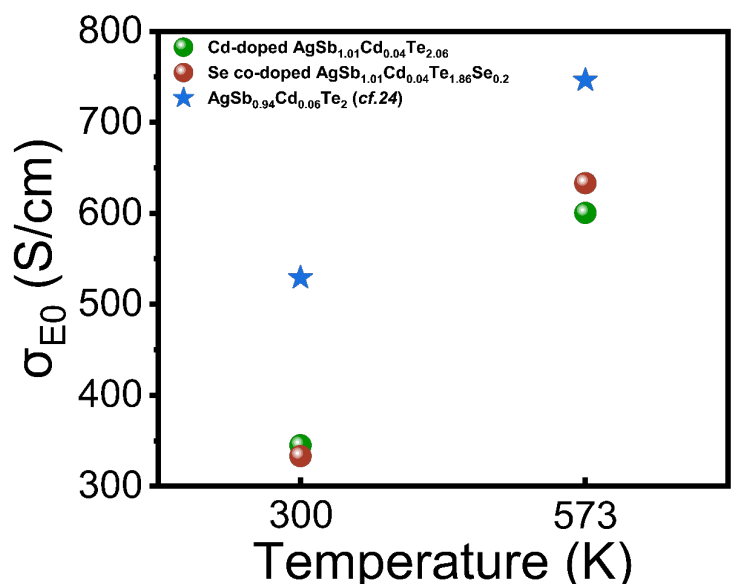

**Figure. S42**. **(A)** Comparison of $\sigma_{E_0}$ of Cd-doped $AgSb_{1.01}Cd_{0.04}Te_{2.06}$ and Se co-doped $AgSb_{1.01}Cd_{0.04}Te_{1.86}Se_{0.2}$ with 6 mol % Cd doped $AgSb_{0.94}Cd_{0.06}Te_2$ [24]

**Supplementary Section S9: Further TDTR Measurements and Boundary determination of thermal mapping.**

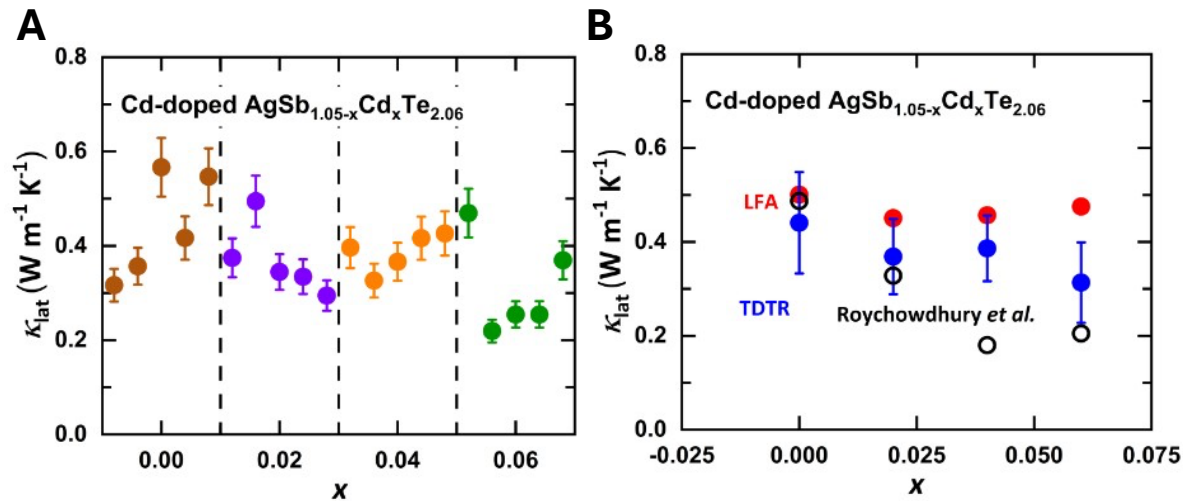

**Figure. S43. More thermal conductivity measurements. (A)** Variation in the lattice thermal conductivity ($\kappa_{lat}$) measured by TDTR. For each sample, TDTR measurements are performed at 4-5 random locations. The uncertainty of each measurement is estimated to be 11%. **(B)** $\kappa_{lat}$ of Cd-doped AgSb$_{1.05-x}$Cd$_x$Te$_{2.06}$ series measured by TDTR (blue) and LFA (red), compared to data from Roychowdhury *et al*. (black).[24] The blue circles and the error bars are the mean values and the standard deviations of 4 – 5 TDTR measurements at random locations, respectively.

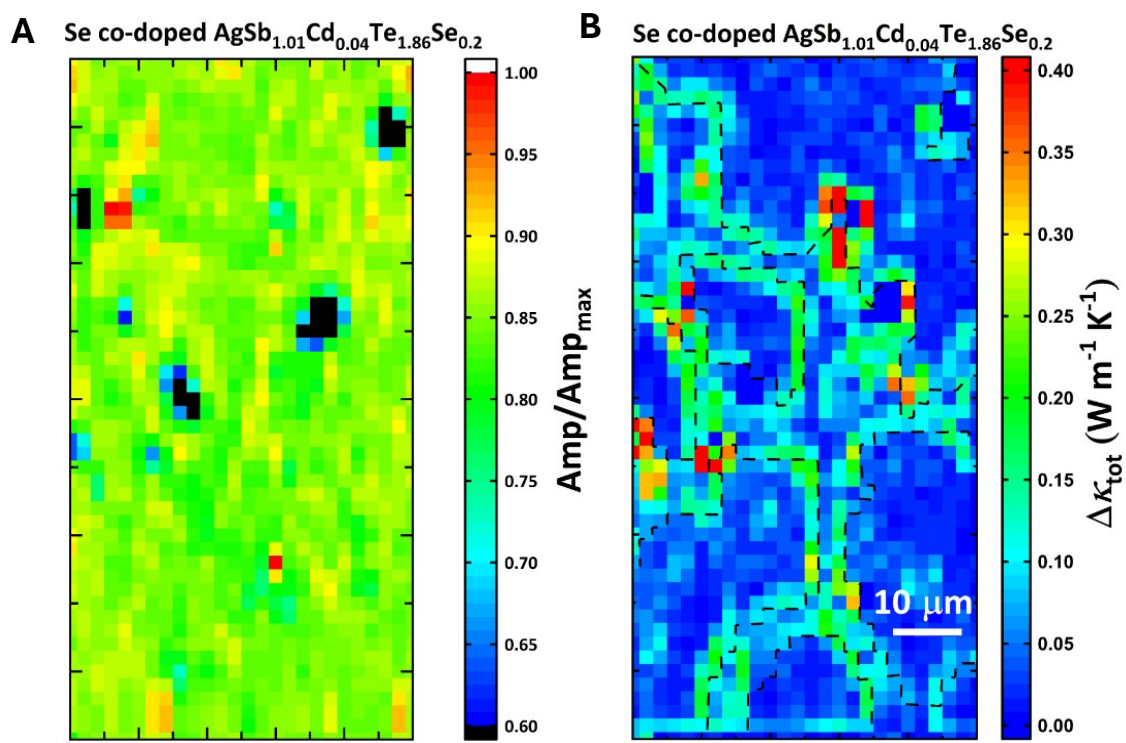

**Figure. S44. Boundary determination of thermal mapping** (A) Percentage of maximum amplitude of the TDTR signal $\sqrt{V_{in}^2 + V_{out}^2}$. Points below 60% (black) are considered affected by surface defects. (B) $\Delta\kappa_{tot}$ mapping of the Se-codoped AgSb$_{1.01}$Cd$_{0.04}$Te$_{1.86}$Se$_{0.2}$ sample. The mapping is derived from Figure. 5D by calculating the maximum absolute difference between each point and its right, top, and top-right neighbours.

To determine the boundaries of the thermal mapping, we first excluded the points affected by surface defects. We analysed the signal amplitude $\sqrt{V_{in}^2 + V_{out}^2}$ of each point, where $V_{in}$ and $V_{out}$ are the in-phase and out-of-phase components of the TDTR signal, respectively. Then, we excluded those that failed to reach 60% of the maximum amplitude. Then, we plotted $\Delta\kappa_{tot}$ mapping, as shown in Figure S44B. In this mapping, we considered the top-right corner of each point with high $\Delta\kappa_{tot}$ as the boundary between the matrix and the ILs. We connected these corners and then transferred this boundary to Figure. 5D to calculate the average $\kappa_{tot}$ of the matrix and the ILs.

**Supplementary Movie Legends**

**Movie S1. Typical DJS process for undoped AgSbTe$_2$.** The corresponding time *vs.* temperature profile for the reaction can be found in Fig. 1C.

**Supplementary References**